\def\RE {I\kern-6pt R    }
\def\Z  {Z\kern-13pt Z   }
\def\be {\begin{equation}}
\def\ee {\end{equation}  }
\def\beq{\begin{eqnarray}}
\def\eeq{\end{eqnarray}  }

\def\bi {\begin{itemize} }

\def\ei {\end{itemize}   }
\def\kkbar {$KK^{^{^{_{\! \! \! \! \! \! \rule{2.4mm}{0.25mm}}}}}$ }
\def\kbar   {$K^{^{^{_{\! \! \! \! \! \! \rule{2.4mm}{0.25mm}}}}}$ }
\def\sech { {\rm sech} }
\newcommand{\scri}{\mathcal {J}}

\documentclass[11pt]{report}         
\usepackage {utthesis}               
\usepackage {latexsym}               
\input epsf
\def\be {\begin{equation}}
\def\ee {\end{equation}  }
\def\ba {\begin{array}}
\def\ea {\end{array}  }
\def\bc {\begin{center}}
\def\ec {\end{center}  }
\def\beq{\begin{eqnarray}}
\def\eeq{\end{eqnarray}  }
\def\bd{\begin{description}}
\def\ed{\end{description}}

\def\CQG{{\it Class.\ Quantum Grav.\ }}
\def\JMP{{\it J Math.\ Phys.\ }}

\def\PRD{{\it Phys.\ Rev.} D\ }
\def\PRL{{\it Phys.\ Rev.\ Lett.\ }}

\def\JCP{{\it  J.\ Comput.\ Phys.\ }}
\def\PL{{\it Phys. Lett.\ }}

\font\bfgreek=cmmib10

\def\bbx{{\hbox{\bfgreek\char'170}}}

 \phdthesis                         

 \doublespace                        

\renewcommand{\thesisauthor}{Ethan Philip Honda}    

\renewcommand{\thesismonth}{August}   

\renewcommand{\thesisyear}{2000}      

\renewcommand{\thesistitle}{ Resonant Dynamics within the nonlinear\\ Klein-Gordon Equation: 
	\hbox{\it Much ado about Oscillons} }

\renewcommand{\thesisauthorpreviousdegrees}{B.S.}

\renewcommand{\thesissupervisor}{Matthew William Choptuik \\ \ \ \ \ \  Philip J. Morrison}

\renewcommand{\thesisauthoraddress}{7507 N. Lowell Ave. \\ Skokie, IL 60076\\ USA}

\renewcommand{\thesisdedication}{To Trish...}

\renewcommand{\thesiscommitteesize}{6}

\begin{document}

\thesiscopyrightpage                 

\thesistitlepage                     

\thesissignaturepage                 

\thesisdedicationpage                

\begin{thesisacknowledgments}        
\small

I sincerely thank my supervisor Matt Choptuik for all of his help, guidance, and 
support over the last four years.
I appreciate the  patience and encouragement he showed me and thank him for 
continuing to supervise me despite the logistical challenges.

I thank Philip J. Morrison for his help on the trial function methods 
and for becoming my advisor-of-record.

I am indebted to my friends (many in the Center for Relativity) 
for sharing their wisdom and knowledge, both about numerical relativity and life.  

Finally, I pay special respect and give all my thanks to my family for their love and 
support over the last twenty-five years.  Mom, Dad, Graham, Grandpa and Gary,  
I would not be here now if it weren't for the sacrifices you have made.
I also 
thank my fiance, Trish, for all her patience, support, and encouragement.

I also would like to acknowledge financial support from the 
National Science Foundation grant PHY9722088, and from a Texas Advanced Research
Projects grant. 
The bulk of the computations described here were carried out 
on the {\tt vn.physics.ubc.ca} Beowulf cluster, which was 
funded by the Canadian Foundation for Innovation, with 
operations support from the Natural Sciences and Engineering 
Research Council of Canada, and the Canadian Institute for 
Advanced Research.
Some computations were also carried out using the 
Texas Advanced Computing Center's
SGI Cray SV-1 {\tt aurora.hpc.utexas.edu} and SGI Cray T3E 
{\tt lonestar.hpc.utexas.edu}.

\normalsize
\end{thesisacknowledgments}          

\begin{thesisabstract}               

This dissertation discusses solutions to the nonlinear Klein-Gordon (nlKG) equation
with symmetric and asymmetric double-well potentials, focusing on the 
collapse and collision of bubbles and critical phenomena found therein.

A new method is presented that allows the solution of massive field equations
on a (relatively) small static grid. 
A coordinate transformation is used that transforms typical 
flatspace coordinates to coordinates that move outward 
(near the outer boundary) at nearly the speed of 
light.  The outgoing radiation is
compressed to nearly the Nyquist limit of the grid where it 
is quenched by dissipation.   
The method is implemented successfully in both spherically symmetric and
axisymmetric codes.

The new method is first used in a code to explore spherically 
symmetric bubble collapse.
New resonant oscillon solutions are found within the solution space of the
nlKG model with a symmetric double-well potential (SDWP).  A time-scaling relation 
is found to exist for the lifetime of each resonance.
The resonant solutions are also obtained independently using 
a set of ordinary differential equations derived from a non-radiative  
periodic ansatz. 
The method is also applied to the nlKG model with an asymmetric double-well potential 
(ADWP); the threshold of expanding bubble formation is 
investigated and a time-scaling law is shown to exist.

The method is then used in an axisymmetric code to simulate bubble collisions.
A technique for boosting arbitrary spherically symmetric finite difference 
solutions is presented and used to generate initial data for the collisions.
The 2D parameter space of bubble width versus collision velocity is
explored and the threshold of expanding bubble formation is again
considered.
On the threshold, there exists a time-scaling law with critical 
exponent similar to the spherically symmetric case.

Lastly, resonant oscillon solutions are constructed using trial function methods
and variational principles.  The solutions are found to be consistent
with the dynamical evolutions.


\end{thesisabstract}                 

\tableofcontents                     
\listoftables                      
\listoffigures                     

\chapter{Introduction}

In our everyday lives, we typically think of bubbles as objects that 
separate two different phases of matter from one another
(like the bubbles you see in boiling water that separate steam from 
liquid water).
Within this work, however, a bubble is something a little more 
abstract: it is any scalar field configuration 
that interpolates between the minima of a double-well
potential.
These scalar field  bubbles are still closely analogous to boiling water,  
where the minima of the double-well potential act like the two different
states of matter.
Although an emphasis is placed on cosmological bubbles, 
the model can be applied to other branches of
physics modeling phenomena ranging  from the polarization states in ferromagnets
to topological defects in superfluids (actually, anything described by 
Landau-Ginzburg theory of phase transitions!).
This work concerns itself primarily with the collapse
and collision of bubbles and to a special type of collapsed bubble 
known as an {\it oscillon}.

The dynamics of these bubbles are governed by the classical flatspace 
{\it nonlinear} Klein-Gordon (nlKG) equation.  
There is a long history in mathematics and physics of trying
to find new (non-trivial) solutions to nonlinear wave equations.
One such type of solution that is of interest to many is the 
soliton.   The first {\it scientific} discussion of 
solitons was due to J. Scott Russell and was published in the
Report of the British Association for the Advancement of Science,
in 1845.  The report describes the creation of a 
surface wave in a narrow shallow water channel following the abrupt stop
of a boat.  Russell followed the wave on horseback until ``after
a chase of {\it one or two miles} [he] lost it in the windings of 
the channel.''
Although he dubbed these localized nonlinear 
waves ``Waves of Translation'', they later came to
be known as solitary waves or solitons\footnote{
Although some people \cite{lee2:1981} consider a soliton to be 
``any spatially confined and nondispersive solution to a classical field 
theory'', many others would call such a solution a solitary wave, reserving
the term soliton to further include the ability for two such solutions to pass
through one another with only a phase shift or time lag \cite{rajaraman:1982}.}.

In the last fifty years, interest in solitons has been revived 
by many mathematicians and physicists.
Most particle physicists, for example, used to believe that for there to 
be bound (particle-like) states in a relativistic field theory, 
quantum theory had to be introduced.
However, this is not the case, since solitons are stable bound states of a 
nonlinear field theory that, heuristically, exist through the balance 
between a nonlinear attraction and a tendency to disperse.  
Few scientists
believe that quantum mechanics and field theory will ever 
be replaced entirely by solitonic interactions within classical 
field theory, but much research has gone into understanding 
how the existence of classical solitons implies the existence of 
a corresponding quantum solution \cite{lee2:1981}.
Although oscillons eventually disperse and are therefore 
{\bf not} solitons (which are stable bound states), they do remain
localized for large times and can pass through one another just like
classical solitons.

The background theory and history of the nlKG equation is the subject of 
Chapter 2.  After presenting the nlKG equation, typical bubble initial data and
the basics of bubble dynamics are discussed.
Previous investigations of kink/antikink soliton interactions
within the (1+1) dimensional $\lambda\phi^4$ model are then presented, and  
the chapter concludes with an introduction to oscillons (their
behavior and history).

Chapter 3 discusses the numerical techniques used throughout the thesis.
Since finite difference methods are used extensively in the dynamic
simulations a brief background is included.  A section on dissipation
is also included since the incorporation of dissipation is integral
to the success of the numerical methods employed.
The reader is then introduced to a new coordinate system that is used
to solve nonlinear wave equations on a (relatively small)
static lattice in one and two spatial dimensions.
The chapter concludes with a brief motivation for presenting the 
coordinates in the ``3+1'' or ADM form, \cite{arnowitt:1962},\cite{choptuik:1998a}.

Chapter 4 contains the discussion of spherically symmetric oscillons.
The finite difference equations used are introduced and the testing
of the code is discussed.  A new resonant solution within the 
symmetric double-well potential (SDWP) model
is presented; the mode structure of the solution is analyzed and a
time scaling law is shown to be present for the critical (resonant)
solution.
Lastly, the threshold of expanding bubble formation is explored within
the asymmetric double-well potential model and another (different) 
time scaling law is also shown to exist.

Chapter 5 discusses axisymmetric evolution of the nlKG equation
in the context of bubble {\it collisions}.  The finite difference
equations used are introduced and the testing of the code is
discussed.
The generation and testing of initial data is presented; 
the initial data is constructed from boosting two rest-frame oscillons 
(like those of chapter 4) at each other.
Parameter space surveys are conducted and the threshold of expanding
bubble formation is found to exhibit a time scaling law.

Trial functions and variational approaches to finding critical oscillon
solutions are the main ideas discussed in chapter 6.
A set of generic ordinary differential equations for critical non-radiative
oscillon solutions used in chapter 4 are rederived using trial function
methods.
With more constrained ansatz (gaussian and hyperbolic secant functions)
the same approach is used to directly obtain (ie. solving algebraic not 
differential equations) a few of the basic attributes of oscillons.

Finally, chapter 7 concludes the thesis with a summary of what was 
accomplished in this work.
An appendix is also included that discusses some 
basic oscillon attributes (with units!): size, shape, lifetimes, 
and whether or not they are expected to form black holes.

\section{Notations, Conventions, and Abbreviations}

Unfortunately, this work mixes a few conventions from different
branches of physics.  
The {\it metric signature} used, for example, is the one 
typical to (modern) general relativity, (-- + + + ).
However, the {\it units} used throughout this thesis are Plankian,
those commonly used by high-energy physicists where 
$\hbar=c=1$ (not those of the typical relativist where 
$G=c=1$!).  
If the lengths, lifetimes, and masses of oscillons considered 
throughout this thesis 
are left in terms of the dimensionful-scale in the model, $m$, with
dimensions $[L]^{-1}$, the
{\it $\hbar$ is not needed}.  However, if one asks for the mass
of a particular oscillon (particularly an early universe oscillon), 
one needs to include appropriate factors of $\hbar$ (or other appropriate 
dimensionful constant).
With luck, any confusion that arises regarding how to reinsert units
used can be dispelled by studying the examples in Appendix A.

Also, the term ``critical bubble'' is used throughout the cosmology
community to refer to a bubble (in a model with an asymmetric double-well
potential) whose radius 
is at, or above, the threshold for expanding bubble formation.  In other
words, the term refers to a bubble large enough that the volume energy 
driving the field to the true vacuum is greater than the surface 
tension trying to collapse the bubble.  Such bubbles will always expand 
and contribute to a phase transition.
However,  we choose to avoid the use of the word ``critical'' in such a 
generic way as it has quite a specific meaning in the study of
critical phenomena, where it describes a solution
that lies {\it exactly} on the threshold of the phase
transition being considered.
Therefore, instead of using the term ``critical bubble'', we tend to use
``expanding bubble''.

The models discussed here are often referred to as ($n$+1) dimensional.
($n$+1) refers to a system with $n$ spatial dimensions and 1 time-like 
dimension.  The symmetry of the model will also be included where possible 
to help distinguish between models with the same dimensionality but
different symmetries, eg. (1+1) plane-symmetric  and (1+1)
spherically symmetric.

Although defined throughout the thesis, 
we also note here the following frequently used abbreviations:
\begin{itemize}
\item{ADWP: asymmetric double-well potential}
\item{SDWP: symmetric double-well potential}
\item{nlKG: nonlinear Klein-Gordon}
\item{KG: Klein-Gordon}
\item{MIB: Monotonically Increasingly Boosted}
\item{FDA: Finite Difference Approximation}
\item{PDE: Partial Differential Equation.}
\item{CN: Crank Nicholson}
\end{itemize}

\chapter{Theory and Background}

This chapter presents a brief background of the history of the 
model to be studied throughout this thesis, the nonlinear 
Klein-Gordon model with double-well potentials.
We focus on the work by Campbell {\it et al} \cite{campbell:1983}, in the 1970's on the resonant 
structure found in the (1+1) dimensional (plane-symmetric) 
kink-antikink scattering with the symmetric double-well potential,
since many attributes of the plane-symmetric model are also found in the 
spherically symmetric case.
Although interesting to this author for their non-linear (mathematical) 
behavior alone, brief descriptions of possible {\it physical} applications 
are included throughout.

\section{The Nonlinear Klein-Gordon Equation and Bubbles\label{theory_bubform}}

Put simply, this thesis is devoted to studying a special type of solution
of the Nonlinear Klein-Gordon Equation,
\begin{equation}
\Box\phi = \frac{\partial V}{\partial \phi},
\end{equation}
where $\phi\equiv \phi(\vec{r},t)$ and the potential is of the form 
$\displaystyle{V(\phi) = -m^2\phi^2 + \alpha\phi^3 +\lambda\phi^4}$,
for $m$, $\alpha$, and $\lambda$ constant.
The potential and some sample initial data for $\alpha=0$ (the symmetric double well)
and for $\alpha\neq 0$ (the asymmetric double well) can be seen in figures 
\ref{fig:fieldprofiles_sdwp} and \ref{fig:fieldprofiles_adwp}, respectively.
\begin{figure}
\epsfxsize=13.5cm
\centerline{\epsffile{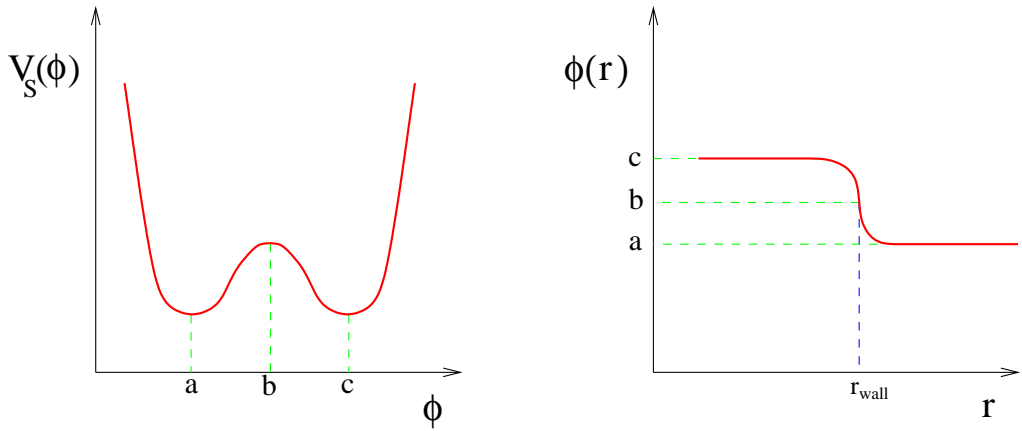}}
\caption[SDWP and kink profile initial data]
{\small \label{fig:fieldprofiles_sdwp}
The symmetric double well potential, $V_S(\phi)$, as a function of $\phi$ (left).
The SDWP has two global minima (degenerate vacua) at (a) and (c) and an unstable local 
maximum at (b).  
The field configuration shown (right) is a kink profile that interpolates between the 
two vacua.  For $r\ll r_{wall}$ the field is at the vacuum point (c) on the potential, while for 
$r\gg r_{wall}$ the field is at the vacuum point (a) on the potential; where constant and in the 
vacuum, the field has no energy.  
For $r\approx r_{wall}$, the field interpolates between the vacua and must leave
the vacuum; the wall therefore has both potential energy and gradient energy.
In spherical symmetry the bubble will always collapse since there is only surface 
tension from the wall and no volume energy within the bubble (due to the 
{\it degenerate} vacua).}
\label{fig:fieldprofiles_sdwp}
\end{figure}
\begin{figure}
\epsfxsize=13.5cm
\centerline{\epsffile{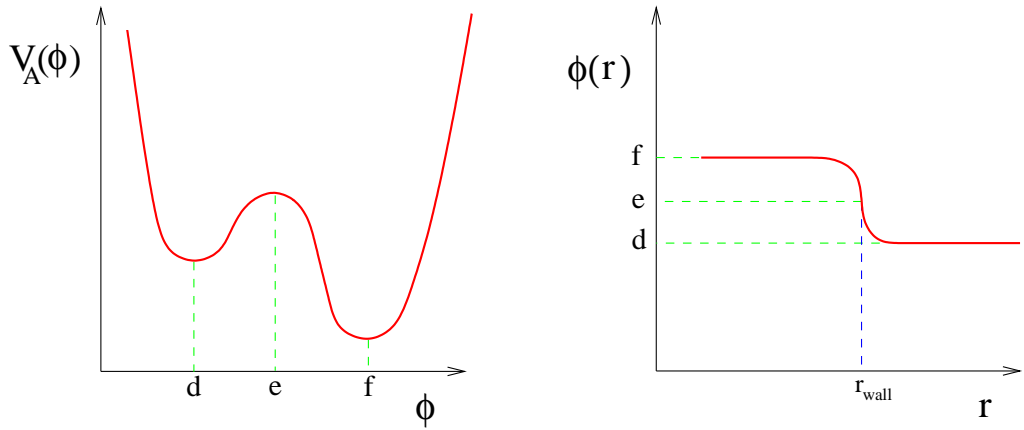}}
\caption[ADWP and kink profile initial data]
{\small \label{fig:fieldprofiles_adwp}
The asymmetric double well potential, $V(\phi)$, as a function of $\phi$ (left).
The ADWP has two local minima (non-degenerate vacua) at (d) and (f) and an unstable local 
maximum at (e).  
Vacuum point (d) is referred to as the {\it false vacuum} since it is not the global minimum 
and is unstable to fluctuations.
Vacuum point (f) is referred to as the {\it true vacuum} since it is the global minimum and
is stable to fluctuations.
The field configuration shown (right) is a kink profile that interpolates between the 
two vacua.  For $r\ll r_{wall}$ the field is in the true vacuum, while for 
$r\gg r_{wall}$ the field is in the false vacuum; where constant and in the 
vacuum, the field has no energy.  
For $r\approx r_{wall}$, the field interpolates between the vacua and must leave
the vacuum; the wall therefore has both potential energy and gradient energy.
In spherical symmetry their is competition between the surface tension trying to 
make the wall collapse and the volume energy (from the more energetically favorable true
vacuum) trying to make the wall expand.  The bubble's fate depends on the balance between
the two forces.
}
\label{fig:fieldprofiles_adwp}
\end{figure}
The first use of the $\lambda\phi^4$ theory to discuss phase transitions is 
usually credited to Landau and Ginzburg, 
\cite{ginzburg:1950}.  The theory is widely applicable and has been 
used to describe many types of phenomena, ranging  from
phase transitions in uniaxial ferrolectrics \cite{krumhansl:1975},
to phase transitions and defect dynamics within superconductors and 
superfluids \cite{ryder:1996}, 
to cosmological phase transitions resulting from the spontaneous breakdown 
of gauge symmetries 
(\cite{coleman:1977},
\cite{coleman:1985},
\cite{fetter:1982}, 
\cite{kirzhnits:1972},
\cite{lee:1981}
\cite{linde:1974},
and 
\cite{weinberg:1967}, 
for general treatment;
and 
\cite{gelmini:1994}, 
\cite{gleiser:1991}, 
\cite{gleiser:1996},
\cite{riotto:1995},
and
\cite{umurhan:1998}, 
for oscillon related studies).
Although in this thesis the particle physics and cosmology vernacular
is mostly used, the results are applicable to any scalar field model
described by the non-linear Klein-Gordon equation with a SDWP or ADWP.

When describing a particle theory, a local minimum describes a vacuum state;
the global minimum is the true vacuum while all other minima are false
vacua. 
For the SDWP in figure \ref{fig:fieldprofiles_sdwp}, 
the sample initial data interpolates between the two degenerate vacuum states.  
For $r > r_{wall}$ the field is in the vacuum state described by point {\it a} 
on the potential, $V(\phi)$, while for 
$r < r_{wall}$ the field is in the vacuum state described by point {\it c}.
At and around $r=r_{wall}$, the field interpolates between the two vacuum 
states and is called a domain wall.   
Almost all of the energy of the field is concentrated in the wall.  
There is no potential energy or gradient energy in the field away from 
$r_{wall}$, since it is in the vacuum;
whereas in the wall the field leaves the vacuum, so it has
both potential energy and energy due to the gradients of the field.
If $r$ is a Cartesian-like coordinate and the field is plane-symmetric then
the wall is a planar domain wall.  
However, if $r$ is a radial (spherical) coordinate,
the wall has the shape of a spherical shell, and is often referred to
as a bubble wall.
The term {\it bubble} is in analogy to bubbles in fluids that are created during
a change of phase, like gas bubbles forming in liquids, only instead of
separating different states of matter, these bubbles separate different
vacuum states of a particle theory.

Although the term {\it bubble} is used for both SDWP and ADWP alike, 
the analogy actually is best suited to the ADWP
(depicted in figure \ref{fig:fieldprofiles_adwp}).  
In the ADWP, the bubble wall separates the volume of space in the
false vacuum from the volume of space in the 
{\it more energetically favorable} true vacuum.  
This is exactly like a superheated liquid undergoing a phase transition
to a gas, where the false vacuum is the liquid and the true vacuum 
is the gas.
In the liquid-gas transition thermodynamic fluctuations cause gas 
bubbles to form in the liquid; the fate of the bubbles is determined by 
the competition between the surface tension in the wall and the 
volume energy within the bubble wall.  For a large enough fluctuation it will be 
energetically favorable for the bubble wall to continue to expand, thus filling up 
space with the more energetically favorable gas.  This is exactly what happens with 
vacuum state 
phase transitions, except instead of thermodynamic fluctuations, the bubbles
are usually nucleated by quantum fluctuations.

\section{The (1+1) plane-symmetric $\left(\phi^2-1\right)^2$ model}

The Klein-Gordon equation  in (1+1) dimensions with the SDWP 
\begin{equation}
\displaystyle{
\frac{\partial^2\phi}{\partial t} 
-
\frac{\partial^2\phi}{\partial x} 
- \phi + \phi^3 = 0}
\label{simpleKGSDWP}
\end{equation}
(where $\phi\equiv \phi(x,t)$)
has been studied extensively over the last century. 
One of the most exciting discoveries was that the model supports
solitary waves, stable localized solutions to 
non-linear field equations.  For equation \ref{simpleKGSDWP} these solitary
waves take the form
\begin{equation}
\displaystyle{
\phi(x,t)_\pm 
= \pm {\rm tanh}\left(\frac{\gamma\left(x-vt\right)}{\sqrt{2}}\right)
}
\end{equation}
where $\gamma=1/\sqrt{1-v^2}$ and $v$ is the velocity of the kink or antikink
(plus or minus sign, respectively).   
Much of the research of this model was motivated by the attempt to understand 
particle physics (meson) scattering experiments. 
In this context, Campbell {\it et al.}\cite{campbell:1983}, 
showed that there is a resonant structure within the kink-antikink
parameter space.  The possible outcomes (varying depending upon the initial  
velocity of the kinks, $\pm v_i$) were {\it reflection}, {\it annihilation}, 
or {\it collapse} to a 
long-lived but unstable bound-state\footnote{Transmission is prohibited 
due to lack of energy conservation}.
A resonance between the translational mode
of the colliding kinks and their individual internal shape 
mode vibrations gives rise to a very intricate structure for the endstate
as a function of initial velocity.  This structure was later shown 
(in the context of cosmological domain wall collisions) to be fractal in
nature, \cite{anninos:1991}.
The fractal structure was found within ``$n$-bounce windows'', where the kinks
collide, reflect off one another, and then collide again after radiating 
away energy (the process repeats itself $n$ times).

\section{What is an Oscillon?}

The term {\it oscillon} has a few different meanings depending on 
context, but here it refers to a time-dependent spherically-symmetric
coherent localized solution to a non-linear field equation (SDWP or ADWP)
which is unstable but long-lived compared to the typical time-scale 
involved in the problem. 
Oscillons, originally called pulsons, were first studied numerically 
with the SDWP in 1977 by Bogolyubskii and Makhankov.  
They showed that for a wide range of initial data, the behavior of 
a collapsing bubble was characterized by three stages: 
\begin{itemize}
\item{{\it Collapse:} The bubble wall collapses toward $r=0$ and the field oscillates
irregularly while radiating a large amount of its initial energy.}
\item{{\it Pseudo-stable Oscillations:} The oscillon settles into a state
where it remains localized (the location of the bubble wall is bounded)
and the field oscillates about the stable vacuum,
radiating very little energy.
The term pseudo-stable is used because although the oscillons are unstable, they 
can last for thousands of times longer than the time predicted by 
linear analysis\footnote{Although the true extent of their longevity was not 
shown convincingly until 1995, \cite{copeland:1995}.} (which also 
is roughly the oscillon's period of oscillation).}
\item{{\it Dispersal:} Eventually, an unstable shape-mode triggers the dispersal 
of the oscillon and the system is left in the original ($r\rightarrow \infty$) vacuum state.}
\end{itemize}
Nearly twenty years later, Copeland {\it et al.} 
\cite{copeland:1995}, performed a 
much more thorough (and computationally rigorous) investigation of oscillons. 
Oscillons were shown to be extremely long-lived for a {\it wide} range of parameter values 
for both the SDWP and the ADWP (with varying degrees of asymmetry).  
The perturbative methods used provided an explanation for two properties of 
oscillons: A) the existence of a minimum radius for oscillon formation created 
from static bubble collapse, and B) the need for the initial energy of the field 
configuration to be above a certain threshold. 

Although Copeland, Gleiser, and M\"uller  \cite{copeland:1995}, were 
the first to really dissect oscillons 
and to start exploring how and why they behave the way they do, 
their investigations (as good ones do) raised as many (or more) questions 
as they answered!
In particular, Copeland {\it et al.} did not explore the fine structure of the parameter space 
as \cite{campbell:1983} 
and \cite{anninos:1991} did for the (1+1) plane-symmetric case, nor did they explore what
effect non-spherical excitations play on the stability of oscillons.
This work explores these two points and others that arose throughout the process.

\chapter{Numerical Analysis and MIB Coordinates}

This chapter reviews the basic numerical methods (both new and old) employed 
throughout 
our research and motivates some of the choices made in the notation 
and form of the equations used.  Although a complete description 
of finite difference equations, stability analysis, and dissipation are well beyond the
scope of this thesis, a basic explanation of one and two dimensional finite difference 
and dissipation operators is provided. 
Finally, a new technique which solves problems associated with solving 
massive field equations 
on a lattice is introduced here. Implementation of this technique is then 
detailed in chapters \ref{chap:1D} and \ref{chap:2D}. 

\section{Finite Differences: Definitions and Notation}

Since the dawn of calculus in the 17th century, there has been 
a desire to be able to solve differential (and partial differential) equations.  
However, for most of the time since Newton and Leibniz, 
the majority of mathematicians and physicists have been limited to solving 
differential equations in closed-form or with various types of perturbative methods.  
These barriers have in large part collapsed since the creation of the computer
in the latter part of the 20th century.
Faster and faster computers, coupled with  
new numerical methods, continue to allow people to solve equations that before were 
far out of their reach.  
There are a great many numerical methods that have been 
developed for the solution of partial differential equations (PDEs), including finite 
differences, finite elements, spectral methods, and more.  A brief explanation of 
finite differences is presented here, focusing on aspects most relevant to this research.
(The following subsections are largely based on the class and lecture notes by Choptuik,
\cite{choptuik:1998b}, \cite{choptuik:1999}). 

\subsection{Discretization}


For problems such as those considered here,
finite differencing provides a very natural and 
straightforward route to the approximate solution of time-dependent partial-differential 
equations.
A finite difference approximation (FDA) to a partial differential equation 
(PDE) is obtained by replacing a continuous differential system of equations
with a discrete system of approximate equations. 
The spacetime domain is represented by a discrete number of points 
on a static uniform mesh that
are labeled by $x_k$ and $t^n$, for integer $k$, $n$.  
The discretization scale is set by $h\equiv \Delta x = \lambda \Delta t$, 
and the error associated with the approximation should go to zero as 
$h$ goes to zero.
For a continuum differential system, 
\begin{equation}
L u = f
\label{eq:contindiff}
\end{equation}
where $L$ is a differential operator, $u$ is the continuous solution, and 
$f$ is the continuous source function, we use 
\begin{equation}
\hat{L} \hat{u} = \hat{f}
\label{eq:discretediff}
\end{equation}
to denote its finite difference approximation, where $\hat{L}$ 
is a discrete difference operator, $\hat{u}$ is the discrete 
solution, and $\hat{f}$ is the discrete source function.
Throughout this chapter, $\hat{}$ will denote
a discrete quantity; for clarity this notation will be dropped in subsequent 
chapters and discrete quantities will be recognized by their 
space and time indexes, $u^n_k$.

\subsection{Residual}

It is useful to rewrite the FDA above as
\begin{equation}
\hat{L} \hat{u} -\hat{f}=0;
\label{eq:resid0}
\end{equation}
for a fully explicit scheme, this equation can be solved 
exactly.  However, for iterative schemes, the right hand side 
of (\ref{eq:resid0}) will actually have a non-zero value that
is representative of how well $\hat{u}$ solves the system.
This leads to the definition of a residual
\begin{equation}
\hat{r} \equiv \hat{L} \tilde{u} -\hat{f},
\label{eq:resid}
\end{equation}
where $\tilde{u}$ is the ``instantaneous approximation'' of
$\hat{u}$, i.e. that which converges to $\hat{u}$ in the limit 
of infinite iteration.
Thus, $\hat{r}$ gives a measure of how well $\hat{u}$ 
satisfies the FDA, and 
an iterative scheme is said to be {\it convergent} if the 
residual is driven to zero in the limit. 

\subsection{Truncation and Solution Errors}

The truncation error of a finite difference approximation is
defined to be
\begin{equation}
\hat{\tau} \equiv \hat{L} u - \hat{f}
\label{eq:tre}
\end{equation}
where we note that the discrete operator $\hat{L}$ acts
on the {\it continuum} solution $u$.  $\hat{\tau}$ often
cannot be obtained exactly since the solution $u$ is not usually
known.

The solution error is defined to be 
\begin{equation}
\hat{e} \equiv u - \hat{u}
\label{eq:solerr}
\end{equation}
and is the direct measure of how different the approximate
solution $\hat{u}$ is to the continuum solution $u$.
Although $u$ must be known exactly 
to know the exact solution error, an approximate solution 
error (the solution error to leading order) 
can usually be obtained numerically (see \ref{sec:convergence} below).

\subsection{Consistency and Order of an FDA}

A difference scheme with discretization scale $h$ 
is a {\it consistent} representation of the continuous 
system  if the truncation error goes to zero as the 
discretization scale goes to zero.
Furthermore, the difference scheme is said to be 
{\it p-th order accurate} if 
\begin{equation}
\lim_{h\rightarrow 0} \hat{\tau} = O(h^p)
\label{eq:order}
\end{equation}
for an integer $p$.  All of the schemes in this thesis
are second-order, so hereafter it is assumed that $p=2$.

\subsection{Richardson Expandability}

For a centered difference scheme $\hat{L}$, with discretization scale
$h$, the solution $u$
is related to the discrete approximation $\hat{u}$ by
\begin{equation}
\hat{u} = u + e_2 h^2 + e_4 h^4 + \cdots
\label{eq:rexpansion} 
\end{equation}
where the $e_i$ {\it are h-independent functions}.  A non-centered
scheme will also have odd terms $e_3 h^3$, $e_5 h^5$, etc.
Equation \ref{eq:rexpansion} is referred to as a Richardson 
expansion.

\subsection{Convergence\label{sec:convergence}}

A finite difference approximation is said to be {\it convergent}
if and only if 
\begin{equation}
\hat{\tau}\rightarrow 0 \ \ {\rm as } \ \ h\rightarrow 0.
\label{eq:convergence}
\end{equation}
Showing convergence is of prime importance in numerical analysis as it is
a statement that the solution obtained numerically really approaches
the continuum solution as the discretization goes to zero.
A useful formula used throughout this thesis describes what 
is called (by Choptuik) the {\it convergence factor},
\begin{equation}
C_f \equiv 
\displaystyle{
\frac{ \hat{u}_{4h} - \hat{u}_{2h}}{\hat{u}_{2h}-\hat{u}_h}},
\label{eq:cfac}
\end{equation}
where $\hat{u}_{4h}$, $\hat{u}_{2h}$, and $\hat{u}_{h}$, are the solutions with 
discretization scales $4h$, $2h$, and $h$, respectively.
Using the expansion (\ref{eq:rexpansion}) it can be shown that for second-order
approximations $C_f\rightarrow 4$ as h$\rightarrow 0$.

\subsection{Difference Operators}

Tables \ref{tab:1Dfdop} and \ref{tab:2Dfdop} 
contain the one and two dimensional second order difference operators
to be used later.  They can be derived by Taylor expanding the solution
using the discretization scale as the expansion parameter.

\begin{table}
\small
\bc
\label{tab:1Dfdop}
\begin{tabular}{rcl}
\hline
Operator & Definition & Expansion \\
\hline
$\Delta^f_r u^n_i$ & $\left( -3u^n_i + 4u^n_{i+1}-u^n_{i+2}\right) /
\left(2\Delta r\right)$ &
$\partial_r u \big\vert^n_i + O\left(h^2\right)$ \\
\\
$\Delta^b_r u^n_i$ & $\left( 3u^n_i - 4u^n_{i-1} + u^n_{i-2}\right)
/\left(2\Delta r\right)$ &
$\partial_r u \big\vert^n_i + O\left(h^2\right)$ \\
\\
$\Delta_r u^n_i$ & $\left( u^n_{i+1}-u^n_{i-1}\right) /\left(2\Delta r\right)$ & 
$\partial_r u \big\vert^n_i + O\left(h^2\right)$ \\
\\
$\Delta_{r^3} u^n_i$ & $\left( u^n_{i+1}-u^n_{i-1}\right) /\left(r^3_{i+1} - r^3_{i-1} \right)$ & 
$\partial_r u \big\vert^n_i + O\left(h^2\right)$ \\
\\
$\Delta_t u^n_i$ & $\left( u^{n+1}_i-u^n_i\right)/\Delta t$ &
$\partial_t u \big\vert^{n+\frac{1}{2}}_i + O\left( h^2\right)$ \\
\\
$\mu^{\rm dis}_r u^n_i$ & -$\epsilon_{\rm dis}[ 6u^n_i + u^n_{i-2}+u^n_{i+2} -$ & 
$\left(\Delta r\right)^3 \partial^4_x u \big\vert^{n}_i + O\left( h^7 \right)$ \\
 & $4\left( u^n_{i-1}+u^n_{i+1}\right)] /\left(16\Delta t\right)$ & \\
\\
$\mu^{\rm ave}_t u^n_i$ & $\left( u^{n+1}_i + u^n_i\right) / 2$ &
$u\big\vert^{n+\frac{1}{2}}_i + O\left( h^2\right)$ \\
\hline
\end{tabular}
\caption{\small \label{tab:1Dfdop}One dimensional two-level FDA operators,
\hbox{$ h = \Delta r = \lambda^{-1} \Delta t $}.
}
\normalsize
\ec
\end{table}

\begin{table}
\bc
\small
\label{tab:2Dfdop}
\begin{tabular}{rcl}
\hline
Operator & Definition & Expansion \\
\hline
$\Delta^u_R u^n_{i,j}$ & $\left( -3u^n_{i,j} + 4u^n_{i+1,j}-u^n_{i+2,j}\right) /
\left(2\Delta R\right)$ &
$\partial_R u \big\vert^n_{i,j} + O\left(h^2\right)$ \\
\\
$\Delta^b_R u^n_{i,j}$ & $\left( 3u^n_{i,j} - 4u^n_{i-1,j} + u^n_{i-2,j}\right)
/\left(2\Delta R\right)$ &
$\partial_R u \big\vert^n_{i,j} + O\left(h^2\right)$ \\
\\
$\Delta_R u^n_{i,j}$ & $\left( u^n_{i+1,j}-u^n_{i-1,j}\right) /\left(2\Delta R\right)$ & 
$\partial_R u \big\vert^n_{i,j} + O\left(h^2\right)$ \\
\\
$\Delta_{R^2} u^n_{i,j}$ & $\left( u^n_{i+1,j}-u^n_{i-1,j}\right) /\left( R^2_{i+1} - R^2_{i-1} \right)$ & 
$\partial_R u \big\vert^n_{i,j} + O\left(h^2\right)$ \\
\\
$\Delta^f_z u^n_{i,j}$ & $\left( -3u^n_{i,j} + 4u^n_{i,j+1}-u^n_{i,j+2}\right) /
\left(2\Delta z\right)$ &
$\partial_z u \big\vert^n_{i,j} + O\left(h^2\right)$ \\
\\
$\Delta^b_z u^n_{i,j}$ & $\left( 3u^n_{i,j} - 4u^n_{i,j-1} + u^n_{i,j-2}\right)
/\left(2\Delta z\right)$ &
$\partial_z u \big\vert^n_{i,j} + O\left(h^2\right)$ \\
\\
$\Delta_z u^n_{i,j}$ & $\left( u^n_{i,j-1}-u^n_{i,j+1}\right) /\left(2\Delta z\right)$ & 
$\partial_z u \big\vert^n_{i,j} + O\left(h^2\right)$ \\
\\
$\Delta_t u^n_{i,j}$ & $\left( u^{n+1}_{i,j}-u^n_{i,j}\right)/\Delta t$ &
$\partial_t u \big\vert^{n+\frac{1}{2}}_{i,j} + O\left( h^2\right)$ \\
\\
$\mu^{\rm dis}_R u^n_{i,j}$ & -$\epsilon_{\rm dis}[ 6u^n_{i,j} + u^n_{i-2,j}+u^n_{i+2,j} - $ &  
$\left( \Delta R \right)^3 \partial^4_R u \big\vert^n_{i,j} + O\left(h^7\right)$\\
 & $4\left( u^n_{i-1,j}+u^n_{i+1,j}\right)] /\left(16\Delta t\right)$ & \\
\\
$\mu^{\rm dis}_z u^n_{i,j}$ & -$\epsilon_{\rm dis}[ 6u^n_{i,j} + u^n_{i,j-2}+u^n_{i,j+2} - $ &  
$\left( \Delta z \right)^3 \partial^4_z u \big\vert^n_{i,j} + O\left(h^7\right)$\\
 & $4\left( u^n_{i,j-1}+u^n_{i,j+1}\right)] /\left(16\Delta t\right)$ & \\
\\
$\mu^{\rm ave}_t u^n_{i,j}$ & $\left( u^{n+1}_{i,j} + u^n_{i,j}\right) / 2$ &
$u\big\vert^{n+\frac{1}{2}}_{i,j} + O\left( \Delta t^2\right)$ \\
\hline
\end{tabular}
\caption{\small\label{tab:2Dfdop}
Two dimensional two-level FDA operators, 
\hbox{$ h = \Delta R = \Delta z = \lambda^{-1} \Delta t $}.}
\normalsize
\ec
\end{table}

\section{Dissipation and Stability\label{sec:dissstab}}

Up to this point, there have been no comments made regarding the
{\it stability} of numerically evolved solutions using finite 
differences.  Much to the dismay of computational physicists,
many of the difference schemes one uses to approximate
solutions to partial differential equations do not work because
they lead to unphysical growing modes.  
To understand this effect (and how to correct it) consider
the discrete analog to the Fourier decomposition of the 
continuous function 
$\tilde{\phi}(k,t) = 
\displaystyle{
\frac{1}{\sqrt{2\pi}} \int^\infty_{-\infty}
e^{-ik x}\ \phi(x,t) dx}$ (for $k$ continuous),
\begin{equation}
\tilde{u}^n(\xi) = 
\displaystyle{
\frac{1}{\sqrt{2\pi}} \sum^\infty_{m=-\infty}
e^{-i\xi m}\ u^n_m
},
\label{diffdecomp1}
\end{equation}
for $-\pi\le \xi \le \pi$ and discrete $m$, \cite{thomas:1995},
\cite{choptuik:1999}. 
A mode analysis can be performed by inserting the decomposition
(\ref{diffdecomp1}) 
into the difference equation
and looking at the behavior of each Fourier mode over time.  
For linear differential equations, it is straightforward to
write the solution to the difference equation in the form
$\tilde{u}^{n+1}(k) = \rho\left( \xi \right) \tilde{u}^n(k)$,
where $\rho$ is a complex function of $\xi$.
It is clear that if  $|\rho|^2>1$ the mode grows over time, 
if $|\rho|^2=1$ the mode remains constant, 
and if $|\rho|^2<1$ the mode decays over time;
$|\rho|^2$ is referred to here as an {\it amplification factor}.
A difference equation is 
{\it dissipative} if no modes grow with time and at least one mode decays,
while it is 
{\it nondissipative} if the modes neither decay nor grow,
or {\it unstable} if any of the modes grow with time. 

One method commonly used to make unstable schemes stable, or to make
nondissipative and dissipative schemes more stable, is to add
dissipation.  
Dissipation can be added in a variety of ways, 
but we add it to our difference scheme by incorporating
higher order spatial derivatives multiplied by 
the grid spacing to some power, $(\Delta x)^n$.  
Dissipation usually lowers the amplification factor for most modes,
and typically affects higher wavenumbers more dramatically.
Put another way, the goal is to dampen high frequency modes while 
maintaining the order of the original difference scheme.  

Using the dissipation operators  in
conjunction with second order CN derivative operators (see 
table \ref{tab:1Dfdop}), the
{\it linear} advection equation 
$\left(\partial_t - \partial_x\right)\phi=0$, 
can be approximated by the following FDA:
\begin{eqnarray}
\displaystyle{
\frac{u^{n+1}_j - u^{n}_j}{dt}} &=&
\displaystyle{
 \frac{u^{n+1}_{j+1} - u^{n+1}_{j-1}}{4 dx} 
+\frac{u^{n}_{j+1} - u^{n}_{j-1}}{4 dx} } \nonumber\\
&&
\displaystyle{
+\frac{\epsilon_{\rm dis}}{16}
\left\{ 6 u^n_j + u^n_{j+2} + u^n_{j-2} 
-4\left(u^n_{j+2} + u^n_{j-2}\right) \right\}
}\, .
\end{eqnarray}
This scheme has an amplification factor $|\rho^2|(\xi)$, where
\begin{equation}
\begin{array}{rcl}
\rho(\xi) &=& 
\displaystyle{
\frac{1 + i\displaystyle{\frac{\lambda}{2}} \sin(\xi) - 
\displaystyle{\frac{\epsilon_{\rm dis}}{8}}\left(
3 + \cos(2\xi) - 4 \cos(\xi)\right)}
{1 - i\displaystyle{\frac{\lambda}{2}} \sin(\xi)}.
}
\end{array}
\end{equation}
Plotting $|\rho|^2$ as a function of wavenumber\footnote{Again,
this is true only for the linear advection equation.},
figure \ref{ampfacs} shows that the CN scheme is nondissipative 
for $\epsilon=0$ and stable for $\epsilon\leq 1$.  
\begin{figure}
\label{ampfacs}
\centerline{
        \hbox{\epsfxsize =12cm\epsffile{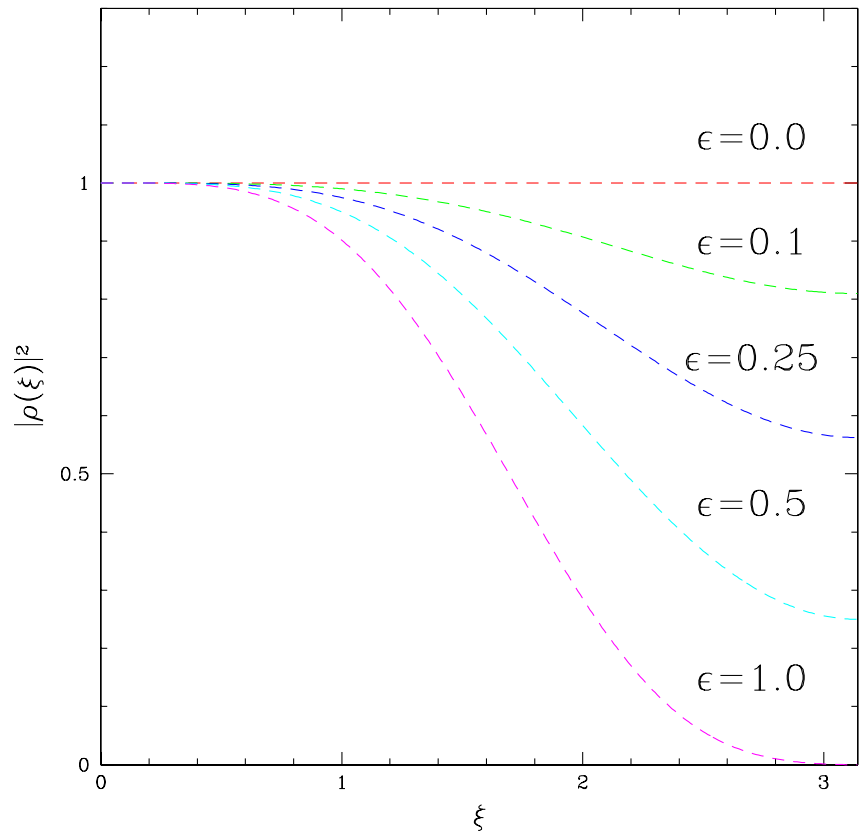}}
	}
\caption[Amplification factors for CN scheme with dissipation.]
{\small \label{ampfacs}
Amplification Factors for CN schemes with and without dissipation,
where $\epsilon$ is the parameter measuring the amount of dissipation.
$\epsilon=0$ corresponds to no dissipation while 
$\epsilon=1$ completely quenches the modes at the Nyquist limit, $\xi=\pi$.  
The CN scheme is marginally stable with no dissipation and 
stable for $\epsilon>0$.
}
\end{figure}

Unfortunately $|\rho|^2(\xi)$ is not so easy to compute for general 
difference schemes. 
In particular, when the equation is nonlinear the stability of a difference 
scheme cannot be easily determined (in closed form), even if the
derivative operators used are known to be stable in the linear case.
We therefore take an empirical approach and include dissipation 
as needed in order to make the scheme stable.
In fact, for evolution equations such as those studied here,
the incorporation of dissipation is often 
{\it essential} to the construction of stable schemes.


\section{Geometry\label{sec:geometry}}

This work is certainly not the first to explore numerical 
solutions to massive scalar field equations.  
The literature is full of work on both 
massive ($m^2 \phi^2$) and nonlinear
($\epsilon\phi^3$, $\lambda \phi^4$, $\sin(\phi)$, etc.) potentials
in one, two, and even three dimensions, 
both coupled to gravity and in flatspace \cite{seidel:1986}, \cite{seidel:1991},
\cite{campbell:1983}. 
One of the well-known problems with solving the nonlinear or massive 
Klein-Gordon (KG) equation (even in flatspace) is that there is no 
closed-form out-going boundary condition.
The massive KG equation is simply 
$\displaystyle{\left(\Box - m^2 \right)\phi(\vec{x},t)=0}$, which 
has a dispersion relation $\omega^2 = k^2 + m^2$.
Therefore, the velocity of the outgoing radiation cannot be uniquely 
determined and a satisfactory outgoing wave condition cannot be applied. 
While some of the radiation that reaches the outer boundary will   
leave the computational domain, significant amounts will also
be reflected back and can contaminate the solution.
If the phenomena being studied is short-lived, the computational domain can be 
made large enough so that radiation that does get reflected off the outer
boundary will not have time to reach the region of interest.
However, for long-lived phenomena (like oscillons or some boson stars)
one must deal with the outgoing radiation more directly.

Many previously used attempts consist of using an approximate
out-going boundary condition and some sort of absorbing region 
near the outer edge of the grid.
\cite{balakrishna:1997}, \cite{marsa:1996}, \cite{seidel:1986}, 
and \cite{seidel:1991} use a {\it sponge filter}, 
which imposes an outgoing radiation condition over a {\it finite
portion} of the computational domain; this allows outgoing radiation
to propagate while attempting to dampen ingoing radiation.
However, due to the
aforementioned unknown radiation velocity, this is done only 
approximately and can be susceptible to back-scatter effects.
Recently Gleiser {\it et al}, \cite{gleiser:1999}, 
have used a method referred to as
{\it adiabatic damping}, where instead of focusing only on the outgoing
radiation (as in the sponge filter methods) with a potential (roughly) of the form
$\gamma\left(\vec{x}\right)\left(\dot{\phi} + \phi'\right)$, 
they use a term of the form $\gamma\left(\vec{x}\right)\dot{\phi}$ and dampen all
scalar radiation.
However, by adding explicit damping terms to the equations of motion 
(as opposed to higher order dissipation added to the difference equations) 
neither of the resulting difference schemes actually reduce to the true 
equations of motion in the limit that the grid spacing goes to 
zero.

This section introduces a new {\it geometric} 
technique that effectively absorbs outgoing radiation, 
has difference equations that reduce to the differential equations in the 
continuum limit, and that is natural and straightforward to implement 
in both spherical and axial symmetry.
The method employed has two parts, the transformation of coordinates
and the incorporation of dissipation into the numerical scheme.
The coordinate system used leaves the interior of the grid alone, while 
transforming the coordinates of the exterior points to be moving outward
at approximately the speed of light relative to the interior or original 
rest frame; the coordinates are {\it monotonically increasingly 
boosted} (MIB).
Characteristic analysis of the wave equation (in MIB coordinates) shows that
both ingoing and outgoing characteristic velocities approach zero in a region
near the outer edge of the grid \cite{choptuik:2000}.
As the field slows down it becomes compressed; 
since the dissipation becomes stronger with increasing wavenumber, 
the field is quenched. 
This is shown to occur in a stable and non-reflective manner in
(1+1) spherical symmetry and (2+1) axisymmetry.

\subsection{Radial MIB Coordinates\label{radialmib}}

In spherical symmetry,
the outgoing radiation is frozen-out by introducing a  
new radial coordinate that smoothly interpolates between the standard polar 
radial coordinate $\tilde{r}$ on Minkowski space
\be
d\tilde{s}^2 = -d\tilde{t}^2 + d\tilde{r}^2 + \tilde{r}^2 d\tilde{\Omega}^2,
\ee
(where $d\tilde{\Omega}^2= d\tilde{\theta}^2 + \sin^2\tilde{\theta}\, d\tilde{\phi}^2$),
and an outgoing null coordinate.  We define 
\begin{equation}
\begin{array}{cccc} 
\tilde{t}=t, &  \tilde{r}=r + f(r) t, & {\rm and}& \tilde{\Omega}=\Omega \\
\end{array} 
\label{eq:1Dtransformation}
\end{equation}
where $f(r)$ is a {\it monotonically increasing} function that interpolates
between $0$ and approximately 1 at some characteristic cutoff, 
$r_{\rm c}$,
\begin{equation}
f(r) \simeq \left\{
\begin{array}{rcr}
0 &{\rm for}& r \ll r_{\rm c}\\
\sim1 &{\rm for}& r \gg r_{\rm c}
\end{array}
\right\}.
\end{equation}

\noindent
In general, these coordinates are not good coordinates everywhere.
However, if $f(r)$ is monotonically increasing, the 
determinant of the Jacobian of the transformation 
is non-zero for all $t > -\max|f'(r)|$
and the coordinate transformation is one-to-one.
Although a coordinate singularity inevitably forms as $t$ approaches 
past timelike infinity, 
this has no effect on the forward evolution of initial data specified at $t=0$.  
We must also demand that $f(0)\!=\!0$ to maintain the condition for 
elementary flatness at the origin.  
This coordinate choice takes the metric to
\begin{equation}
\begin{array}{rcl}
ds^2 &=& \left(-1 + f^2(r)\right)dt^2  + 2 f(r)\left(1 + f'(r)t\right) dt dr \\
     & &   +\left( 1 + f'(r)t\right)^2 dr^2 + 
            \left(r + f(r)t \right)^2 d\Omega^2
\end{array}
\end{equation}
or in a more familiar (3+1) form 
(\cite{arnowitt:1962},\cite{choptuik:1986},\cite{choptuik:1998a}) to
\begin{equation}\label{eq:1Dshiftmet} 
ds^2=\left(-\alpha^2+a^2\beta^2\right) dt^2 + 2 a^2\beta dt dr + a^2 dr^2 +
r^2b^2 d \Omega^2, 
\end{equation}
where 
\be \label{eq:ADM_1Dauxvars}
\begin{array}{rclcrcl}
a(r,t)   &\equiv& 1+f'(r)t & \ \  &  b(r,t)    &\equiv& \displaystyle{1 + f(r)\frac{t}{r}} \\
\vspace{-0.275in} \\
\alpha(r,t)&\equiv& 1           & \ \  & \beta(r,t) &\equiv& \displaystyle{\frac{f(r)}{1 + f'(r)t}}. 
\\
\end{array}
\end{equation}
\begin{figure}
\centerline{
	\hbox{\epsfxsize= 12cm\epsffile{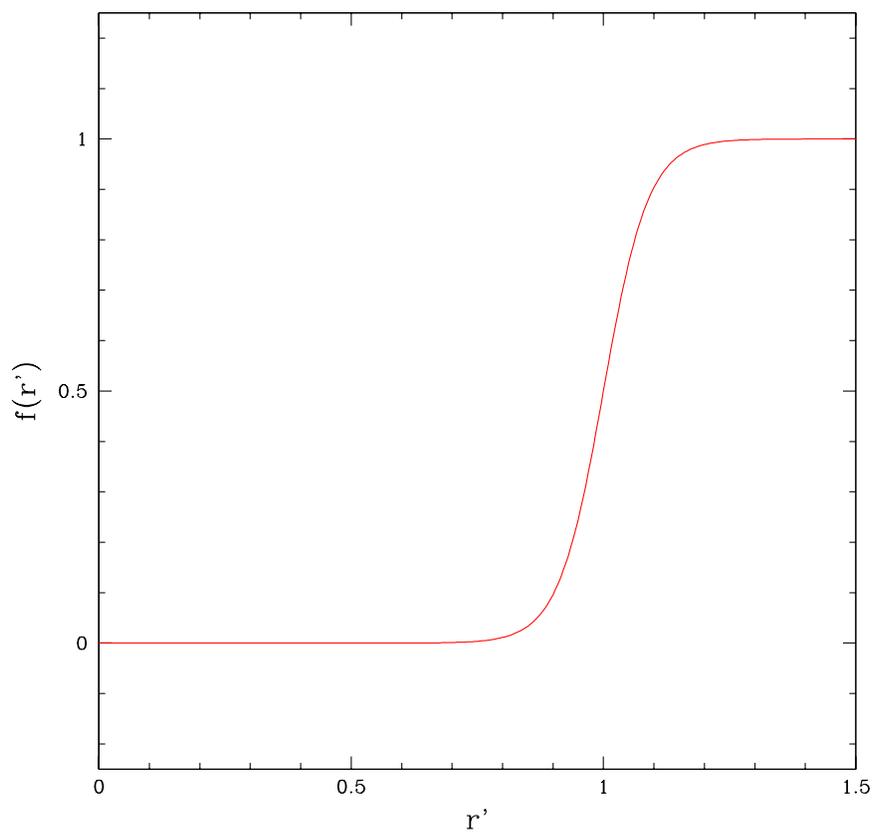}}
}
\caption[Interpolating function radial MIB coordinates]
{\small \label{fig:mibradf} 
Interpolating function, $f(r')$, for radial MIB coordinates,
where $r'$ is in units where $r_c=1$.
$\delta$ is taken to be $\delta\rightarrow \delta/r_c \approx 0.0893$ 
(corresponding to the system to be used in chapter 4).
Wherever \hbox{$f=0$}, the new radial coordinate is left unchanged
with respect to the old coordinate, $\tilde{r}$.
Where $f > 0$ the new radial coordinate is being ``shifted'' 
with respect to $\tilde{r}$ with velocity \hbox{$\beta=f/(1+f't)$}. 
}
\label{fig:mibradf}
\end{figure}
Assuming the following specific form for $f$ (see figure \ref{fig:mibradf}):
\be \label{eq:fdef} 
f(r) = \left[ 1 + \tanh\left(\left(r-r_c\right)/\delta\right)\right]/2 
+ \epsilon(r_c,\delta) \, ,
\ee
where $\epsilon(r_c,\delta) = -\left[ 1 + \tanh\left(\left(-r_c\right)/\delta\right)\right]/2 $,
\begin{figure}
\centerline{ 
	\hbox{\epsfxsize =12cm\epsffile{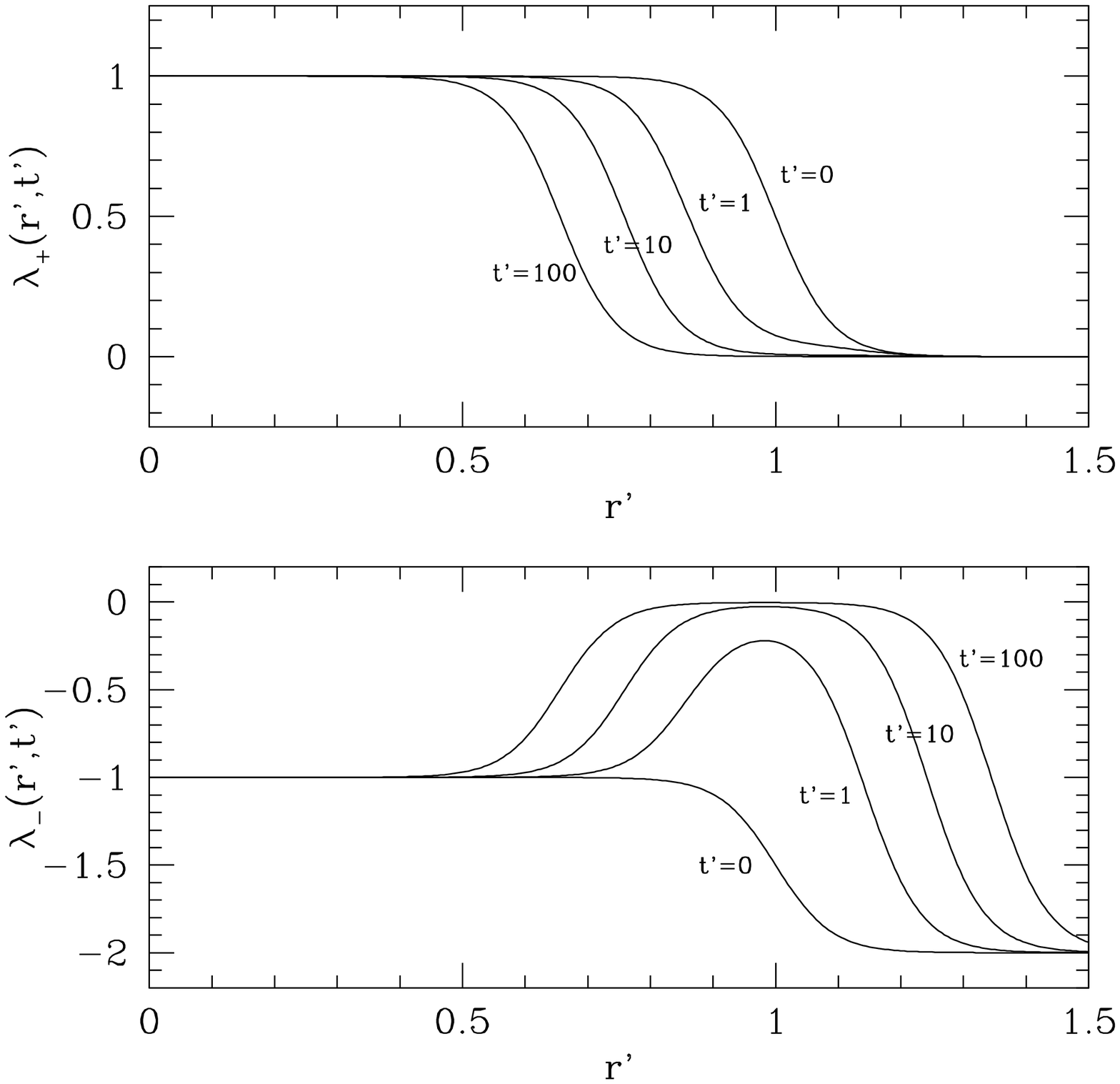}}
	}
\caption[Characteristics for radial MIB system]
{\small \label{fig:1Dchars} 
Plot of the characteristic velocities as a function of $r'$ and 
$t'$, where $r'$ and $t'$ are radial MIB coordinates in units where $r_c$ is set to 
unity, and $\lambda_+$ and $\lambda_-$ are the outgoing and ingoing characteristics,
respectively.  $\delta$ is taken to be $\delta\rightarrow \delta/r_c \approx 0.0893$ 
(corresponding to the system to be used in chapter 4).
Characteristic velocities are plotted for times $t'=0$, $1$, $10$, and $100$
($t'=100$ is larger than the lifetime of the longest lived solution studied in this work).
Both characteristic velocities approach zero around $r'=1$ as $1/{t'}$.
}
\label{fig:1Dchars}
\end{figure}
\begin{figure}
\centerline{ 
	\hbox{\epsfxsize =12cm\epsffile{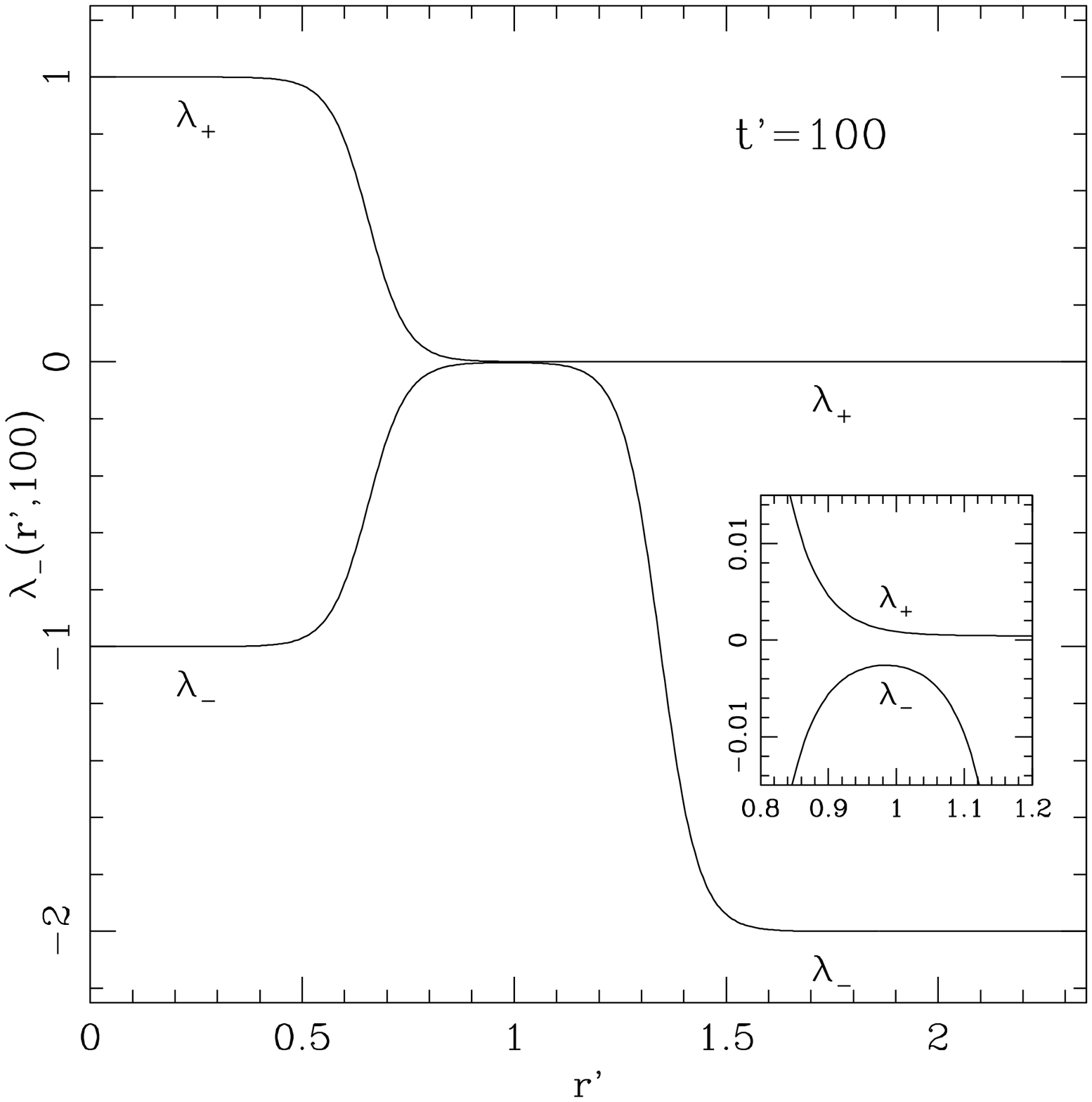}}
	}
\caption[Coincidence of ingoing/outgoint $r$:$t$ characteristics]
{\small \label{fig:1Dchars_coin} 
Plot of characteristic velocites, $\lambda_\pm(r',100)$, 
where $r'$ and $t'$ are radial MIB coordinates in units where $r_c$ is set to 
unity, and $\lambda_+$ and $\lambda_-$ are the outgoing and ingoing characteristics,
respectively.  $\delta$ is taken to be $\delta\rightarrow \delta/r_c \approx 0.0893$ 
(corresponding to the system to be used in chapter 4), and $t'=100$ is larger than 
the lifetime of the longest lived solution discussed in this thesis.
Both characteristic velocities approach zero around $r'=1$ as $1/{t'}$.
}
\label{fig:1Dchars_coin}
\end{figure}
we can obtain the actual metric variables,
the characteristic velocities, 
and the conformal structure of the new hypersurfaces.

The characteristic analysis of the Klein-Gordon equation
with metric (\ref{eq:1Dshiftmet}) yields characterstics
\begin{equation}
\lambda_\pm = -\beta \pm \frac{\alpha}{a},
\label{eq:1dchars}
\end{equation} 
where $\lambda_+$ and $\lambda_-$ are the outgoing and ingoing characteristics,
respectively \cite{choptuik:1986}, \cite{courant:1962}.
The MIB system behaves 
like the old ($\tilde{t}$,$\tilde{r}$) coordinates for $r\ll r_{\rm c}$, 
but the outgoing radiation gets frozen out in the $r\approx r_{\rm c}$ region.
Figure \ref{fig:1Dchars} and
equations (\ref{eq:1dchars}) and (\ref{eq:ADM_1Dauxvars})
show that around $r\approx r_c$,  {\it both} the ingoing and the outgoing
characteristic velocities go to zero as $t\rightarrow \infty$ 
(as the inverse power of $t$).  
It is this property 
that is responsible for the ``freezing-out'' of the outgoing 
radiation~\cite{choptuik:2000}.
We call these coordinates ($t$,$r$) monotonically increasingly 
boosted (MIB) radial coordinates.  

The conformal structure (figure \ref{fig:confdiag})
is obtained by applying equations (\ref{eq:1Dtransformation}) to 
the standard conformal compactification on Minkowski space,
$\tilde{t} \pm \tilde{r}  = \tan \left( \frac{T\pm R}{2}\right)$
(where here $T$ and $R$ are the axes in the conformal diagram,
see \cite{hawking:1973} or \cite{wald:1984}), 
and then plotting curves of constant $r$ and $t$.
\begin{figure}
\epsfxsize=14cm
\centerline{\epsffile{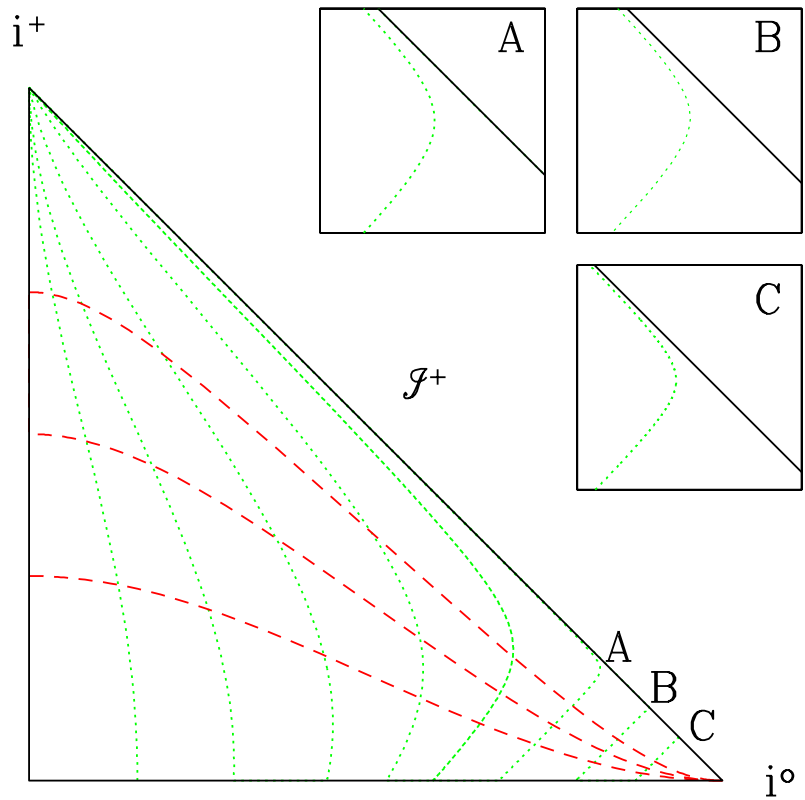}}
\caption[Conformal Diagram for radial MIB coordinates.]
{\small \label{fig:confdiag} 
Conformal diagram showing surfaces of constant $r$ (dotted lines)
and lines of constant $t$ (dashed lines).
Lines of constant $t$ look exactly like the constant-$t$
hypersurfaces of Minkowski space, whereas the lines of constant $r$ 
behave much differently.
For $r>r_c$, it appears as if the constant-$r$ surfaces 
are null.  This occurs since the coordinates are being shifted outwards 
at {\it nearly} the speed of light.
However, as insets A, B, and C show, the constant-$r$ lines do not
become null (do not intersect future null infinity), and are everywhere 
timelike.  This is actually easy to understand as the radial 
coordinate is never moved out at the speed of light (not even at 
spatial infinity, because of the $\epsilon$ term in equation 
\ref{eq:fdef}). 
}
\label{fig:confdiag}
\end{figure} \noindent
The constant-$t$ hypersurfaces are 
everywhere spacelike, and all reach spatial infinity, ${\rm i^o}$.
Although constant-$r$ surfaces for $r > r_{\rm c}$ appear at 
first glance to be null (or as if $\left(\frac{\partial}{\partial t}\right)^a$
``tips over'' to become a null vector), closer examination (see insets of Fig.
\ref{fig:confdiag}) reveals that they are 
indeed everywhere timelike and do {\em not} reach future null
infinity, {${\scri^+}$}.
This behaviour follows from the fact that the interpolating function, $f(r)$,
only reaches unity {\em asymptotically}, at ${\rm i^o}$.

\subsection{Axi-Symmetric (2-D) MIB Coordinates\label{aximib}}

\begin{figure}[h]
\epsfxsize=12cm
\centerline{\epsffile{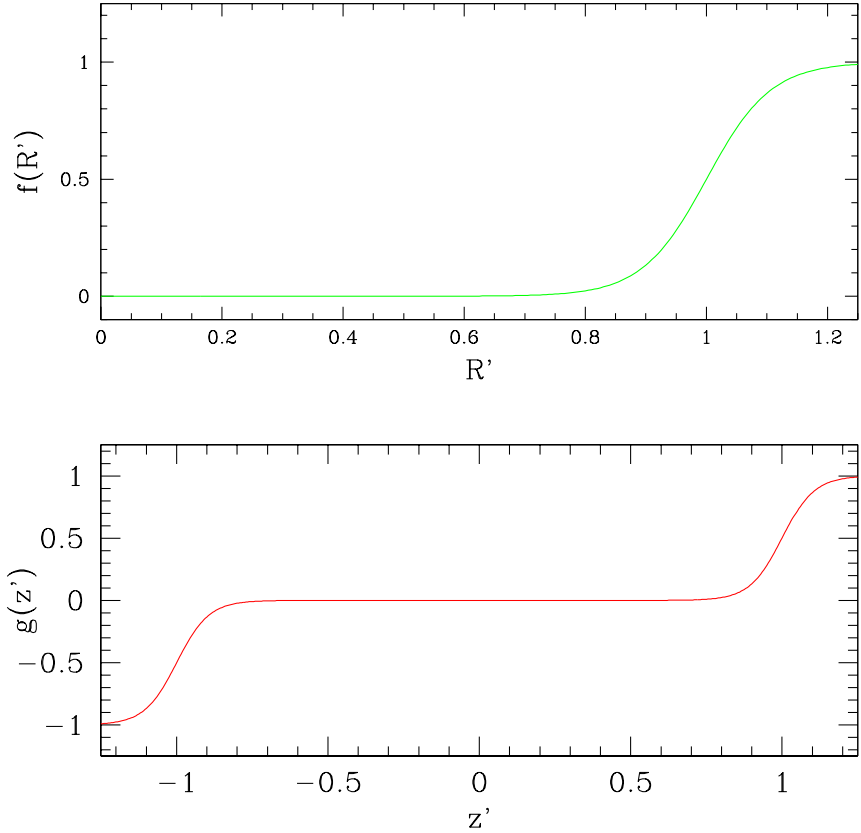}}
\caption[Interpolating functions for axisymmetric MIB coordinates]
{\small 
Interpolating functions, $f(R')$ and $g(z')$, for axisymmetric MIB coordinates,
where $R'$ and $z'$ are in units where $R_c$ and $z_c$ are equal to one.
$\delta_R$ and $\delta_z$ are taken to be approximately $0.1067$ 
(corresponding to the system to be used in chapter 5).
Wherever $f=0$, the new radial coordinate, $R$, is left unchanged with respect to the old
coordinate,  $\tilde{R}$.  For $f>0$ the new radial coordinate is being shifted outward
with respect to $\tilde{R}$ with velocity 
$\beta^R=f/(1+\partial_Rf t)$.
Wherever $g(z')=0$, the new $z$ coordinate is left unchanged with respect to the old
coordinate $\tilde{z}$.  For $g>0$ the $z$ coordinate is being shifted outward with respect to
$\tilde{z}$ with velocity
$\beta^z= g/(1 + \partial_zgt)$; note however that the sign of $g$ 
(and consequently $\beta^z$) is negative for $z<0$ as the coordinate is moving outward 
toward more negative z.
}
\label{fig:mibaxi_fg}
\end{figure}

The (1-D) MIB coordinates introduced in the previous subsection also
have a straightforward implementation in two dimensions.
The (2-D) problems discussed in this work all model (3+1) dimensional
spacetimes with axial symmetry.  
(2-D) MIB coordinates are obtained by again starting from the Minkowski metric,
but now in cylindrical coordinates
\be
d\tilde{s}^2 = -d\tilde{t}^2 + d\tilde{R}^2 
+\tilde{R}^2 d\tilde{\theta}^2 + d\tilde{z}^2 .
\ee
These coordinates are modified just like the radial coordinate in spherical symmetry 
(section \ref{radialmib}), except now both the axisymmetric coordinates $R$ and $z$ 
interpolate between the old ($\tilde{R}$,$\tilde{z}$) coordinates 
and ``null coordinates at  ${\rm i^o}$'':
\begin{equation}
\begin{array}{ccccc} 
\tilde{t}=t, &  \tilde{R}=R + f(R) t,& \tilde{z}=z + g(z) t, & {\rm and}& \tilde{\theta}=\theta \, . \\
\end{array} 
\label{eq:2Dtransformation}
\end{equation}
Here, $f(R)$ and $g(z)$ are monotonically increasing functions:
$f(R)$ interpolates 
between (approximately) $0$ and $1$ at a characteristic cutoff, $R_c$;
$g(z)$ interpolates  (approximately)
between $-1$ and $1$ at characteristic cutoffs, 
$\pm z_c$\footnote{We could have chosen to have to distinct cutoffs,
$(z_{c})^+$ and $(z_c)^-$, but by performing the ``experiment'' around 
$z=0$ this simplification works perfectly well.},
\begin{equation}
\label{2D_interpfunk}
\begin{array}{lcr}
f(R) \simeq \left\{
\begin{array}{rcc}
0 &{\rm for}& R \ll R_c\\
1 &{\rm for}& R \gg R_c
\end{array}
\right\}
& \ \ &
g(z) \simeq \left\{
\begin{array}{rcc}
-1 &{\rm for}& z \ll -z_c\\
0 &{\rm for}&  -z_c< z < z_c\\
1 &{\rm for}& z \gg z_c
\end{array}
\right\}.

\end{array}
\end{equation}
This coordinate choice takes the metric to 
\begin{equation}
\begin{array}{rcl}
ds^2 &=& \left(-1 + f(R)^2 + g(z)^2\right)dt^2 
	+ 2f(R)\left(1 + \partial_Rf(R)t\right) dt dR \\
&&
	+ 2g(Z)\left(1 + \partial_zg(z)t\right) dt dz 
	+ \left( 1 + \partial_Rf(R) t\right)^2 dR^2 \\
&&
	+ \left( R + f(R)t\right)^2 d\theta^2
	+ \left( 1 + \partial_zg(z) t\right)^2 dz^2
\end{array}
\end{equation}
or in a more familiar (3+1) form to 
\begin{equation}\label{eq:axishiftmet}
\begin{array}{rcl} 
ds^2&=&\left(-\alpha^2+a^2{\beta^R }^2 + b^2{\beta^z }^2\right) dt^2 
+ 2 a^2\beta^R dt dR 
+ 2 b^2\beta^Z dt dZ \\
&&
  + a^2 dR^2 
  + \tilde{R}^2d\theta^2
  + b^2 dz^2 ,
\end{array}
\end{equation}
where 
\be \label{eq:2DADM_auxvars}
\begin{array}{rclcrcl}
a(R,t)     &=& 1+\partial_Rf(R)t & \ \  &  b(z,t)    &=& 1 + \partial_zg(z)t \\
\vspace{-0.5cm}\\
\beta^R(R,t)  &=& \displaystyle{\frac{f(R)}{1 + \partial_Rf(R)t}} & \ \ & 
\beta^z(z,t)  &=& \displaystyle{\frac{g(z)}{1 + \partial_zg(z)t}}
\vspace{0.25cm}\\
\alpha(R,t)&=& 1           & \ \  & \tilde{R}(R,t) &=& R + f(R) t .
\end{array}
\end{equation}
In accordance with (\ref{2D_interpfunk}),
the interpolating functions, $f$ and $g$, are taken to be
\begin{eqnarray}
f(R) &=& \left[ 1 + \tanh\left(\left(R-R_c\right)/\delta_R\right)\right]/2 
+ \epsilon(R_c,\delta_R) \\
g(z) &=& \tanh\left(\left(z-z_c\right)/\delta_z\right)/2 + 
      \tanh\left(\left(z+z_c\right)/\delta_z\right)/2, 
\end{eqnarray}
where $\epsilon(R_c,\delta_R)=-\left[ 1 + \tanh\left(\left(-R_c\right)/\delta_R\right)\right]/2$,
(see Figure \ref{fig:mibaxi_fg}).
 \noindent
\begin{figure}
\centerline{ 
	\hbox{\epsfxsize =12cm\epsffile{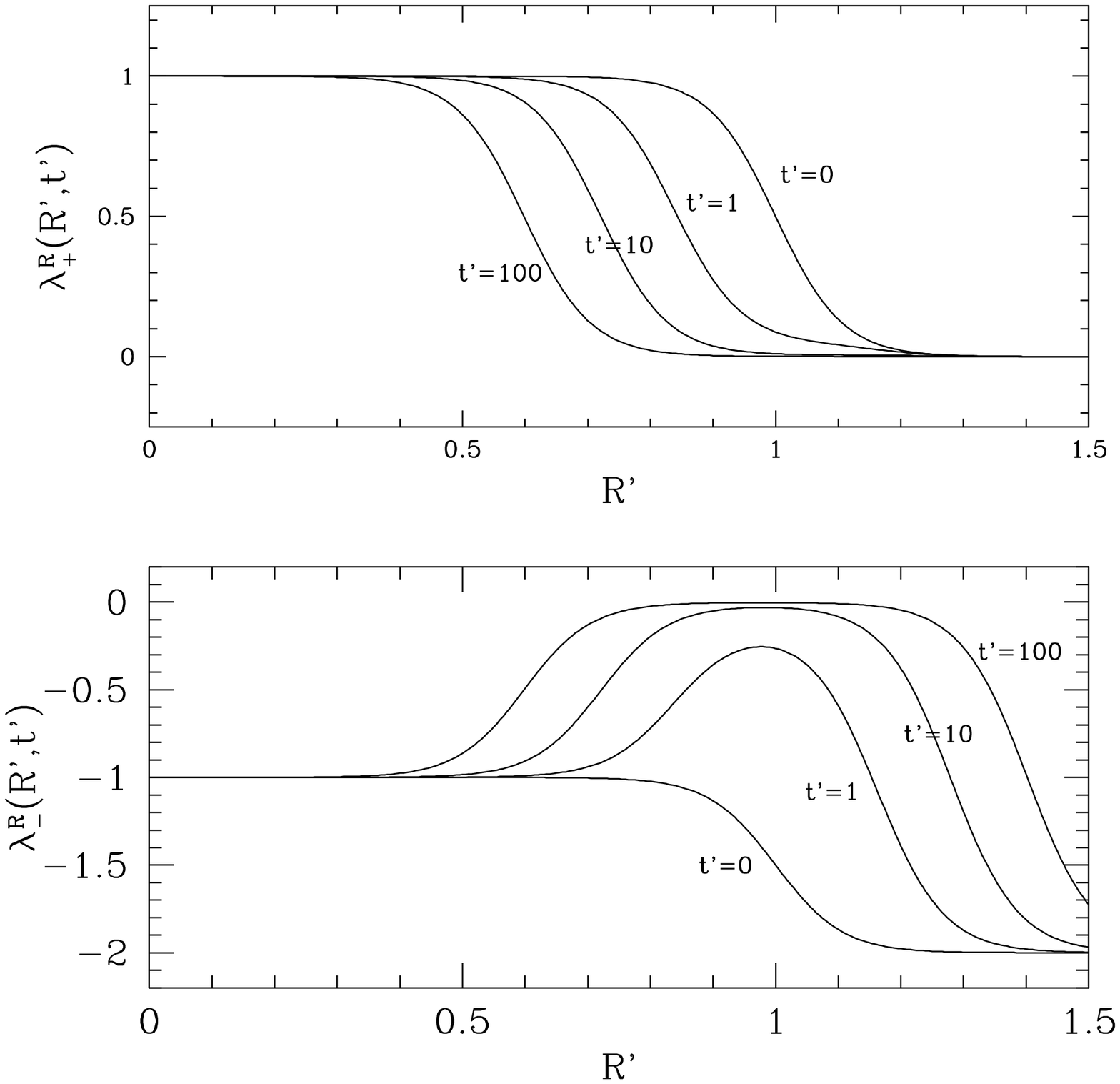}}
	}
\caption[$dR$:$dt$ characteristics for axisymmetric MIB system]
{\small \label{fig:2DRchars} 
Plot of the characteristic velocities, $\lambda^R_\pm$, as a function of $R'$ and 
$t'$, where $R'$ and $t'$ are axisymmetric MIB coordinates where 
$R_c$ is set to unity,
and $\lambda_+$ and $\lambda_-$ are the outgoing and ingoing characteristics,
respectively.  $\delta$ is taken to be $\delta_r\rightarrow \delta_r/R_c \approx 0.1067$ 
(corresponding to the system to be used in chapter 5).
Characteristic velocities are plotted for times $t'=0$, $1$, $10$, and $100$, 
and where $t'=100$ is larger than the lifetime of the longest of the 
longest lived solution studied in this work.
Both characteristic velocities approach zero around $r'=1$ as $1/{t'}$.
}
\label{fig:2Dzchars}
\end{figure}
\begin{figure}
\centerline{ 
	\hbox{\epsfxsize =12cm\epsffile{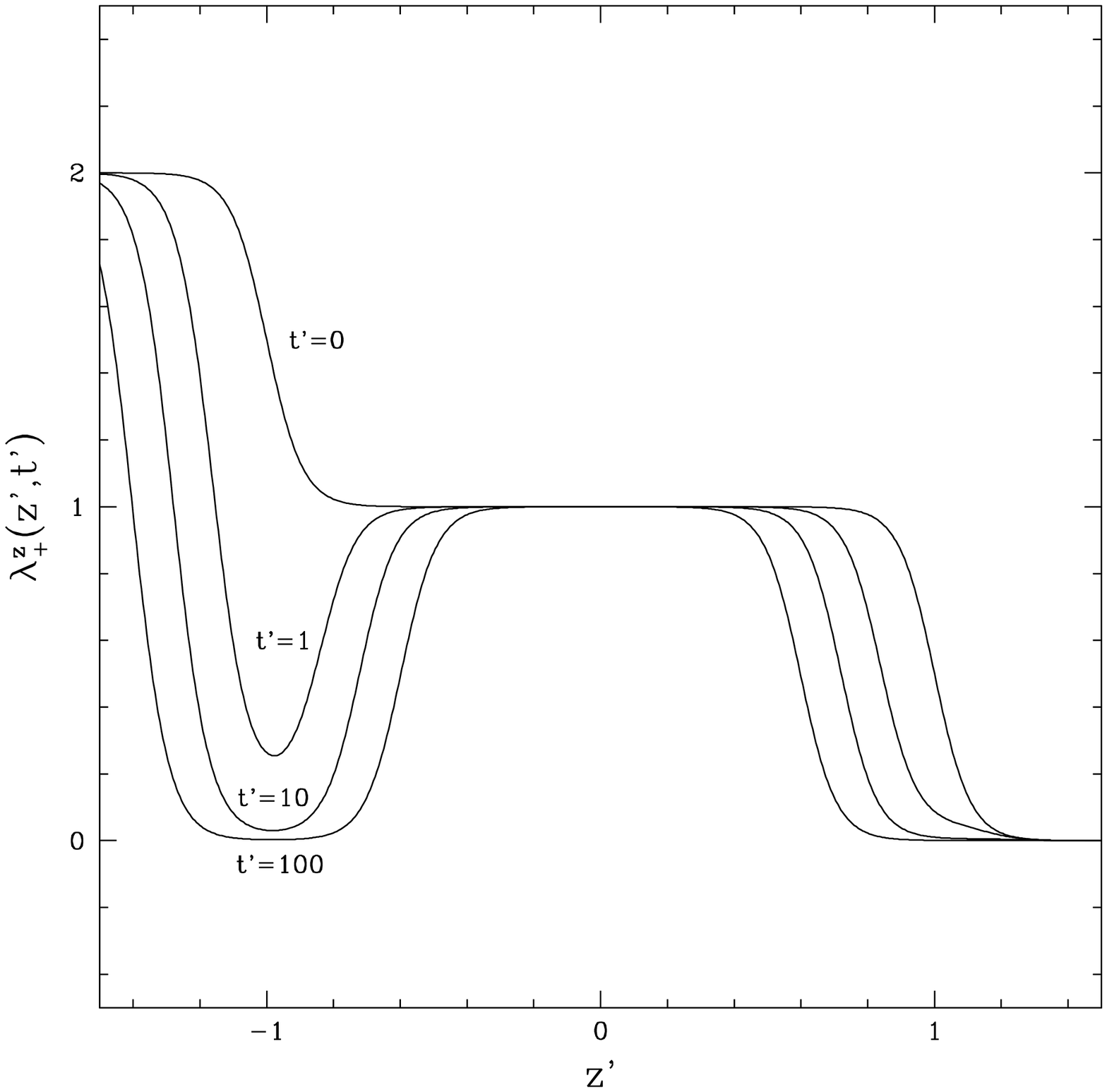}}
	}
\caption[$dz$:$dt$ characteristics for axisymmetric MIB system]
{\small \label{fig:2Dz1chars} 
Plot of the characteristic velocities, $\lambda^z_\pm$, as a function of $z'$ and 
$t'$, where $z'$ and $t'$ are axisymmetric MIB coordinates where $z_c$ is set to 
unity, and $\lambda_+$ and $\lambda_-$ are the outgoing and ingoing characteristics,
respectively.  $\delta_z$ is taken to be $\delta_z\rightarrow \delta_z/z_c \approx 0.1067$ 
(corresponding to the system to be used in chapter 5).
Characteristic velocities are plotted for times $t'=0$, $1$, $10$, and $100$
($t'=100$ is larger than the lifetime of  the longest lived solution studied in this work).
Both characteristic velocities approach zero around $z'=\pm 1$ as $1/{t'}$.
}
\label{fig:2Dz1chars}
\end{figure}
\begin{figure}
\centerline{ 
	\hbox{\epsfxsize =12cm\epsffile{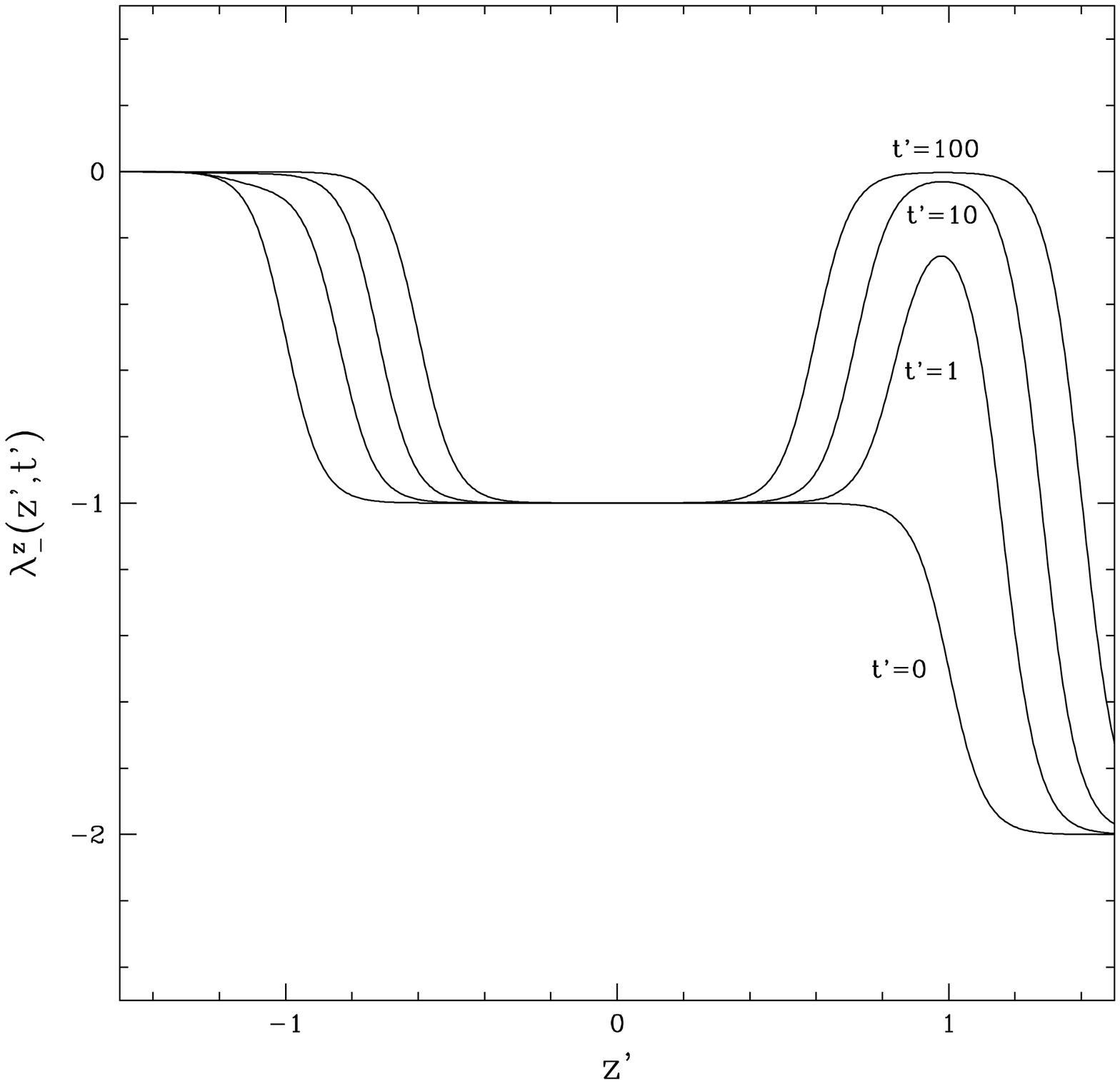}}
	}
\caption[$dz$:$dt$ characteristics for axisymmetric MIB system]
{\small \label{fig:2Dz2chars} 
Plot of the characteristic velocities, $\lambda^z_\pm$, as a function of $z'$ and 
$t'$, where $z'$ and $t'$ are axisymmetric MIB coordinates where $z_c$ is set to 
unity, and $\lambda_+$ and $\lambda_-$ are the outgoing and ingoing characteristics,
respectively.  $\delta_z$ is taken to be $\delta_z\rightarrow \delta_z/z_c \approx 0.1067$ 
(corresponding to the system to be used in chapter 5).
Characteristic velocities are plotted for times $t'=0$, $1$, $10$, and $100$
($t'=100$ is larger than the lifetime of the  longest lived solution studied in this work).
Both characteristic velocities approach zero around $z'=\pm 1$ as $1/{t'}$.
}
\label{fig:2Dz2chars}
\end{figure}

The characteristics of the system are
\begin{eqnarray}
\lambda^R_\pm &=& -\beta^R \pm 1/a \\
\lambda^z_\pm &=& -\beta^z \pm 1/b,
\label{eq:2dchars}
\end{eqnarray}
where $\lambda^R_\pm$ and $\lambda^z_\pm$ are the characteristics in the $R$
and $z$ directions, respectively.
As with the spherically symmetric case, the system behaves 
like the old ($\tilde{R}$, $\tilde{z}$, and $\tilde{t}$) coordinates 
for $R< R_c$ and $-z_c< z < z_c$, while around 
$R\approx R_c$ and  $z\approx \pm z_c$, the characteristic
velocities,
$\lambda^R_\pm\rightarrow 0$ and 
$\lambda^z_\pm\rightarrow 0$,
as $t\rightarrow \infty$ (as the inverse power of $t$).
This again has the effect of freezing-out the outgoing radiation.

\begin{figure}
\centerline{ 
	\hbox{\epsfxsize =12cm\epsffile{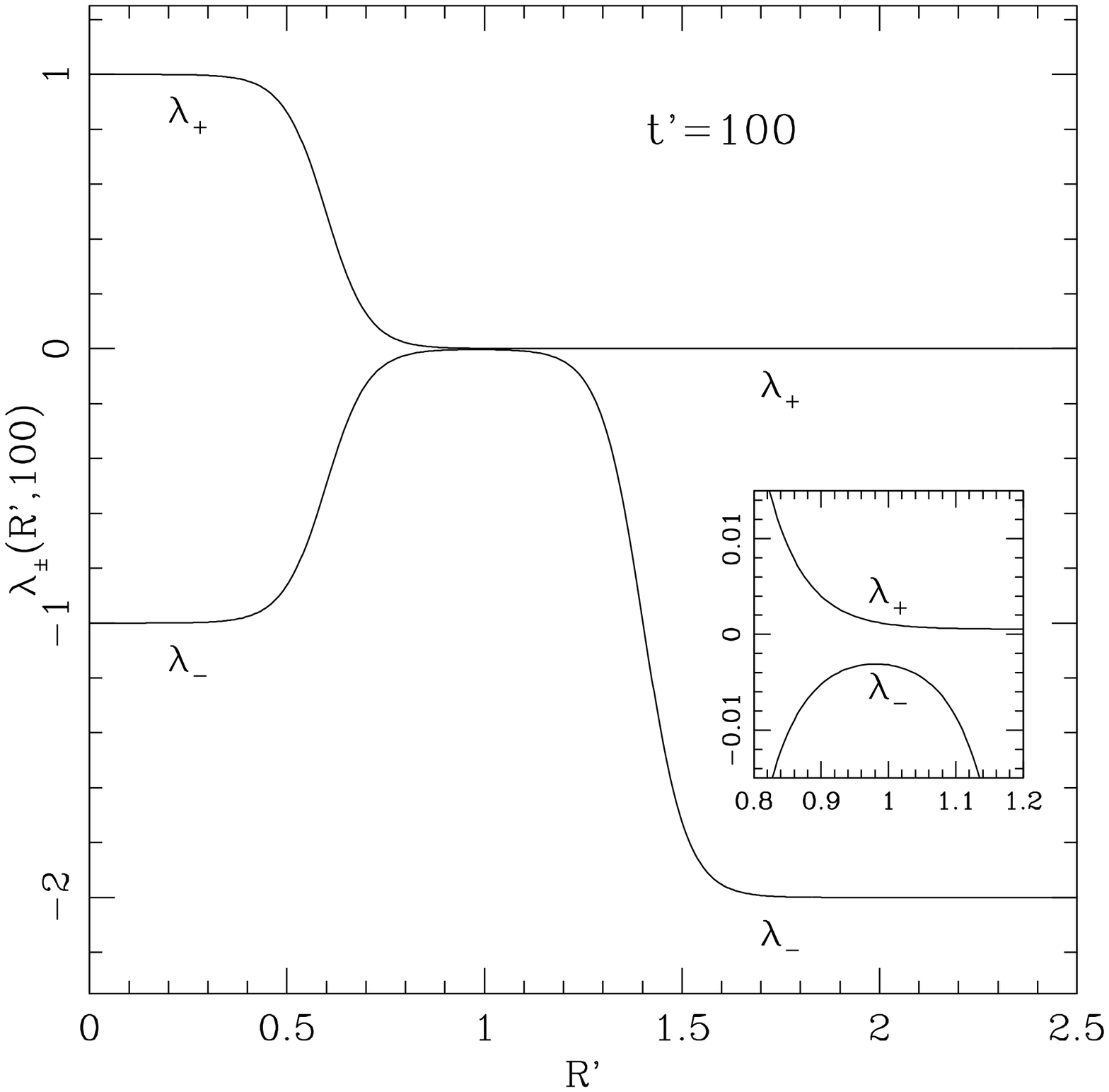}}
	}
\caption[Coincidence of ingoing/outgoint $R$:$t$ characteristics]
{\small \label{fig:2Drchars_coin} 
Plot of characteristic velocites, $\lambda_\pm(R',100)$, 
where $R'$ and $t'$ are axisymmetric MIB coordinates in units where $R_c$ is set to 
unity, and $\lambda_+$ and $\lambda_-$ are the outgoing and ingoing characteristics,
respectively.  $\delta$ is taken to be $\delta\rightarrow \delta/R_c \approx 0.1067$ 
(corresponding to the system to be used in chapter 5), and $t'=100$ is larger than 
the lifetime of the longest lived solution discussed in this thesis.
Both characteristic velocities approach zero around $r'=1$ as $1/{t'}$.
}
\label{fig:2Drchars_coin}
\end{figure}
\begin{figure}
\centerline{ 
	\hbox{\epsfxsize =12cm\epsffile{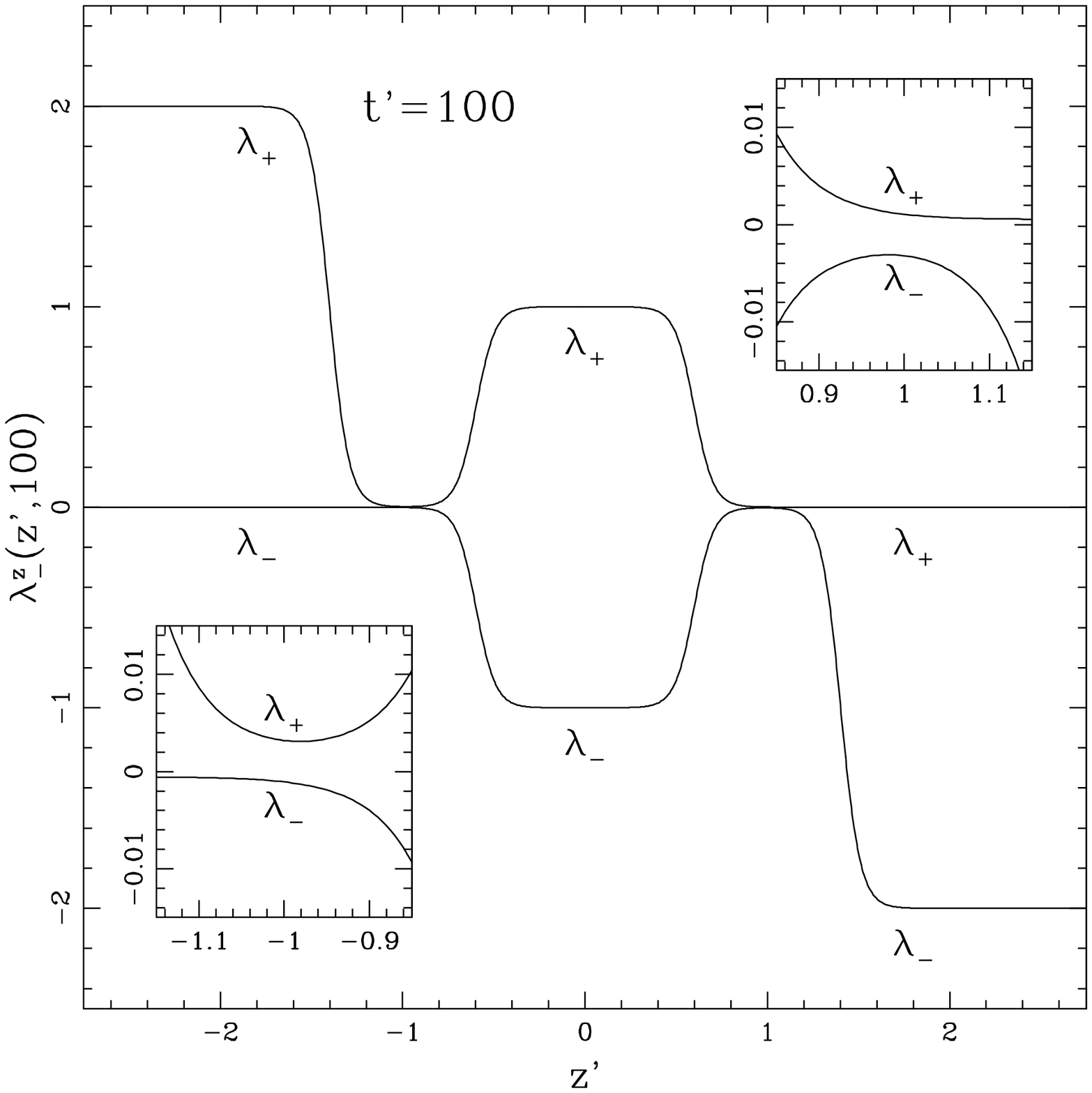}}
	}
\caption[Coincidence of ingoing/outgoint $z$:$t$ characteristics]
{\small \label{fig:2Dzchars_coin} 
Plot of characteristic velocites, $\lambda_\pm(z',100)$, 
where $z'$ and $t'$ are axisymmetric MIB coordinates in units where $z_c$ is set to 
unity, and $\lambda_+$ and $\lambda_-$ are the outgoing and ingoing characteristics,
respectively.  $\delta$ is taken to be $\delta\rightarrow \delta/z_c \approx 0.1067$ 
(corresponding to the system to be used in chapter 5), and $t'=100$ is larger than 
the lifetime of the longest lived solution discussed in this thesis.   
Both characteristic velocities approach zero around $r'=1$ as $1/{t'}$.
}
\label{fig:2Dzchars_coin}
\end{figure}

\section{Dissipation \& MIB Coordinates {\it Working Together}}

The primary reason dissipation was introduced and stressed in this
chapter is because it is integral to the stability of the MIB scheme.
Since MIB coordinates cause all the outgoing radiation to accumulate in
a very small region of the grid (creating large gradients in the 
fields), when MIB coordinates are used without 
dissipation the scheme is {\it always} unstable.
However, with dissipation the field is quenched in a stable 
and non-reflective manner. 
This can happen
because the amplification factor for the Crank-Nicholson scheme 
(analagous to that seen in figure \ref{fig:dissplot2}) is significantly 
less than one for high frequency solution components.
Therefore, the more the radiation gets spatialy compressed, the more it 
is dissipated.
Details of this process are discussed in chapter
\ref{chap:1D} in the context of spherically collapse. 

Note that although the above explanation is self-consistent and explains 
how the system {\it can} be stable, it certainly provides 
no convincing argument that the scheme {\it should} be stable (or non-reflective
for that matter). In fact, many techniques explored throughout 
the creation of the one discussed here were very unstable and reflected 
significant amounts of radiation!
In addition to the sponge-filter approach (mentioned in 
section \ref{sec:geometry}), 
there has been much research devoted to the study of
outgoing radiation and outgoing boundary conditions.
For the sake of completeness, we briefly review some of this
research here.

Cauchy-Characteristic Matching (CCM) is a
way to solve Einstein's Equations that matches typical 
Cauchy evolution on the interior of the computational domain
to characteristic evolution on the exterior,
\cite{bishop:1996}.
The original motivation was to measure the outgoing radiation
that reaches null infinity, however, in the process the method
eliminates the need for an outgoing radiation boundary condition.
The technique was developed for massless radiation although some
success has been made recently with the incorporation of matter
\cite{bishop2:1996},
\cite{lehner:2000}.
CCM is reported to be more efficient than previously 
used techniques, but still requires separate evolution
of two domains and calculations  to match them together;
the methods employed here involve a single Cauchy evolution.

Null-cone evolution \cite{gomez:1994} and other purely 
characteristic evolution techniques  (ie. 
where the entire evolution is performed
in characteristic coordinates,
\cite{gomez:1992}, \cite{winicour:1988}),
have also been successful in numerical relativity and 
seemed like a very promising way to evolve the nlKG 
equation.
Unfortunately, 
using characteristic coordinates with a truncated grid is still 
problematic due to the lack of a satisfactory outgoing boundary 
condition.
However, unlike typical ($r$,$t$) coordinates, compactification
is natural with null evolution and generally preserves the
smoothness properties of radiation~\cite{winicour:1988}.
Furthermore, in compactified characteristic coordinates 
exact boundary conditions can be set on the outer boundary 
of the grid, $\scri^+$. Numerical implementations 
can usually be evolved stably and radiation can 
be measured at $\scri^+$.
However, this tends to work well only for radiation that travels
along null characteristics (ie. massless fields).
When the fields are {\it massive} the field does not
propagate along null characteristics (and therefore 
does not reach null infinity), and there is either leakage out to 
${\scri^+}$ of massive field, or instabilities that arise from steep 
gradients due to compression of the outgoing radiation,
\cite{winicour:1999}. 

Similar problems occur evolving massive radiation 
with the conformal compactification methods 
developed by Friedrich, \cite{friedrich:1998}, and implemented 
numerically by Hubner, 
\cite{hubner:1996}
and Frauendiener,
\cite{frauendiener:1998a},
\cite{frauendiener:1998b},
\cite{frauendiener:2000};
the conformal compactification of the spacetime brings null infinity
to the outermost grid point and still causes problems with {\it massive}
radiation \cite{hubner:1996}.

\chapter{Spherically Symmetric Oscillons\label{chap:1D}}

\section{The Klein-Gordon Equation in MIB Coordinates with SDWP}
\label{sec:tex_eqs}

In this chapter we introduce the notation, equations of motion,
and discuss results from a computer code which uses 
spherically symmetric MIB coordinates.

We are interested in massive scalar field theory 
described by the (1+1) spherically symmetric action
\begin{equation} \label{eq:action}
S[\phi] = \int d^4x\sqrt{|g|}\left(
-\frac{1}{2} g^{\mu\nu}\nabla_\mu\phi \nabla_\nu\phi - V(\phi)
\right)
\end{equation}
where $\phi\equiv\phi(r,t)$, $V(\phi)$ is a symmetric double well
potential\footnote{This is identical to using $\displaystyle{V(\phi) = 
\frac{\lambda}{4}\left(
\phi^2 - \frac{m^2}{\lambda}\right)^2}$ and introducing dimensionless 
variables $r= \tilde{r} m$, $t = \tilde{t} m$, and 
$\psi=\frac{\sqrt{\lambda}}{m}\phi$.} (see figure \ref{fig:sdwp}), 
\hbox{$\displaystyle{V_{S}(\phi) = \frac{1}{4}\left(\phi^2 -1 \right)^2}$},
$g_{\mu\nu}$ is the flatspace metric in spherically symmetric 
MIB coordinates, (\ref{eq:1Dshiftmet}), and $g$ is the determinant 
of  $g_{\mu\nu}$.
\begin{figure}
\label{fig:sdwp}
\epsfxsize=12cm
\centerline{\epsffile{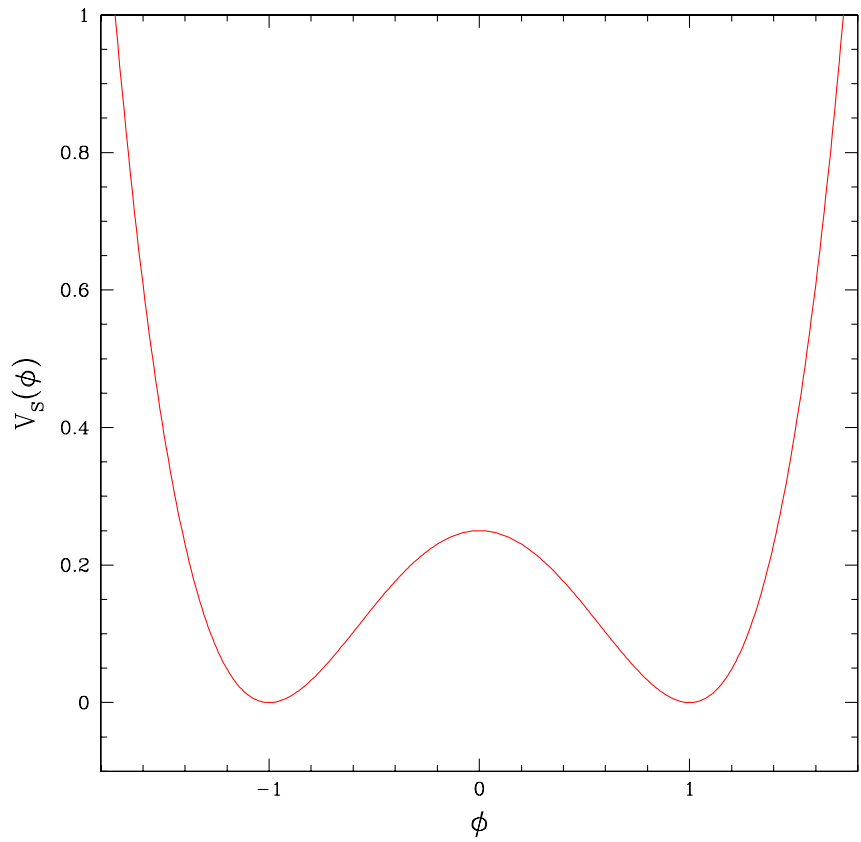}}
\caption[Symmetric double well potential (SDWP)]
{\small \label{fig:sdwp}
Symmetric double well potential, 
\hbox{$\displaystyle{V_{S}(\phi) = \frac{1}{4}\left(\phi^2 -1 \right)^2}$}.
The potential has two degenerate minima at $\phi=\pm 1$ and an unstable 
local maximum at $\phi=0$.
See figure \ref{fig:fieldprofiles_sdwp}
for detailed description of features and interpretation of kink profile 
(bubble) initial data.
}
\label{fig:sdwp}
\end{figure}
The equation of motion for the action (\ref{eq:action}) is
\begin{equation} \label{eq:boxphi}
\frac{1}{\sqrt{|g|}}\partial_\mu\left( \sqrt{|g|} g^{\mu\nu} 
\partial_\nu\phi \right) = \phi\left(\phi^2 -1 \right)
\end{equation}
which with (\ref{eq:1Dshiftmet}), (\ref{eq:ADM_1Dauxvars}), and 
the definitions 
\begin{eqnarray}
\Pi(r,t) & \equiv & \frac{a}{\alpha}\left( 
 	\partial_t{\phi} - \beta \partial_r\phi\right), \\
\Phi(r,t) & \equiv & \partial_r\phi,
\end{eqnarray}
yields
\beq 
\dot{\Pi}  &=& \frac{1}{r^2b^2}\left( r^2 b^2 \left( \frac{\alpha}{a} \Phi 
+ \beta \Pi \right)\right)'  
- 2 \frac{\dot{b}}{b}\Pi - \alpha a \phi \left( \phi^2 - 1\right)
\label{eq:EOM_Pi}\\
\dot{\Phi} &=& \left( \frac{\alpha}{a} \Pi + \beta \Phi \right)' 
\label{eq:EOM_Phi}\\
\dot{\phi} &=& \frac{\alpha}{a} \Pi + \beta \Phi 
\label{eq:EOM_phi}
\eeq
where \ $\dot{} \equiv \partial_t$ and \ ${}' \equiv\partial_r$. 
These equations are the familiar (3+1) form for
the spherically symmetric Klein-Gordon field coupled
to gravity.  However, instead of having a truly dynamical geometry, here
$a(r,t)$, $b(r,t)$, $\alpha(r,t)$, 
and $\beta(r,t)$ are known functions of $(r,t)$,
which result from the MIB coordinate transformation of {\it flatspace}.

\section{Finite Difference Equations}
\label{app:FDE}

Equations (\ref{eq:EOM_Pi},\ref{eq:EOM_Phi},\ref{eq:EOM_phi}) 
are solved using two-level second order (in both space and time) 
finite difference approximations on a static uniform grid
with $N_r$ grid points.
The scale of discretization is set by $\Delta r$ and $\Delta t =
\lambda \Delta r$, where we fixed the Courant factor, $\lambda$,
to $0.5$ as we changed the base discretization.

Using the operators from Table \ref{tab:1Dfdop}, 
$\partial_r\tilde{r}=a$, 
\hbox{$\partial_r=nr^{n-1}\partial_{r^n}$}, 
and $rb=\tilde{r}$,
the interior difference equations ($3 \le i \le  N_r -2$) are
\beq
\displaystyle{ \Delta_t \Pi^n_i } &=& 
\displaystyle{ 3\mu_t\left[a \Delta_{\tilde{r}^3}\left(
\tilde{r}^2 \left( \frac{\alpha}{a}\Phi + \beta\Pi\right)\right)
\right]^n_i} \nonumber \\
&& 
\displaystyle{-2\mu_t\left(\frac{\dot{b}}{b} \Pi
-\alpha a \phi\left(\phi^2 - 1\right)\right)^n_i} 
+ \mu_t^{\rm diss}\Pi^n_i,\\
\Delta_t\Phi^n_i &=& \mu_t\Delta_r\left( 
\frac{\alpha}{a}\Pi + \beta\Phi
\right)^n_i+ \mu_t^{\rm diss}\Phi^n_i,\\
\Delta_t\phi^n_i &=& \mu_t\left( 
\frac{\alpha}{a}\Pi + \beta\Phi
\right)^n_i+ \mu_t^{\rm diss}\phi^n_i.
\eeq
and include dissipation.
On the two grid points adjacent to the boundaries,
$i=2$ and $i=N_r-1$, the same equations are
solved without dissipation (since the {\it two} points needed
on either side are unavailable):
\beq
\displaystyle{ \Delta_t \Pi^n_i } &=& 
\displaystyle{ 3\mu_t\left[a \Delta_{\tilde{r}^3}\left(
\tilde{r}^2 \left( \frac{\alpha}{a}\Phi + \beta\Pi\right)\right)
\right]^n_i} \nonumber \\
&& 
\displaystyle{-2\mu_t\left(\frac{\dot{b}}{b} \Pi
-\alpha a \phi\left(\phi^2 - 1\right)\right)^n_i},\\
\Delta_t\Phi^n_i &=& \mu_t\Delta_r\left( 
\frac{\alpha}{a}\Pi + \beta\Phi
\right)^n_i,\\
\Delta_t\phi^n_i &=& \mu_t\left( 
\frac{\alpha}{a}\Pi + \beta\Phi
\right)^n_i. \label{eq:phidot}
\eeq 
These equations are solved using an iterative scheme, which 
is iterated until the $\ell_2$-norms of the solutions at $t^{n+1}$ 
converge to one part in $10^{8}$, where the $\ell_2$-norm is defined
for a vector $x_i$ to be
\begin{equation}
\displaystyle{
||{\bf{\bbx}}||_2 = \left( \frac{1}{N} \sum^N_i|x_i|^2\right)^{1/2}
}
\label{eq:l2norm}
\end{equation} 
Since the functions $a$, $b$, $\dot{b}$, $\alpha$, $\beta$, 
and $\tilde{r}$ are explicitly known functions, when discretized
we simply evaluate the given function at  
\hbox{$r_i=(i-1)\Delta r$} and \hbox{$t^n=(n-1)\Delta t$}.
On the inner ($r=0$) boundary we applied conditions necessary for  
regularity at the origin:
\begin{equation}
\Phi^n_i = 0
\label{eq:Phieq0}
\end{equation}
\begin{equation}
\mu_t\left( \Delta^f_i\Pi  -
\displaystyle{ \frac{a'}{a} \Pi }\right)^n_i =0
\label{eq:PiPrmeq0}
\end{equation}
where (\ref{eq:Phieq0}) is a statement of regularity in the 
scalar field, and (\ref{eq:PiPrmeq0}) results from equation
(\ref{eq:phidot}) and the commutation of partial derivatives.
The field $\phi$ is evolved using 
\begin{equation}
\Delta_t\phi = \mu_t\left( 
\frac{\alpha \Pi}{a} + \beta\Phi
\right)^n_i.
\end{equation}
For the outer boundary, our choice of equations makes very little
difference since the {\it physical} radial position  
($\tilde{r}$) corresponding to the outermost point is moving out at
nearly the speed of light, therefore none of the outgoing field 
ever reaches this gridpoint.  Nevertheless, we employed the typical
massless scalar field outgoing boundary condition 
for $\Pi$ and $\Phi$, and evolved $\phi$ with its equation of motion:
\begin{equation}
\Delta_t \Pi^n_i + \mu_t \left( 
\Delta^b_r\Pi + \frac{\Pi}{r}
\right)^n_i=0,
\end{equation}
\begin{equation}
\Delta_t \Phi^n_i + \mu_t \left( 
\Delta^b_r\Phi + \frac{\Phi}{r}
\right)^n_i=0,
\end{equation}
\begin{equation}
\Delta_t\phi = \mu_t\left( 
\frac{\alpha \Pi}{a} + \beta\Phi
\right)^n_i.
\end{equation}

\section{Testing the MIB Code}
\label{sec:codetest}

One might think that freezing out the outgoing radiation 
while keeping a static uniform mesh  
would lead to a ``bunching-up'' of the outgoing radiation from 
the oscillating source which would cause a loss of resolution,
numerical instabilities, and eventually crash the code.
However, as already discussed, this turns out not to be the case;
all outgoing radiation is stably ``frozen-out'' around
$r\approx r_{\rm c}$ and the steep gradients that should form 
in this region are smoothed out and quenched by the dissipation
(see figure \ref{fig:dissplot2}).
As the oscillations approach $r\approx r_c$ their wavenumber
approaches the Nyquist limit,
$\xi\rightarrow \pi$, and since the amplification factor for 
these modes is significantly less than one they are rapidly quenched
(analogously to the solution of the advection equation using the 
CN scheme shown in figure \ref{ampfacs}).
Put simply, the coordinate system causes the oscillations
to ``bunch-up'', while the higher-order dissipation quenches the 
resultant high-frequency modes.

\begin{figure}
\epsfxsize=14cm
\centerline{\epsffile{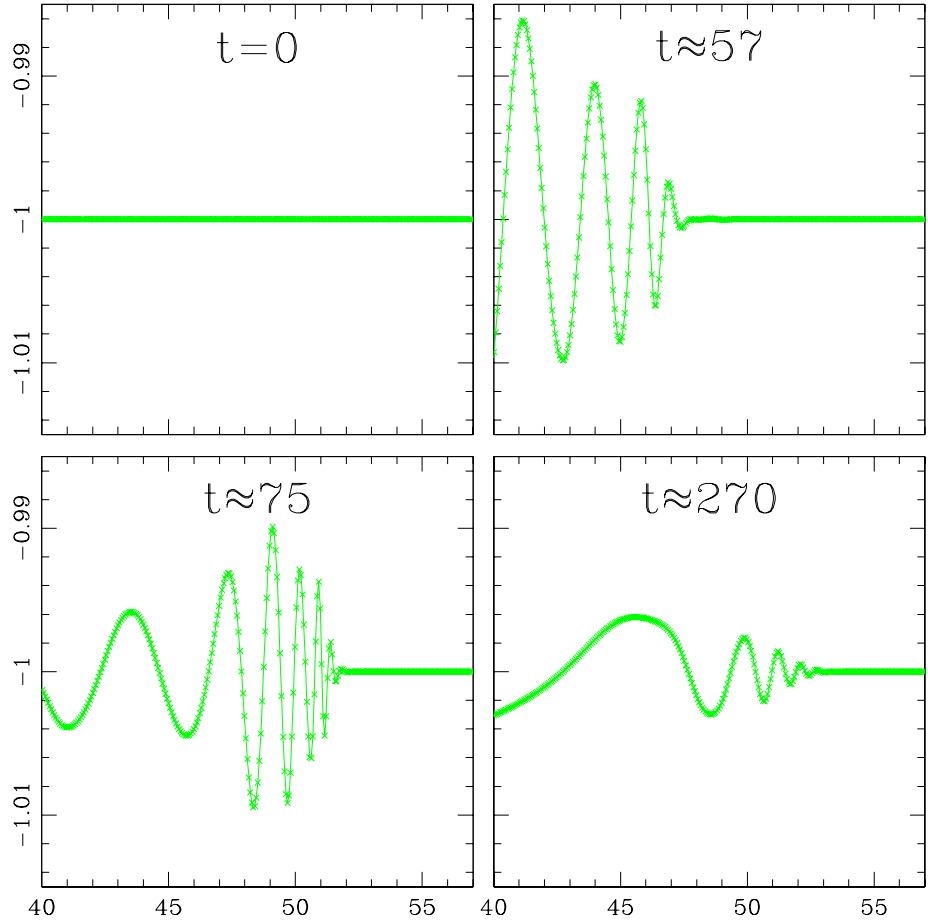}}
\caption[Outgoing radiation in freeze-out region]
{\small \label{fig:dissplot2}
Fundamental field $\phi(r)$ in the freeze-out region 
(cutoff actually at $r_c=56$)
at $t=0,\ 57,\ 75,\  {\rm and}\ 270$.
As the characteristic velocities of the radiation go to zero around $r_c$
(as seen in figure \ref{fig:1Dchars_coin}),
the wavelength of the radiation 
is shortened to the Nyquist limit on the lattice, $2\Delta r$.
The higher-order dissipation added to the system 
is responsible for quenching the field;
figure \ref{ampfacs} shows the amplification factor $|\rho|^2$ 
as a function of wavenumber is significantly less than one as the wavenumber 
approaches the Nyquist limit (the plot of $|\rho(\xi)|^2$ 
is actually for the advection equation, but is qualitatively similar).
}
\label{fig:dissplot2}
\end{figure}

There {\it is} a loss of resolution and
second order convergence directly around $r_{\rm c}$, but this does
not affect stability or convergence of the solution for $r \ll r_{\rm c}$.  
Figure (\ref{fig:convtest}) shows a convergence test for the field
$\phi$ for $r < r_{\rm c}/2$ over roughly six crossing times.
\begin{figure}
\epsfxsize=14cm
\centerline{\epsffile{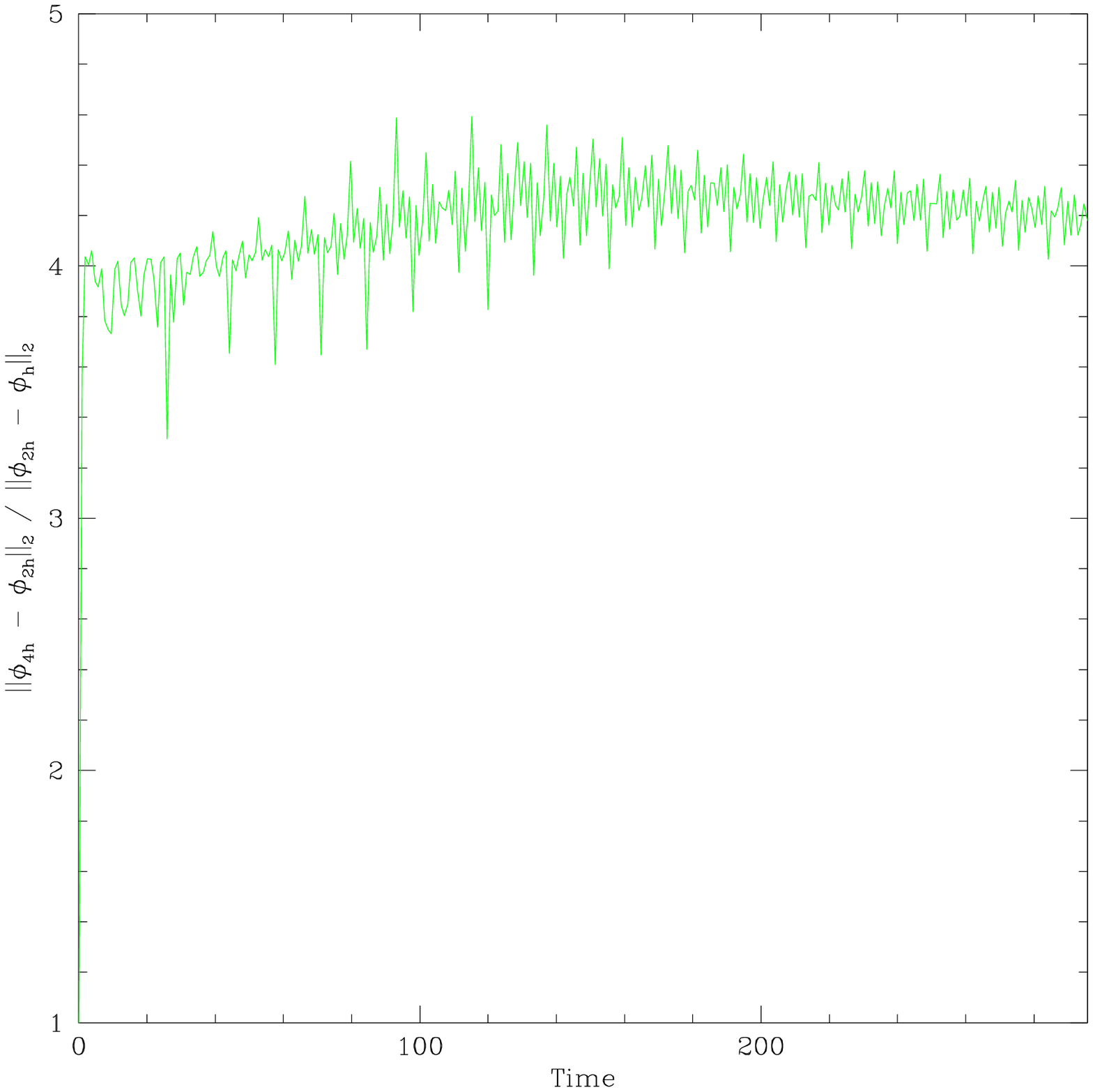}}
\caption[Convergence test of fundamental field, $\phi$]
{\small \label{fig:convtest}
Convergence factor, $C_f=||\phi_{4h} - \phi_{2h}||_2/||\phi_{2h}-\phi_{h}||_2$, 
for the field $\phi$
composed from the solution at three different discretizations 
(value of 4 indicates 2nd order convergence, as shown in 
section \ref{sec:convergence}).
}
\label{fig:convtest}
\end{figure}
Since we are solving equation (\ref{eq:boxphi}) in flatspace, it is very
simple to monitor energy conservation.  The spacetime 
admits a timelike Killing vector, $t^\nu$, so we have 
a conserved current, $J_\mu\equiv t^\nu T_{\mu\nu}$.  We monitor the flux of $J_\mu$ through
a gaussian surface constructed from two adjacent spacelike hypersurfaces
for $r\leq r_{\rm c}$ (with normals $n_\mu = (\pm 1,0,0,0)$),
and an ``endcap'' 
at $r=r_{\rm endcap}$ (with normal
$n_r^\mu = (0,a^{-1},0,0))$.
To obtain the the conserved energy at a time, $t$, 
the energy contained in the bubble,
\be
E_{\rm bubble}(t) = 4\pi\hspace{-0.15in}\int\limits_{0}^{\ \ \ r_{endcap}} 
\hspace{-0.125in}r^2b^2 \left(
\frac{\Pi^2 + \Phi^2}{2 a^2} + V(\phi)\right) dr, 
\label{eq:ebub}
\ee
(where the integrand is evaluated at time $t$) 
is added to the total radiated energy, 
\be 
E_{\rm rad}(t) = 4\pi\hspace{-0.05in}\int\limits_{0}^{\ t}
\hspace{-0.05in} r^2b^2 
\frac{\Pi \Phi}{a^2} dt'
\label{eq:erad}
\ee
(where the integrand is evaluated at $r=r_{\rm endcap}$). 
The sum, $E_{\rm total} = E_{\rm bubble} + E_{\rm rad}$, remains
conserved to within a few tenths of a 
percent\footnote{A few hundredths of a percent if measured from after 
the initial radiative burst from the collapse.}
through almost two thousand oscillation periods
($t\approx 8000 m^{-1}$ or up to a  quarter million iterations, see Fig. (\ref{fig:econ_a})).
  
\begin{figure}
\epsfxsize=14cm
\centerline{\epsffile{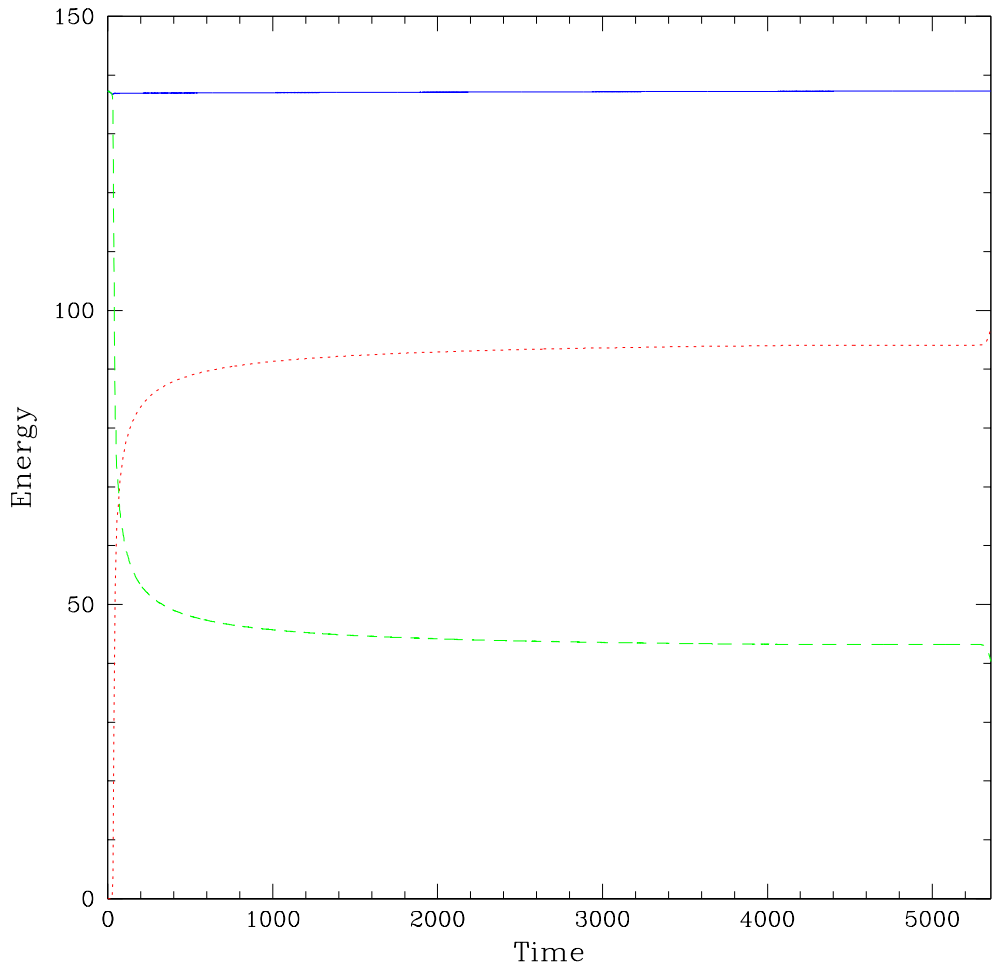}}
\caption[Energy conservation test of 1D bubble code]
{\small \label{fig:econ_a}
Plot of energy contained in oscillon (dashed green lines), 
energy radiated (dotted red lines),
and total energy (solid blue line).  
The total energy of the system is a constant of motion and is numerically
conserved to within a few tenths of a percent (hundredths of a percent if 
measured from after the initial radiative burst from the collapse).
The energy contained within the oscillon drops rapidly during the 
initial radiative phase and plateaus around $E\approx 43m/\lambda$ 
during the pseudo-stable regime of the oscillon.  
}
\label{fig:econ_a}
\end{figure}

\begin{figure}
\epsfxsize=14cm
\centerline{\epsffile{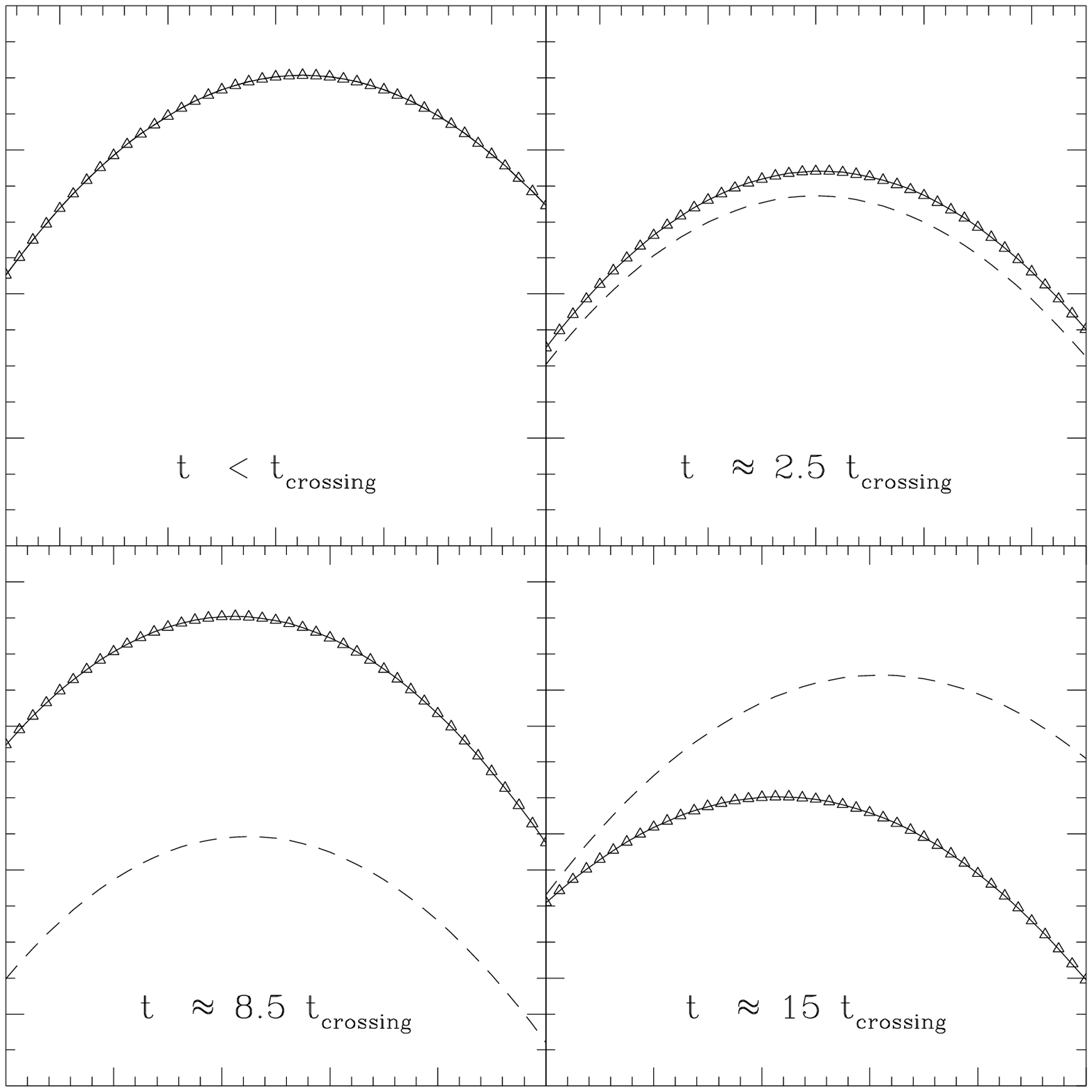}}
\caption[Contamination test for 1D bubble code, $\phi(0,t)$
 after many crossing times]
{
$\phi(0,t)$ versus time 
for the large-grid solution (blue triangles),
MIB solution (green solid curves), and OBC solution (red dashed curves).
The solutions all agree before $2 t_{\rm crossing}$, but the
OBC solution begins to drift away from the ideal solution
after $2 t_{\rm crossing}$.  
The error in the OBC solution is due to radiation that is
reflected off of the outer boundary (hence needing two crossing 
times to return to $r=0$ to contaminate the oscillon).
All pictures span the same area, 
$\Delta \phi = 0.075$ by $\Delta t = 0.5$. 
}
\label{fig:crossingtimes}
\end{figure}

Although monitoring energy conservation is a very important test,
it says little about whether there is reflection off of the 
outer boundary or the $r\approx r_{\rm c}$ region.  
To check this we compare output from the MIB code to two 
other types of numerical solutions.
The first type of reference solution involves evolution of 
equation (\ref{eq:boxphi}) 
in ($\tilde{r}$,$\tilde{t}$) 
coordinates, but on a grid so large that radiation never reaches
the outer boundary (large-grid solutions).  These solutions serve
as our ``ideal'' reference solutions (for a given discretization), since we are
guaranteed that they are free of contamination from reflections
off the outer boundary.  The second type of reference solution involves
evolution on spatial grid comparable in extent to the MIB grid, but using 
an outgoing
boundary condition (OBC solutions).  Since we
know these solutions {\em do} have error resulting from reflection 
off of the outer boundary, they demonstrate clearly what can go
wrong when the solution becomes contaminated by reflected 
radiation.
\begin{figure}
\epsfxsize=14cm
\centerline{\epsffile{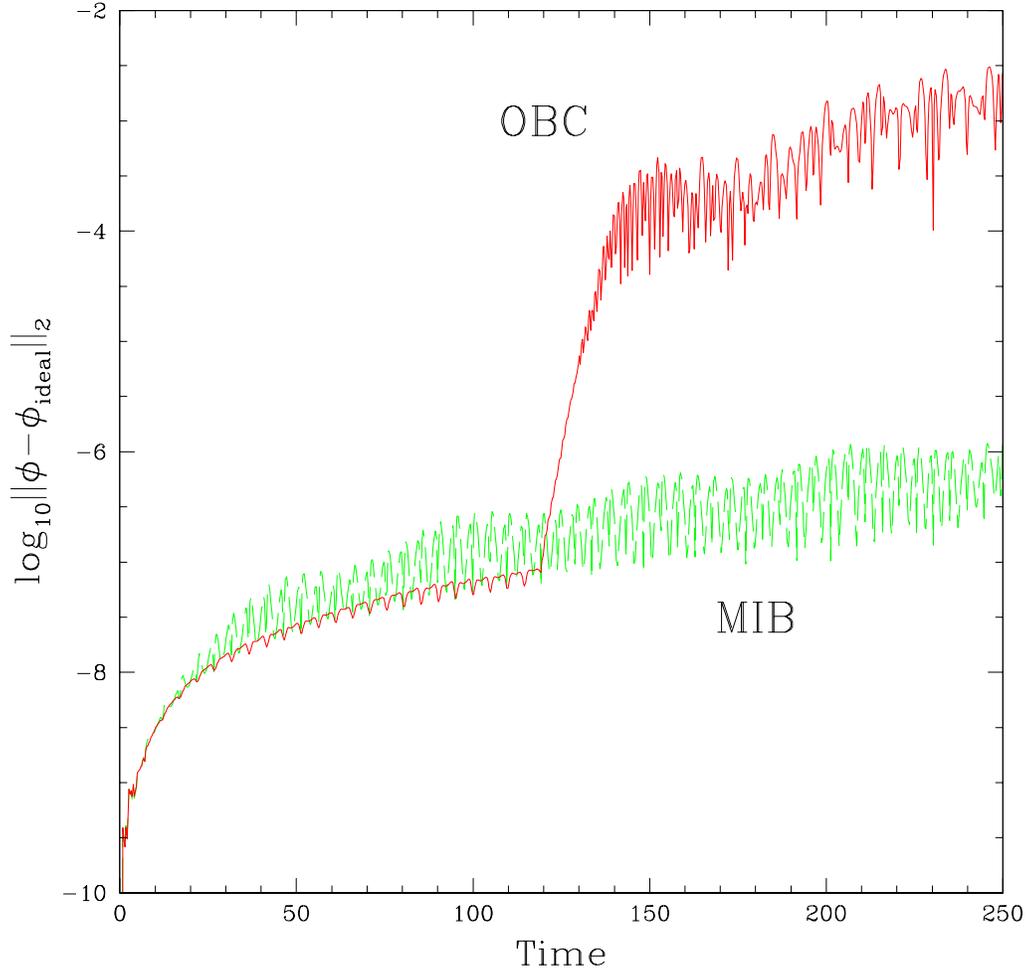}}
\caption[Contamination test of 1D bubble code]
{ \small \label{fig:lerr}
Plot comparing the OBC (solid red) and MIB (dashed green) 
solutions to an ``ideal'' solution.
The OBC solution is obtained using a massless outgoing
boundary condition, the MIB solution is obtained by solving the 
system in spherical MIB coordinates, and the ideal solution is 
obtained by evolving the solution in standard ($r$,$t$) coordinates on 
a grid large enough to ensure no reflection off the outer boundary.    
The error estimates are obtained from the $\ell_2$-norm of the difference 
between the trial solutions (OBC or MIB) and the ideal solution, 
$||\phi-\phi_{\rm ideal}||_2$.
Contamination of the OBC solution is observed at two crossing times, $t\approx 120$,
where the error estimate increases over three orders of magnitude.
The MIB solution error grows slowly and steadily as expected when solving
a continuum equation with two {\it different} finite difference 
techniques.
}
\label{fig:lerr}
\end{figure}
Treating the large-grid solution as ideal, Fig. (\ref{fig:lerr}) 
compares typical $\log_{10}||\phi-\phi_{\rm ideal}||_2$ for 
the MIB and OBC solutions.
There is a steep increase in the OBC 
solution error (three orders of magnitude) 
around $t=125$, which is at roughly
two crossing times.  This implies that some radiation emitted from 
the initial collapse reached the outer boundary and reflected
back into the region $r<r_{\rm endcap}$.  There is no such behavior
found in any MIB solutions.
Lastly, for a more direct look at the field itself, we can see
$\phi(0,t)$ for large-grid (triangles), MIB (solid curves), and OBC 
(dashed curves) solutions 
in Fig. (\ref{fig:crossingtimes}).
Initially, both the MIB and OBC solutions agree with the large-grid 
solution extremely well, while after two crossing times the OBC solution 
starts to diverge.  

Since the MIB solution conserves energy to second 
order, converges quadratically (in the domain of interest), and 
is equivalent to the 
large-grid (again, within second order error), it
is an acceptable means of solving equations 
(\ref{eq:EOM_Pi},\ref{eq:EOM_Phi},\ref{eq:EOM_phi}) while being 
a great deal more computationally efficient 
than other previously used techniques.  In fact, assuming that 
dynamical grid methods add more points linearly with time, 
$N_r(t) \propto t$, the 
total computational work, $W\propto t\,N_r(t) $, grows as $t^2$.
With the MIB method, which uses a static 
uniform mesh, the computational demand grows {\it linearly} with the 
oscillon lifetime, $W\propto t\,N_r$ for $N_r$ constant.  
Thus, particularly for long integration times, there is 
signifcant cost benefit in using the MIB system instead of a 
dynamically-growing-grid technique.

\section{The Resonant Structure of Oscillons \\
\hbox{({\it 1D Critical Phenomena I})}}
\label{sec:tex_fine}

 
Copeland {\it et al.} \cite{copeland:1995}, showed quite clearly that 
oscillons formed for a wide range of initial bubble radii.
By collapsing bubbles of many different radii,
they even caught a glimpse of resonant structure (which 
in large part motivated this study) but they did not explore 
the parameter space in detail.
With the efficiency of our new code, we are able to explore
much more of the parameter space for a given amount of computational 
resources than with either large or dynamically growing grids.

Following \cite{copeland:1989} we use a gaussian profile for initial 
data where the field at the core and outer boundary values are 
set to the vacuum values ($\phi_c=1$ and $\phi_o=-1$ respectively) 
and the field interpolates between them at a characteristic radius,
$r_0$:
\be
\phi(r,0) = 
\phi_o + \left(\phi_c - \phi_o\right) \exp\left( 
- r^2 / r_0^2 \right).
\ee
By keeping 
$\phi_c$ and $\phi_o$ constant but varying $r_0$, we have a one
parameter family of initial data to explore.
Figure \ref{fig:lifetime1} shows the behavior of oscillon lifetime 
as a  function of $r_0$.
\begin{figure}
\epsfxsize=14cm
\centerline{\epsffile{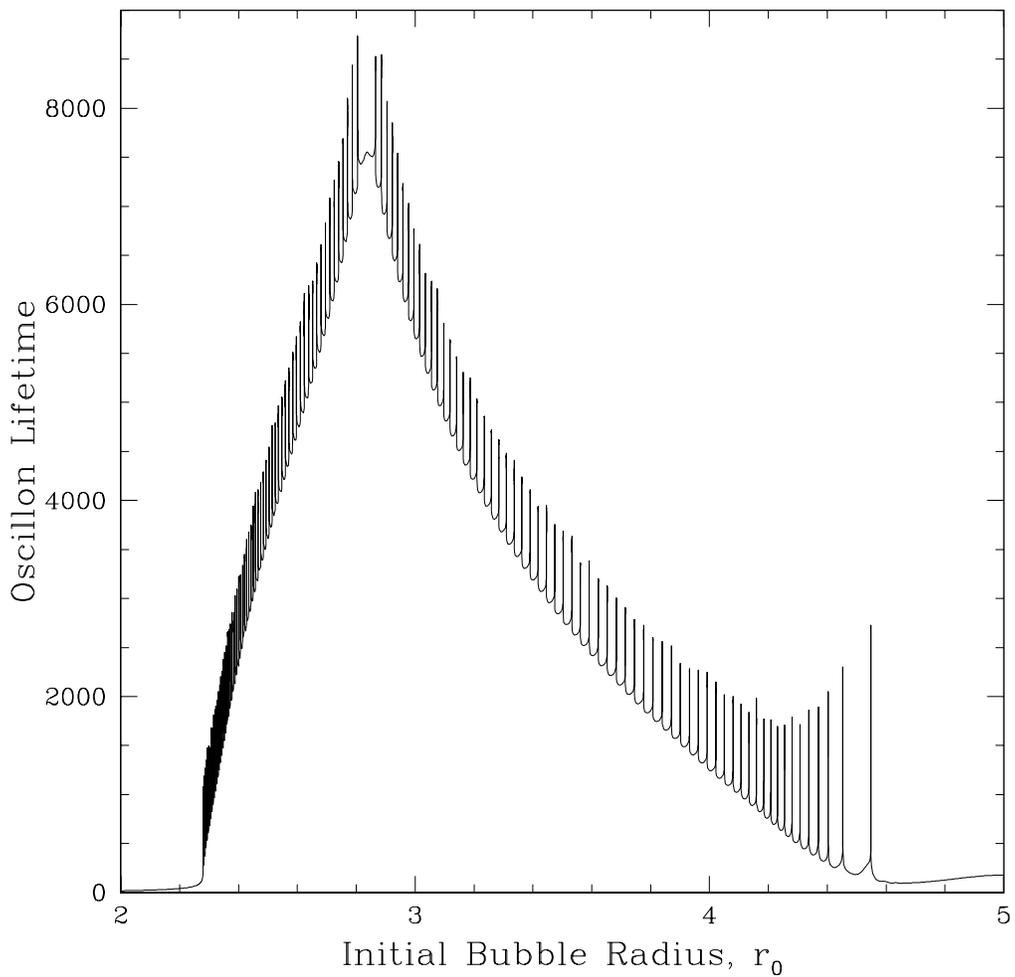}}
\caption[Oscillon lifetime versus initial bubble radius]
{\small  \label{fig:lifetime1}
Plot of Oscillon lifetime versus initial bubble radius
for $2.0 \le r_0 \le 5.0$.  
With the high resolution parameter space survey the 125 resonances
become apparent.  The survey was conducted at (relatively)
coarse resolution to reveal the location of the resonances.
Once a resonance was found, it was resolved efficiently 
first by using a three-point routine that maximizes
the lifetime as a function of initial bubble radius, and then by
bisecting (to one part in $10^{14}$) 
on the bifurcate modulation behavior discussed below
(and seen in figure \ref{fig:phenv_energy}). 
}
\label{fig:lifetime1}
\end{figure} 
We discuss three main findings that are distinct from 
previous work: the existence of resonances and their 
time scaling properties, 
the mode structure of the resonant solutions, and the 
existence of oscillons outside the region $2\leq r_0\leq 5$.

\subsection{Time Scaling}

\begin{figure}
\epsfxsize=14cm
\centerline{\epsffile{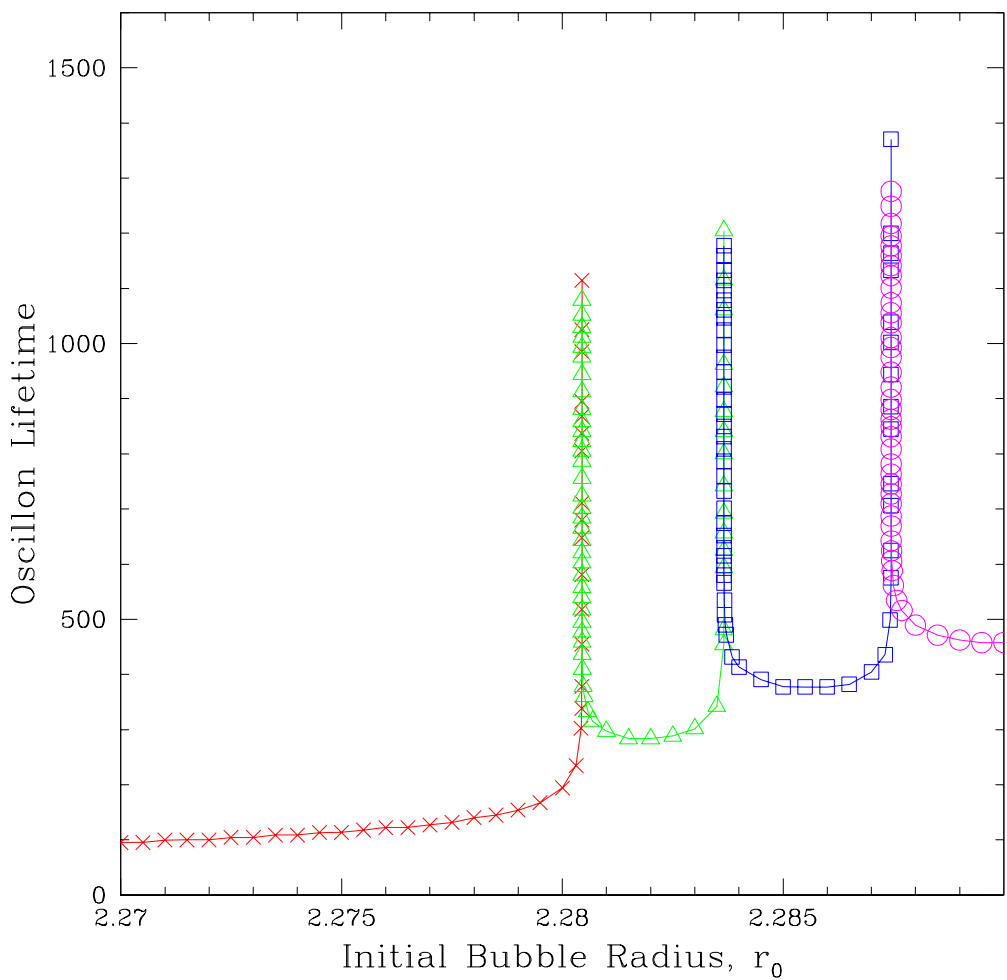}}
\caption[Oscillon lifetime versus initial bubble radius]
{
Plot of oscillon lifetime versus initial bubble radius
for  $2.27\le r_{initial}\le 2.29$.
The three resonances shown occur at 
$r^*_0\approx 2.2805$,
$r^*_1\approx 2.2838$,
and
$r^*_2\approx 2.2876$.
Each resonance separates the parameter space into regions
with $n$ and $n+1$ modulations; the red points correspond
to oscillons with no modulations, the green points to oscillons
with one modulation, the blue to two modulations, and the 
violet to three modulations.
}
\label{fig:life_1449.r0=2.27_2.29}
\end{figure}

A new feature is observed in the lifetime profile 
of collapsing bubbles;
figure \ref{fig:lifetime1} displays 
the appearance of 125 resonances.
These resonances (also seen in Fig.
\ref{fig:life_1449.r0=2.27_2.29})
become visible only after carefully resolving the parameter space.
Upon fine-tuning to one part in $\sim 10^{14}$ we 
noticed interesting bifurcate behavior about the resonances
(figure \ref{fig:phenv_energy}, top).
The field oscillates with a period
$T \approx 4.6$, 
so the individual oscillations 
cannot be seen in the plot, but it is the modulation that is 
of interest here\footnote{This would be
$\tilde{T} = 4.6m^{-1}$ in the original coordinate system.  To recover
the proper dimensions, lengths and times are multiplied by 
$m^{-1}$ and energies by $m \lambda^{-1}$.}.  
The top figure shows the envelope of  $\phi(0,t)$ 
on both sides of a resonance (dotted and solid curves).  We see that the 
large period 
modulation that exists for all typical oscillons disappears late in the
lifetime of the oscillon as $r_0$ is brought closer to a resonant
value (which we define to be $r^*_0$).
On one side of $r^*_0$ the modulation returns before the oscillon 
disperses (solid curve),
where on the other side, the modulation does not return and the the
oscillon just disperses (dotted curve).  
This behavior does not manifest itself until $r_0$ is quite close
to $r^*_0$.  So in practice, we used a three point maximization 
(bracketing interval of $\sim\!0.62$) 
routine to get $r_0$ close to $r^*_0$ and then we bisected on the 
bifurcate behavior thereafter (bracketing interval of $0.5$). 
This is {\it much} more efficient than a uniformly sampled parameter
space; if the parameter space were sampled {\it uniformly} 
at the finest resolution used in figure \ref{fig:lifetime1},
the survey would contain approximately $3\times 10^{14}$ points,
whereas we used only 7605.
\begin{figure}
\epsfxsize=14cm
\epsfysize=14cm
\centerline{\epsffile{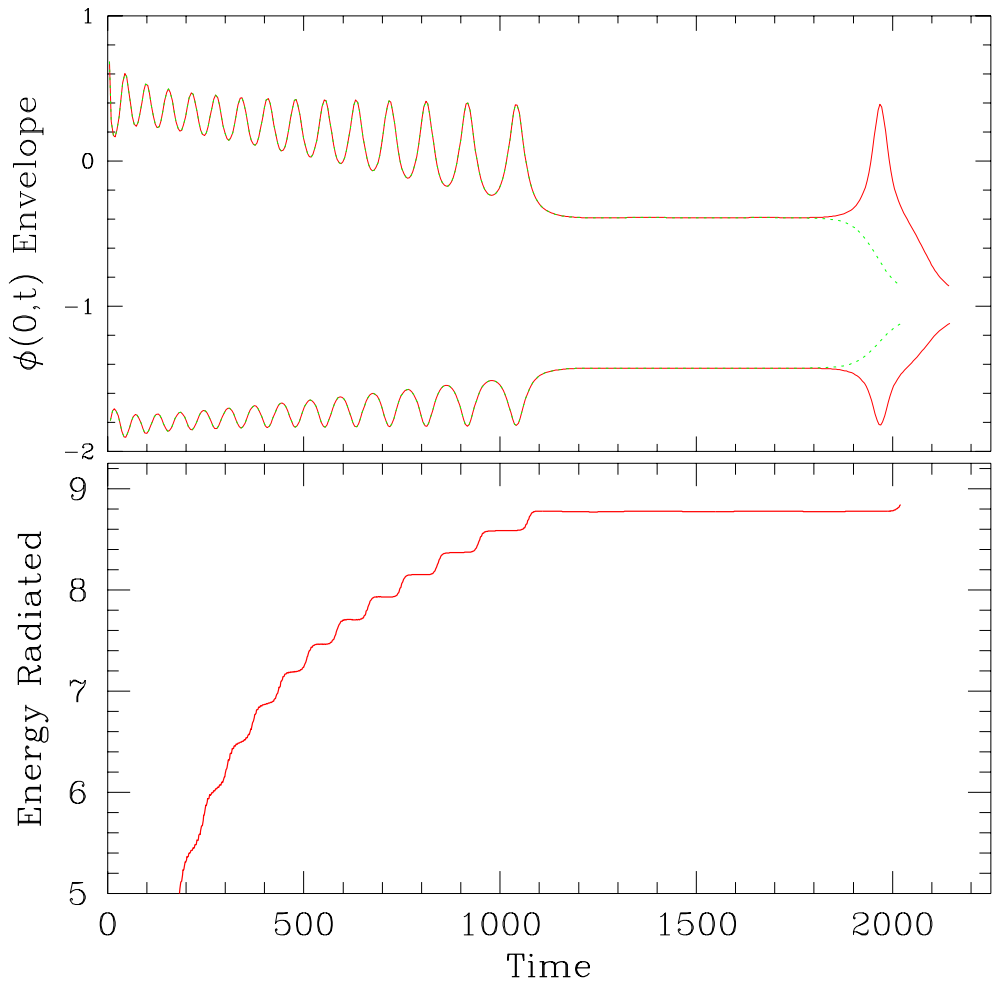}}
\caption[Field envelope barely above/below resonance and 
power radiated versus time. ]
{\small \label{fig:phenv_energy}
Top plot shows the envelope of $\phi(0,t)$ for $r^*_0\!\pm\!\Delta r_0$
displaying bifurcate behavior around the 
$r^*_0 \approx 2.335$ resonance ($\Delta r_0 \sim 10^{-14}$);
the solid red curve is the envelope barely above resonance while
green dotted line is the envelope barely below resonance.
Bottom plot shows the energy radiated out of the gaussian surface
containing the oscillon as a function of time.  The increases in the
energy radiated are synchronized with the modulation in the field.
}
\label{fig:phenv_energy}
\end{figure}
Although we can see from Fig. \ref{fig:phenv_energy} that
the modulation is directly linked to the resonant solution, 
it is not obvious why this is so.
However, if we look at the relationship between the modulation in the 
field (top) to the power radiated by the oscillon (bottom), 
we see that they are clearly synchronized.  

This should really be no surprise for one familiar
with the 1+1 \kkbar scattering studied with the
same model \cite{campbell:1983}.  Campbell, et al., 
showed in a study of \kkbar collisions that 
after the ``prompt radiation'' phase (the initial release
of radiation upon collision)
the remaining radiation emitted is from the decay of what they 
referred to as ``shape'' oscillations.  The ``shape modes'' 
were driven by the contribution to the field ``on top'' of the 
$K$ and \kbar soliton solutions.  Since the exact closed-form solution for 
the ideal non-radiative \kkbar interaction is not known, initial data
are only an approximation and the ``leftover'' field is responsible for 
exciting these shape modes.  The energy stored in the 
shape modes slowly decays away as the kink and antikink interact
and the solution eventually disperses.

In our case, we believe the large period modulation is 
signaling the excitation of a similar ``shape mode'' on top
of a periodic, non-radiative, localized, oscillating
solution. On either side
of a resonance in the $r_0$ parameter space, the solution is on the 
threshold of having one more shape mode oscillation.
If this is the case, and we are ``tuning in'' this unstable 
shape mode, we might expect to see a scaling law arise.
\begin{figure}
\epsfxsize=14cm
\centerline{\epsffile{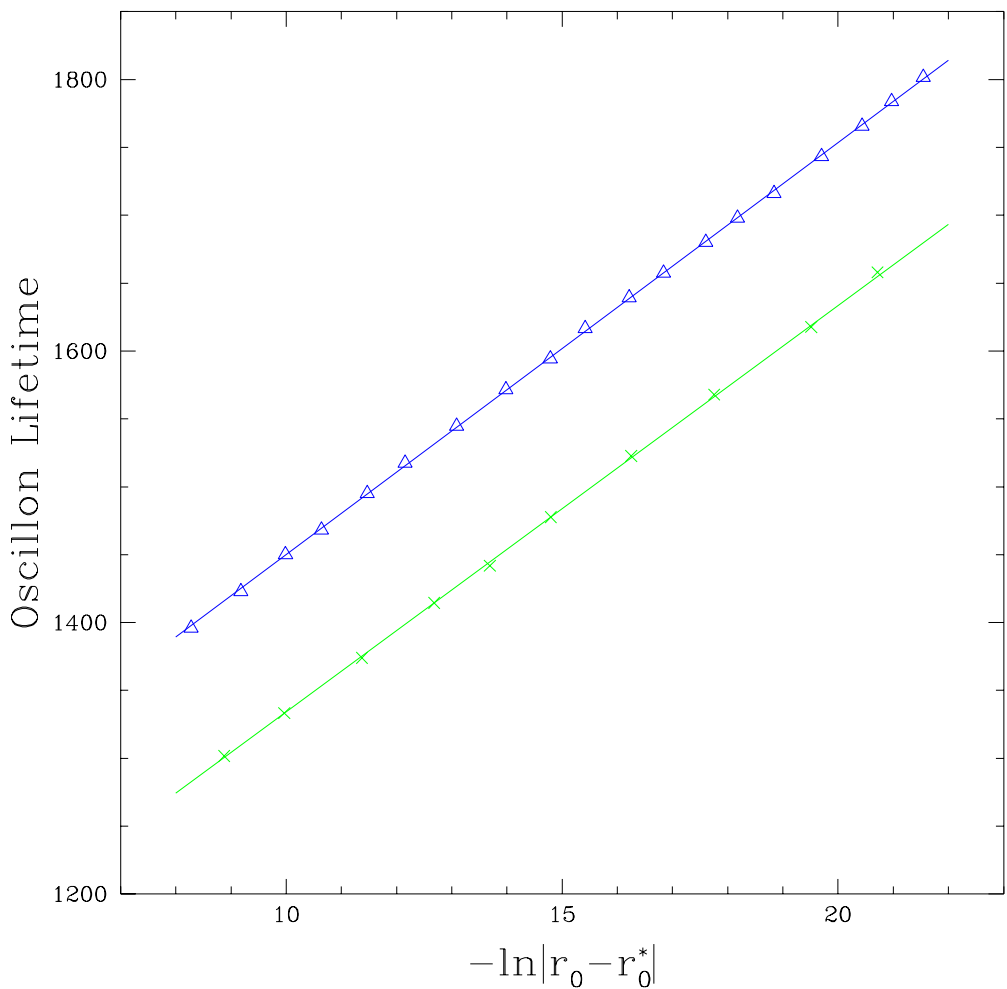}}
\caption[Time scaling, $T$ versus $-\ln|r_0 - r_0^*|$ about a resonance]
{\small \label{fig:tscaling}
Plot of time scaling, $T$ versus $\ln|r_0 - r_0^*|$ for the
 $r_0\approx 2.335$ resonance.
The top (blue) line displays the scaling behavior for $r_0>r_0^*$ 
while the bottom (green) line for $r_0<r_0^*$.
The exponents (measured by the slope) are both approximately equal
to $\gamma = 30$.
Although each resonance has a scaling law, the exponents vary from
resonance to resonance;
a plot of the scaling exponent, $\gamma$, versus the critical initial
bubble radius can be seen in figure \ref{fig:exponents}.
}
\label{fig:tscaling}
\end{figure}
Figure \ref{fig:tscaling} shows a plot of oscillon lifetime versus 
\hbox{$\ln|r_0-r_0^*|$} (for the $r_0\approx 2.335$ resonance) and we can see quite 
clearly that there is a scaling law, $T\sim\gamma\ln|r_0-r_0^*|$, 
for the lifetime of the solution on each side of the resonance.
We denote $\gamma_{\rm HI}$ for the scaling exponent on the 
$r_0>r_0^*$ side, and $\gamma_{\rm LO}$ for the scaling exponent on the 
$r_0<r_0^*$ side.  A plot of both scaling parameters 
as a function of $r_0^*$ is seen in Fig. \ref{fig:exponents}.
We observe that for almost all of the resonances that 
$\gamma_{\rm HI}\approx\gamma_{\rm LO}$.
\begin{figure}
\epsfxsize=14cm
\centerline{\epsffile{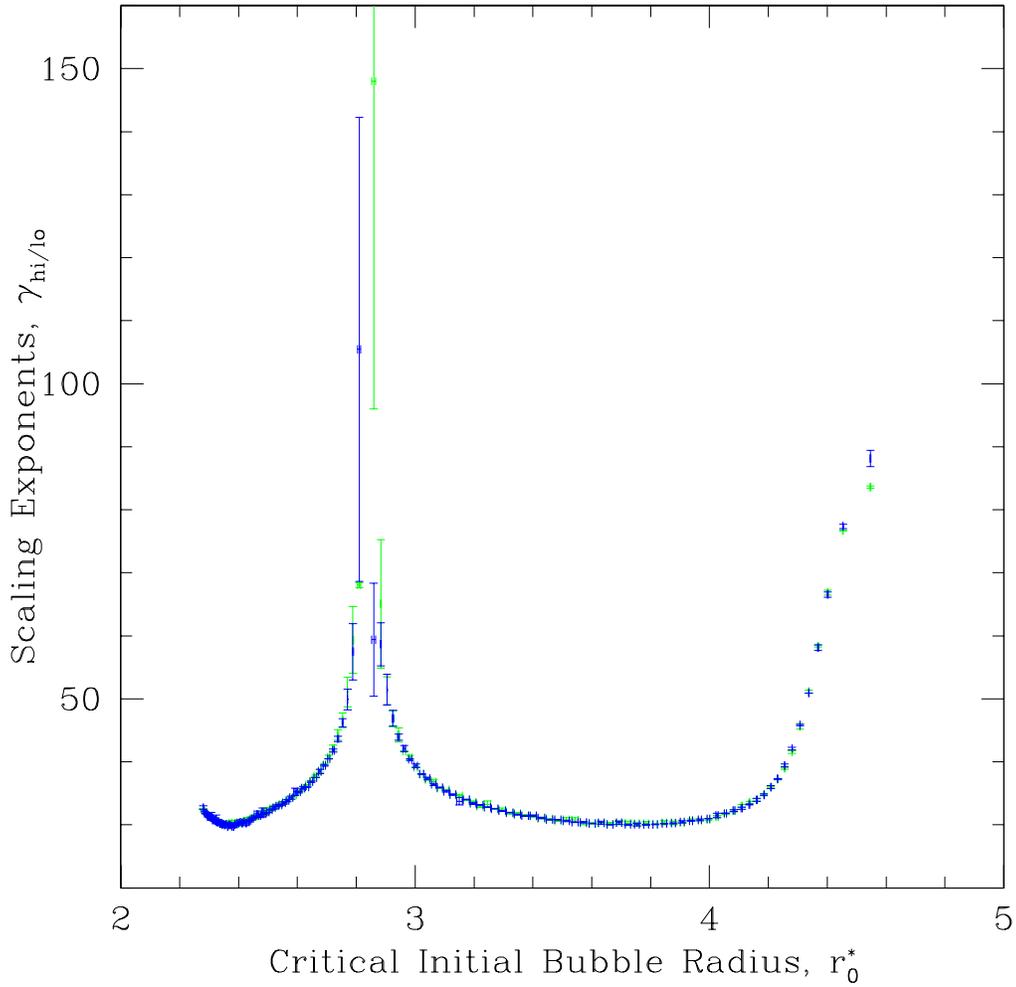}}
\caption[Critical exponents for each resonance]
{\small \label{fig:exponents}
Plot of critical exponents for each resonance. 
There are two values of $\gamma$ for each resonance. The green values
are the $\gamma_{\rm hi}$ values measured on the $r_0>r_0^*$ side of the 
resonance while the blue values are the $\gamma_{\rm lo}$ values measured 
on the $r_0<r_0^*$ side of the resonance.
The uncertainties are estimated from running the {\it entire} parameter
space surveys at two resolutions and estimating the error, 
$\Delta\gamma = |\gamma_{(N_r=1449)} -\gamma_{(N_r=1025)}|$. 
}
\label{fig:exponents}
\end{figure}

\subsection{Mode Structure\label{subsec:modestruct}}

If there does exist a 
periodic non-radiative solution to \hbox{equation
\ref{eq:boxphi}}, we should be able to construct it by applying an 
{\em ansatz} of the form
\be
\phi(r,t) = -1 + \phi_0(r) + \sum_{n=1}^{\infty}\phi_n(r) \cos\left(n\omega t\right)
\label{eq:ansatz}
\ee
to the equations of motion 
and solving the resulting system of ordinary differential equations
obtained from matching $\cos\left(n\omega t\right)$ terms:
\begin{equation}
\begin{array}{rcl}
r^{-2} \left( r^2 \phi_0' \right)'
&=& \phi_0\left( \phi_0 - 1\right)\left(\phi_0-2\right) \\
&&+ \frac{3}{2} \left( \phi_0 -1\right) \sum\limits_n \left( \phi_n\right)^2\\
&&+ \frac{1}{4}\sum\limits_{n,p,q}\phi_n\phi_p\phi_q\left(
\delta_{n, \pm p \pm q} \right) ,
\end{array}
\label{modestruct0}
\end{equation}
\begin{equation}
\begin{array}{rcl}
r^{-2}\left( r^2 \phi_n' \right)' 
&=& \left( 3 \left(\phi_0-1\right)^2 -
\left( n^2 \omega^2 +1\right)\right)\phi_n\\
&&+\frac{3}{2} \left(\phi_0-1\right) \sum\limits_{p,q}\phi_p\phi_q 
\left(\delta_{n,\pm p \pm q}\right) \\
&&+\frac{1}{4}\sum\limits_{m,p,q}\phi_m\phi_n\phi_q\left(
\delta_{n,\pm m \pm p \pm q}\right),
\end{array}
\label{modestructn}
\end{equation}
where 
$\delta_{n, \pm p \pm q} =\left(
 \delta_{n, + p + q}
+\delta_{n, + p - q}
+\delta_{n, - p + q}
+\delta_{n, - p - q} \right)$,
and likewise for $\delta_{n,\pm m \pm p \pm q}$, but with nine terms.
Equations \ref{modestruct0} and \ref{modestructn} can also be 
obtained by inserting {\em ansatz} \ref{eq:ansatz} into the action 
and varying with respect to the $\phi_n$
\cite{morrison:1999} (see chapter \ref{chap:trial}).
This set of ODE's can be solved by ``shooting'', where the 
$\phi_n(0)$ are the shooting parameters. 
Unfortunately, we were unable to construct a method that 
determined $\omega$ on its own; the best we could do was to 
solve equations \ref{modestruct0} and \ref{modestructn}
for a given $\omega$ that we measured from the dynamic solution.  
To easily compare the shooting method to the dynamic data, we 
Fourier decomposed the dynamic data.  This was done by taking the solution
during the interval of time when the large period 
modulation disappears ($1200<t<1800$ 
for the oscillon in Fig. \ref{fig:phenv_energy}, for example)
and constructing FFTs from the field at each gridpoint, $r_i$.  
The amplitude of each Fourier mode was obtained at each $r_i$ from the 
FFT using a window of 4096 time-slices (bins).  These amplitudes were
then used to piece together the 
$\phi_n$ (Fig. \ref{fig:modstruct}).  

\begin{figure}
\epsfxsize=14cm
\centerline{\epsffile{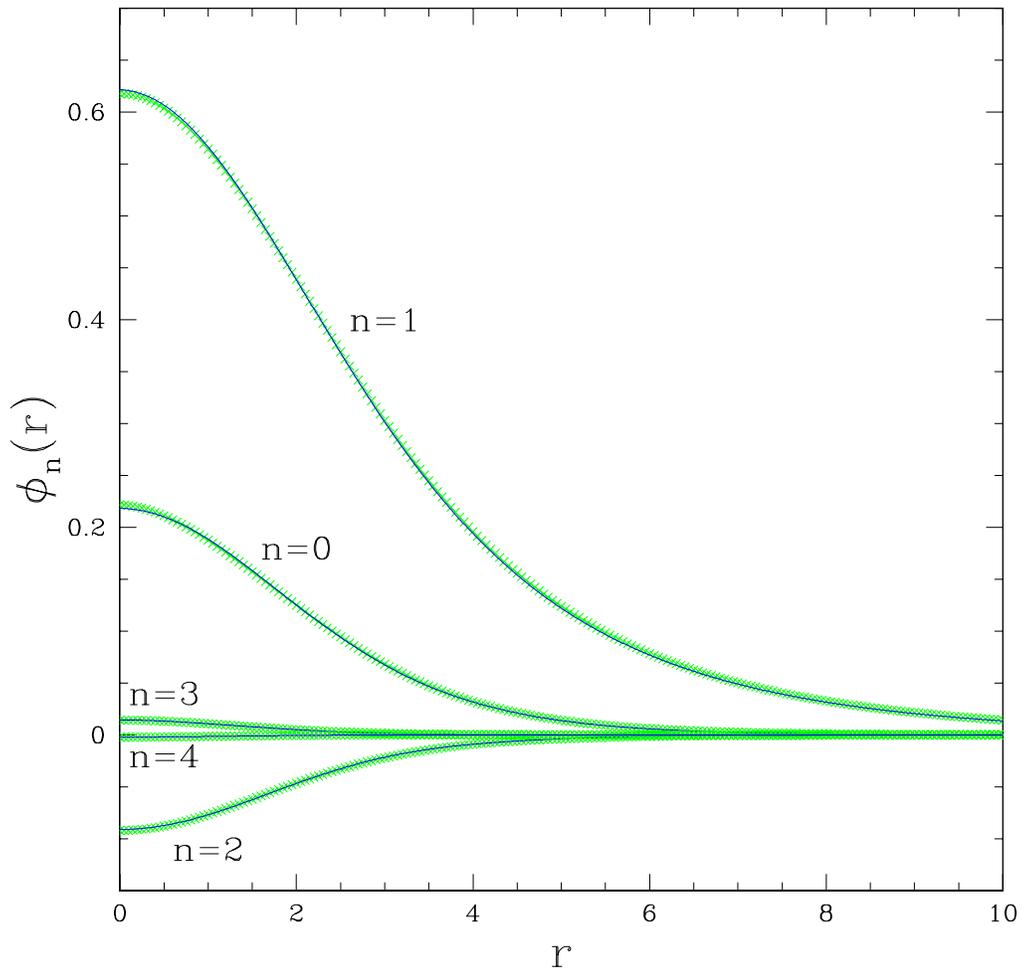}}
\caption[Mode decomposition of critical oscillon]
{\small \label{fig:modstruct}
Critical solution  $\phi_n(r)$ (for $n=0,1,2,3,4$) obtained 
from the Fourier decomposed dynamic data (green x's)
overlayed with $\phi_n(r)$ obtained by 
shooting equations \ref{modestruct0} and \ref{modestructn} 
(blue solid curves).
The dynamic data was Fourier decomposed by forming
a time series for each spatial gridpoint, $r_i$, while 
the oscillon entered its non-radiative (no modulation) phase.
FFT's were then constructed for each time series and the amplitude of
each mode  (the green x's) was measured.
}
\label{fig:modstruct}
\end{figure}
Keeping only the first five modes, 
we compare the Fourier decomposed dynamic data with the
shooting solution (see Fig. \ref{fig:modstruct}).
The value for $\omega$ was determined from the dynamic solution
and the shooting parameters (the $\phi_n(0)$) were varied 
so as to maximize $r_{\rm max}$, the distance before {\it any} mode 
diverged to $\phi\rightarrow \pm\infty$.
The correspondence strongly suggests that the resonant solutions
(ie. in the limit as $r_0\rightarrow r_0^*$) 
observed in the dynamic simulations 
are indeed 
the periodic non-radiative oscillons obtained from 
ansatz (\ref{eq:ansatz})

\begin{figure}
\epsfxsize=14cm
\centerline{\epsffile{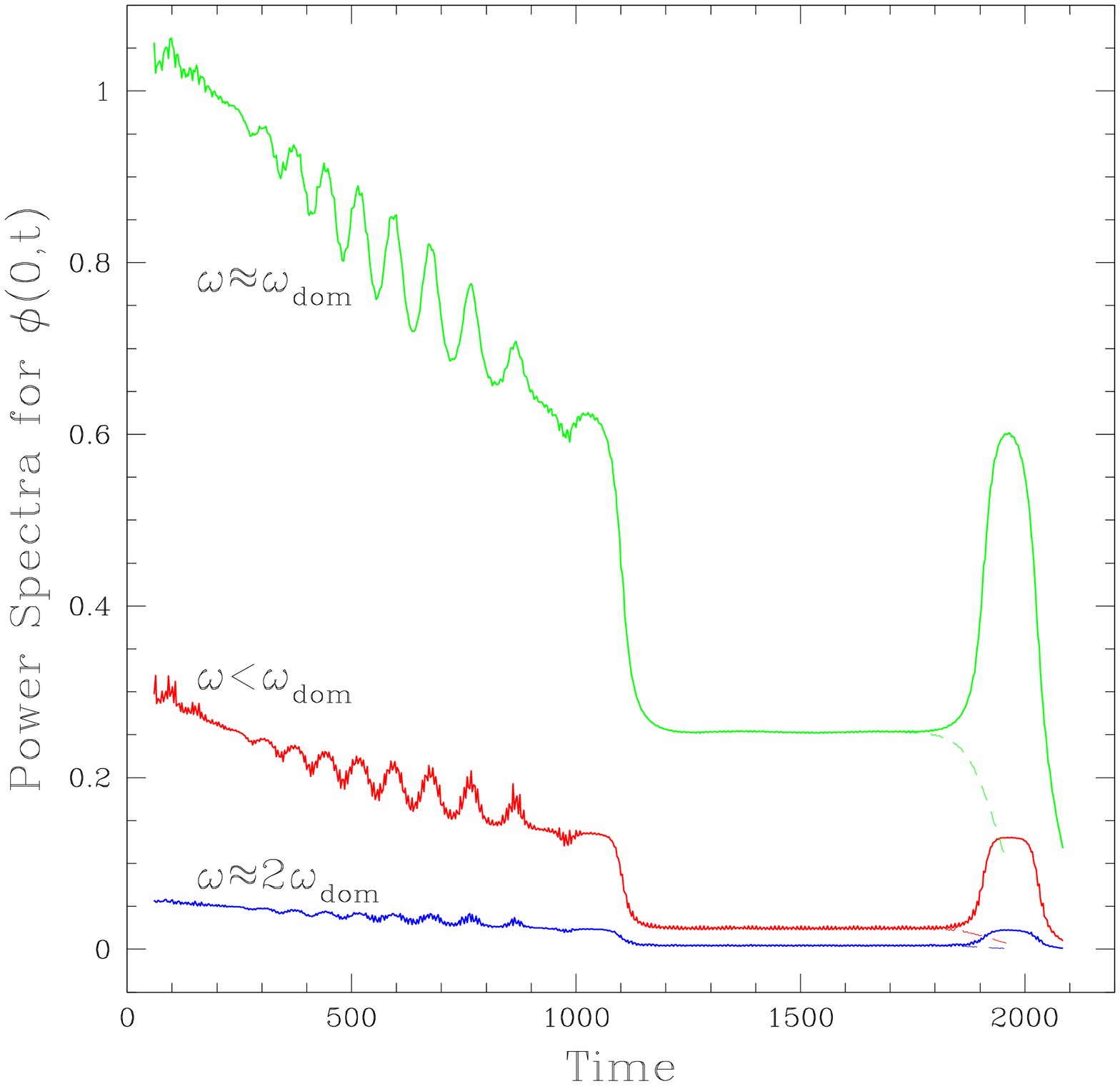}}
\caption[Power spectra of core amplitude versus time at resonance]
{
Power spectra of the core amplitude, $\phi(0,t)$, 
for the oscillons barely above and below the 
$r_0\approx2.335$ resonance.  
The power measured in each frequency regime slowly diminishes
as the oscillon radiates away much of its energy until 
approximately $t=1100$ where the oscillon enters a non-radiative 
state and all the components of the power spectrum become 
constant.
}
\label{fig:modrad}
\end{figure} 

Looking at the three most dominant components of the 
power spectrum of $\phi(0,t)$, Fig. \ref{fig:modrad},
we can see that during the period of no-modulation, the 
amplitude of each Fourier mode becomes constant.  Although the 
plot is for the core amplitude, $r=0$, this behavior 
holds for all $r$.  This means that a ``resonant'' oscillon
solution latches onto a non-radiative periodic 
solution as an intermediate attractor.  
Each resonance, however, does have a different
critical solution; figure \ref{fig:exponents} shows distinct
critical exponents for each resonance.
This is reminiscent of the Type I critical phenomena 
studied in the massive Einstein-Klein-Gordon (EKG) model, 
\cite{brady:1997},
\cite{brady2:1997},
\cite{chambers:1997},
\cite{choptuik:1998}, where a {\it family} of (apparently)
periodic solutions exist
on the threshold of black hole formation.
The behavior observed here is thus 
in contradistinction to Type I Einstein-Yang-Mills
critical collapse,
\cite{choptuik:1996},
where the critical solution is found to be static and 
{\it universal}.
In any case, instead of existing on the threshold of black hole formation,
the critical oscillon solution is an oscillating 
{\it time-dependent} intermediate attractor
on the threshold of having one more shape mode oscillation.

\subsection{(Bounce) Windows to more Oscillons}

Lastly, we look at the region of the parameter space beyond $r_0\sim 5$.
The oscillons explored by Copeland, et al, were restricted to the 
domain of roughly $2\leq r_0 \leq 5$.  It was even wondered if oscillons could
form for larger initial bubble radii.  We found that oscillons can
form in this region and do so by a rather interesting mechanism.  

\begin{figure}
\epsfxsize=14cm
\centerline{\epsffile{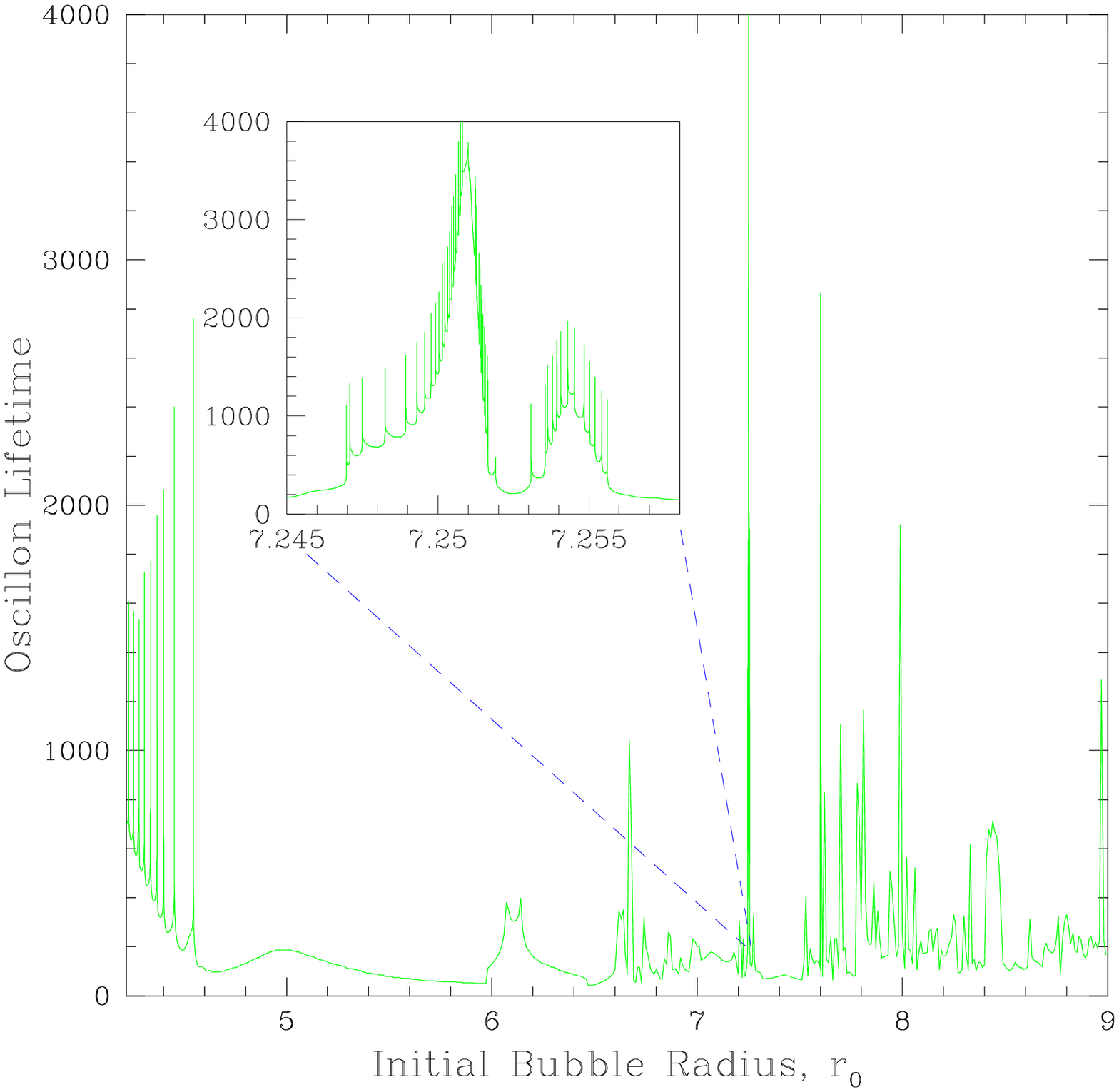}}
\caption[Oscillon lifetime versus initial bubble radius in bounce regime]
{
Plot of oscillon lifetime versus initial radius of bubble
for $4.22\le r_0\le 9$.
Although there seem to be no oscillons within $4.6\le r_0\le 6$, it is clear that 
oscillons {\it do} exist for higher initial bubble radii.
Furthermore, within these ``windows'' of parameter space that support
oscillon formation, there also exist resonances similar to those 
observed throughout the $2\le r_0\le 4.6$ region (see inset to observe
resonances within $7.245\le r_0\le 7.257$).
}
\label{fig:lifefine}
\end{figure}

Again, going back to the 1+1 dimensional \kkbar scattering of
Campbell, et al, it is well known that the kink and antikink 
often ``bounce'' many times before either dispersing or falling 
into an (unstable) bound state.
A bounce occurs when the kink and antikink reflect off one another, 
stop after a short distance, and recollapse.  
We find that this happens in the (1+1) spherically symmetric case as well, 
where the unstable bound state is an oscillon.
For larger $r_0$, instead of the bubble wall remaining within 
$r\leq 2.5$ (as occurs for $2\leq r_0\leq 5$), 
after reflecting off the origin the bubble wall can travel out to 
larger r (typically, 
$3\le r\le 6$), stop, and then recollapse, shedding away large amounts of
energy in the process;  
\begin{figure}
\epsfxsize=14cm
\centerline{\epsffile{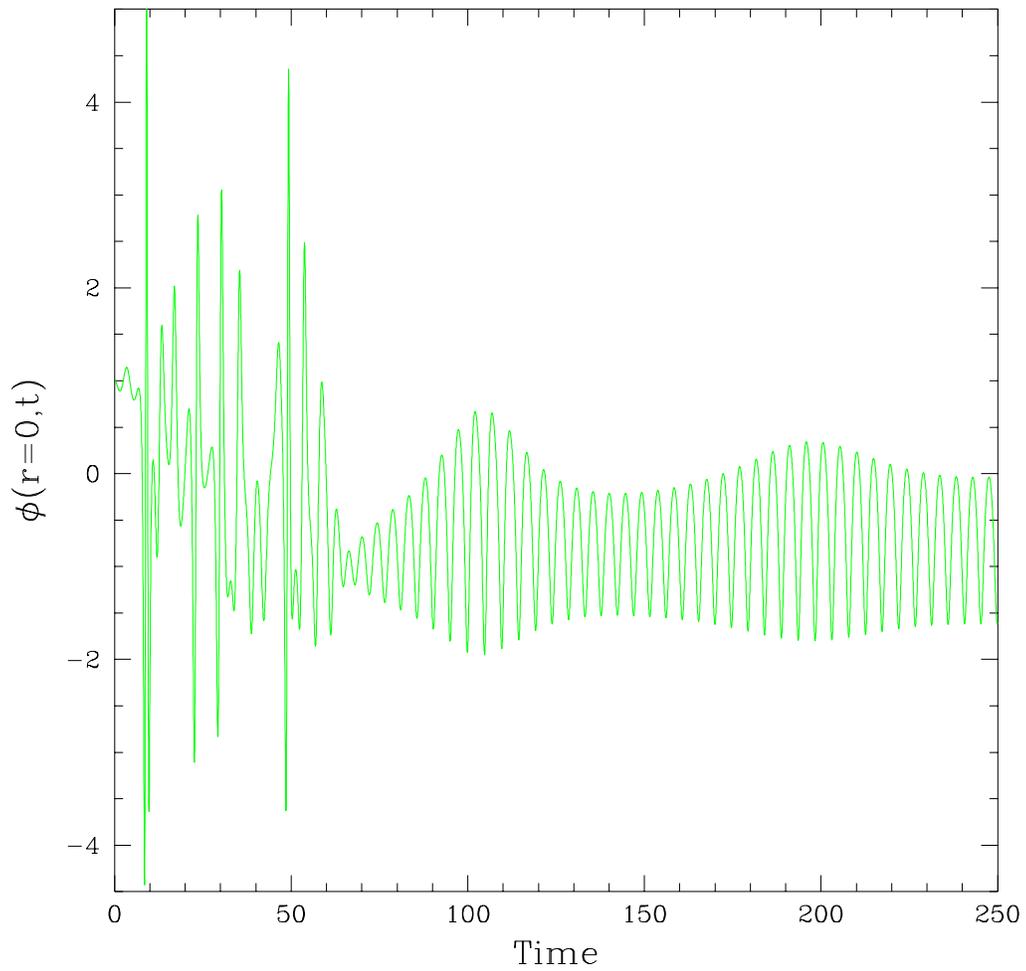}}
\caption[Chaotic behavior of $\phi(0,t)$ for ``bouncing'' bubble]
{
Plot of $\phi(0,t)$ for $r_0=7.25$ displaying extremely nonlinear and
unpredictable behavior during the ``bouncing'' phase (for $t< 60$), 
after which the field settles into a typical oscillon evolution.  
Once in the oscillon regime, the period
is approximately $\omega \approx 4.5$, and the first two modulations
of the field can be seen (envelope 
maxima at $t\approx 105$ and $t\approx 200$).
}
\end{figure}
we refer to these regions of parameter space as ``bounce windows''.
In the recollapse the simulation is effectively starting over with new 
initial data (albeit a different shape) but with much less energy (smaller $r_0$).  
In Fig. \ref{fig:lifefine} we see a lifetime 
profile and its resonances for a typical bounce window.

\section{The Klein-Gordon Equation in MIB Coordinates with ADWP ({\it 1D Critical Phenomena II})\label{sec:1dcritII}}

This chapter concludes with time evolution of spherically 
symmetric bubbles in the context of a slightly different action than the one
introduced in section \ref{sec:tex_eqs}.
The action used is the same as equation (\ref{eq:action}), 
\begin{equation} \label{eq:ADWPaction}
S[\phi] = \int d^4x\sqrt{|g|}\left(
-\frac{1}{2} g^{\mu\nu}\nabla_\mu\phi \nabla_\nu\phi - V(\phi)
\right),
\end{equation}
except with an asymmetric potential,
\hbox{$\displaystyle{V(\phi) \equiv V_{A}(\phi) = \frac{1}{4}\phi^2\left(\phi^2 
- 4\left(1+\delta\right)\phi +4 \right)}$}
where again $\phi\equiv \phi(r,t)$, 
and $\delta$ is a measure of the asymmetry of the potential.
When $\delta=0$, 
$\displaystyle{V_{A}(\phi)}$ has the same shape as 
the SDWP potential of Chapter \ref{chap:1D}, but is shifted 
so that the vacuum states are at $\phi=0$ and $\phi=2$. 
With $\delta \neq0$, the false and true vacuum states 
are $\phi_F=0$ and 
$\displaystyle{\phi_T=\frac{3}{2}\left(1+\delta\right)
+ \frac{1}{2}\sqrt{ 1+ 18\delta + \delta^2}}$, 
respectively. 
\begin{figure}
\label{fig:adwp}
\epsfxsize=12cm
\centerline{\epsffile{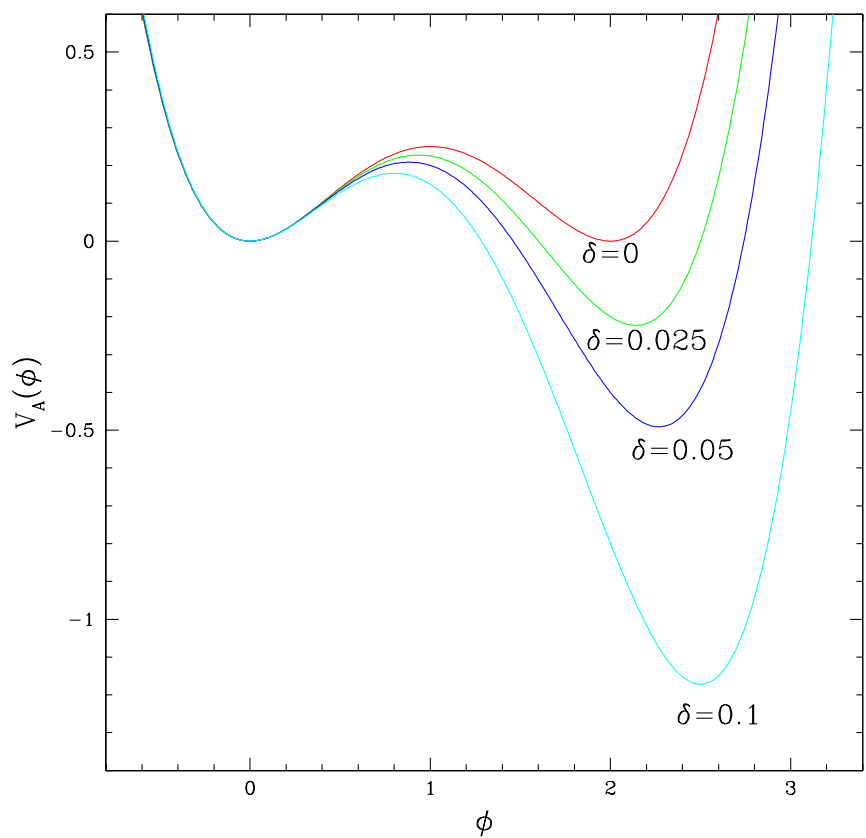}}
\caption[Asymmetric double well potential (ADWP)]
{\small \label{fig:adwp}
Asymmetric double well potential 
$\displaystyle{V_{A}(\phi) = \frac{1}{4}\phi^2\left(\phi^2 
- 4\left(1+\delta\right)\phi +4 \right)}$ for various values
of asymmetry parameter, $\delta$.
Minima are clearly {\it not} degenerate for $\delta \neq 0$ and are at
$\phi_F=0$ and 
$\displaystyle{\phi_T=\frac{3}{2}\left(1+\delta\right)
+ \frac{1}{2}\sqrt{ 1+ 18\delta + \delta^2}}$ (the
false vacuum and true vacuum states, respectively).
}
\label{fig:adwp}
\end{figure}

The spherically symmetric MIB code is (trivially) modified by replacing 
the potential terms in the finite difference equations with the new potential,
leaving $\delta$ as a run-time parameter. 
Instead of performing extensive parameter space surveys with this 
potential like in section (\ref{sec:tex_fine}),
a different type of phenomena is explored\footnote{Brief parameter surveys 
were conducted and resonances were also observed, but no additional 
new {\it resonance} phenomena were observed.}.
With the introduction of the {\it asymmetric} potential which has non-degenerate
vacuum states, the threshold of expanding bubble formation can be examined.
As discussed in section (\ref{theory_bubform}), the initial bubble radius can be 
used to define a one-parameter family of initial data that transitions from 
bubble collapse (and eventual dispersal) to expanding bubble formation.
As one familiar with critical phenomena might expect, there exists a time
scaling law for the lifetime of the bubble lying on this threshold.  
The time scaling exponent is observed to be, $\gamma \simeq 2.1$, where 
$T \propto -\gamma \ln|\sigma - \sigma^*|$.

\begin{figure}
\label{fig:adwp1dtscale}
\epsfxsize=12cm
\centerline{\epsffile{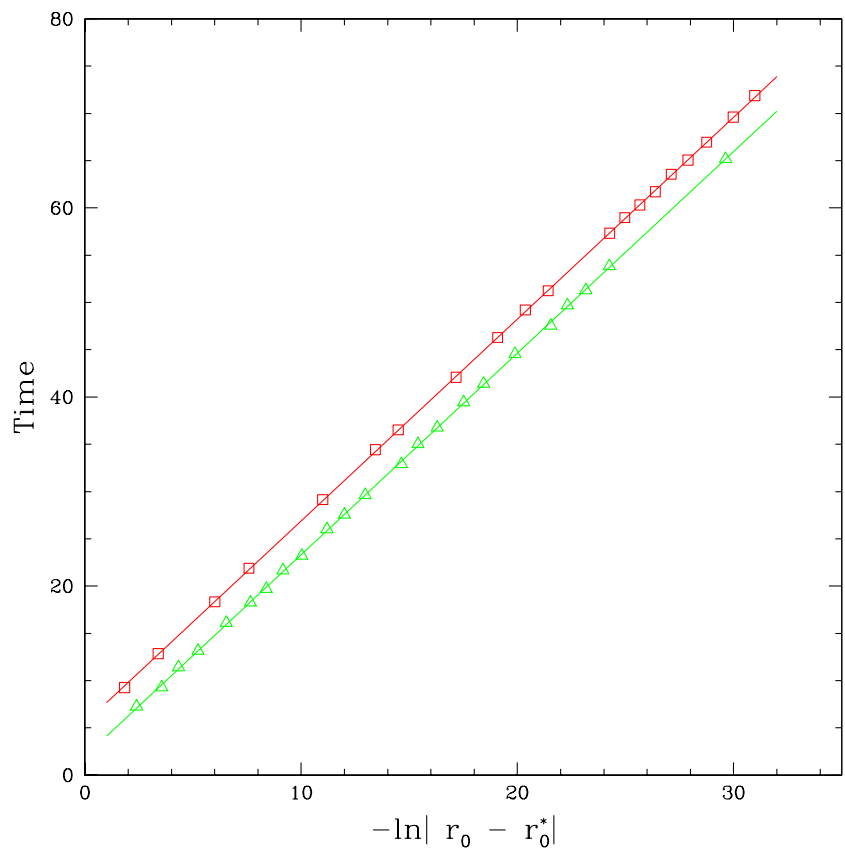}}
\caption[Time Scaling for Spherically Symmetric ADWP]
{\small \label{fig:adwp}
Plot of oscillon lifetime as a function of $-\ln|r_0-r_0^*|$ for the
collapse (or expansion) of static bubble initial data using the ADWP.
The lifetimes on the collapsing side of criticality are represented 
by green triangles, while the lifetimes on the expanding side of 
criticality are represented by red triangles.
There is a single time scaling law with scaling exponent (slope) 
$\gamma\approx 2.1$.
}
\label{fig:adwp1dtscale}
\end{figure}

\chapter{\hbox{Axi-Symmetric Oscillon Dynamics}\label{chap:2D}}

This chapter discusses our investigations of the non-linear 
Klein-Gordon model in axisymmetry.
The equation of motion is written in 2-D MIB coordinates and a
finite difference version of the equations is presented. 
The convergence properties of the code which solves the 
difference equations are shown, and evidence to support
the validity of the MIB system as an effective absorbing outer-boundary
is presented.
A simple numerical technique is presented that allows for the 
arbitrary Lorentz boosting of spherically symmetric data.  
Finally, the code is used to collide two oscillons and a time-scaling
phenomenon analogous to that seen in chapter \ref{sec:1dcritII}
is observed.

\section{The Klein-Gordon Equation in Axi-Symmetric MIB Coordinates}

This section discusses the implementation of 2-D MIB coordinates 
in the solution of the nonlinear Klein-Gordon model 
with an asymmetric double-well (ADWP) potential.
The model studied is the same ADWP model just used to study the threshold of 
expanding bubble formation in the previous chapter.
The resultant PDE we must solve is still simply the Klein-Gordon 
equation
\begin{equation} \label{eq:2Dboxphi}
\frac{1}{\sqrt{|g|}}\partial_\mu\left( \sqrt{|g|} g^{\mu\nu} 
\partial_\nu\phi \right) = \phi\left(\phi^2 
		-3 \left( 1-\delta\right) +2 \right)
\end{equation}
but now with 
$\phi\equiv \phi(R,z,t)$. 
Defining
\begin{eqnarray}
\displaystyle{\Pi(R,z,t)} &\equiv& 
\displaystyle{ab\left( \dot{\phi} - \beta^R\Phi_R 
- \beta^z\Phi_z \right)}, \\
\displaystyle{
\Phi_R(R,z,t)}  &\equiv & \displaystyle{\partial_R\phi,}\\
\displaystyle{
\Phi_z(R,z,t) } &\equiv & \displaystyle{\partial_z\phi,} 
\end{eqnarray}
we have
\beq
\dot{\Pi} &=& 2a\frac{\partial}{\partial\left(\tilde{R}^2\right)}
\left[  \tilde{R} \left( \beta^R \Pi + \frac{b}{a}\Phi_R \right)\right]
+ \frac{\partial}{\partial z}\left( \beta^z \Pi + \frac{a}{b} \Phi_z\right) \nonumber \\
& & -\frac{\dot{\tilde{R}}}{\tilde{R}}\Pi  
- ab \frac{\partial V}{\partial \phi} \label{eq:2Dpidot}\\
\dot{\Phi}_R &=& \frac{\partial}{\partial R}
\left( \frac{1}{ab} \Pi  + \beta^R \Phi_R  + \beta^z\Phi_z \right) \label{eq:2Dphirdot}\\
\dot{\Phi}_z &=& \frac{\partial}{\partial z}
\left( \frac{1}{ab} \Pi  + \beta^R \Phi_R  + \beta^z\Phi_z \right) \label{eq:2Dphizdot}\\
\dot{\phi}  &=& \frac{1}{ab} \Pi  + \beta^R \Phi_R  + \beta^z\Phi_z \label{eq:2Dphidot},
\eeq
for $a$, $b$, $\beta^R$, $\beta^z$, $\alpha$, and 
$\tilde{R}$ functions of $R$, $z$, and $t$,
as defined in equations (\ref{eq:2DADM_auxvars}).
As with the spherically symmetric case, these equations are similar in 
form to the Klein-Gordon equation coupled to gravity, except that,
again, the 
``geometric variables'' are known functions of our coordinates.

\section{Finite Difference Equations}

Equations 
(\ref{eq:2Dpidot}),
(\ref{eq:2Dphirdot}),
(\ref{eq:2Dphizdot}), and
(\ref{eq:2Dphidot}) are solved using two-level second order (in both space and time) 
finite difference approximations on a static uniform mesh with
$N_R$ by $N_z$ grid points in the $R$ and $z$ directions, respectively.  
The scale of discretization is set by $\Delta R$, $\Delta z$, and 
$\Delta t = \lambda \min(\Delta R,\Delta z)$\footnote{Although we usually take
$\Delta z=\Delta R$}.  
The field variables were first updated without dissipation everywhere (yielding
the $\hat{}$ variables), and then
dissipation was added were possible\footnote{The sole purpose of this
was to streamline the computer code.
This allows the complete update to be execute using fewer loops. 
} (yielding the $\tilde{}$ and final quantities).
Using the difference operators from Table \ref{tab:2Dfdop},
$\partial_R\tilde{R}=a$, and 
\hbox{$\partial_R=nR^{n-1}\partial_{R^n}$}, the 
interior ($2\leq i \leq N_R-1$, $2\leq j \leq N_z-1$) difference equations are:
\\
\vbox{
\beq
\Delta_t\hat{\Pi}^n_{i,j} &=& 
2\mu^{ave}_t\left[
	a^n_{i,j} \Delta_{\tilde{R}^2}\left( 
		\tilde{R}\left(
			\displaystyle{\beta^R\Pi + \frac{b}{a}\Phi_R}
		\right)
	\right)
	+ \Delta_z\left( 
			\displaystyle{\beta^z\Pi + \frac{a}{b}\Phi_z}
		\right)
	\right]^n_{i,j} \nonumber \\
&& - \displaystyle{ 
	\mu^{ave}_t \left(
         \frac{\dot{\tilde{R}}}{\tilde{R}}\Pi
	-
	 a b \phi\left( \phi^2 - 3\left( 1+\delta \right) \phi +2\right)
    	    \right)^n_{i,j} },\label{2D_pidiff_int}\\
\Delta_t{\left(\hat{\Phi}_R\right)}^n_{i,j} &=& 
\mu^{ave}_t\Delta_R\left( 
	\frac{1}{ab}\Pi + \beta^R\Phi_R + \beta^z\Phi^z
	\right)^n_{i,j},\label{2D_phirdiff_int}\\
\Delta_t{\left(\hat{\Phi}_z\right)}^n_{i,j} &=& 
\mu^{ave}_t\Delta_z\left( 
	\frac{1}{ab}\Pi + \beta^R\Phi_R + \beta^z\Phi^z
	\right)^n_{i,j},\label{2D_phizdiff_int}\\
\Delta_t\hat{\phi}^n_{i,j} &=& 
\mu^{ave}_t\left( 
	\frac{1}{ab}\Pi + \beta^R\Phi_R + \beta^z\Phi^z
	\right)^n_{i,j}\label{2D_phidiff_int}.
\eeq
}
For the inner $R=0$ boundary ($i=1$, $2\leq j \leq N_z-1$), 
the conditions for regularity at 
the origin are applied for $\Phi_R$ and $\Pi$ (analogous to
equations (\ref{eq:Phieq0}) and (\ref{eq:PiPrmeq0}) for spherical 
symmetry):
\beq
\mu^{ave}_t\left[ 
	\Delta^f_R \hat{\Pi} - \frac{a'}{a}\hat{\Pi} 
\right]^n_{i,j} &=& 0\\
\left(\hat{\Phi_R}\right)^{n+1}_{i,j} &=& 0
\eeq
while evolving $\hat{\Phi}_z$ and $\hat{\phi}$ as in the interior of the grid
\beq
\Delta_t{\left(\hat{\Phi}_z\right)}^n_{i,j} -
\mu^{ave}_t\Delta_z\left( 
	\frac{1}{ab}\Pi + \beta^R\Phi_R + \beta^z\Phi^z
	\right)^n_{i,j} &=& 0\\
\Delta_t\hat{\phi}^n_{i,j} - 
\mu^{ave}_t\left( 
	\frac{1}{ab}\Pi + \beta^R\Phi_R + \beta^z\Phi^z
	\right)^n_{i,j} &=& 0.
\eeq
For the outer boundaries, as with the spherically symmetric case, the equations
used make very little difference since the physical position 
($\tilde{R}$,$\tilde{z}$) corresponding to the outermost gridpoint is moving out at 
nearly the speed of light.  Therefore none of the outgoing field actually
reaches the outer ($R$,$z$) boundary.  Nevertheless, the typical massless 
scalar field outgoing boundary condition is imposed (on three edges and four
corners):
\beq
\Delta_t \hat{\Pi}^n_{i,j} &=& \mu^{ave}_t\left[ 
\displaystyle{	\frac{R \Delta^b_R\Pi + z\Delta_z\Pi + \Pi}{\sqrt{R^2 + z^2}} }
	\right]^n_{i,j} \\
\Delta_t \left(\hat{\Phi}_R\right)^n_{i,j} &=& \mu^{ave}_t\left[ 
\displaystyle{	\frac{R \Delta^b_R\left(\Phi_R\right) + z\Delta_z\left(\Phi_R\right) + \left(\Phi_R\right)}{\sqrt{R^2 + z^2}} }
	\right]^n_{i,j} \\
\Delta_t \left(\hat{\Phi}_z\right)^n_{i,j} &=& \mu^{ave}_t\left[ 
\displaystyle{	\frac{R \Delta^b_R\left(\Phi_z\right) + z\Delta_z\left(\Phi_z\right) + \left(\Phi_z\right)}{\sqrt{R^2 + z^2}} }
	\right]^n_{i,j} \\
\Delta_t\hat{\phi}^n_{i,j} &=& 
\mu^{ave}_t\left[
	\frac{1}{ab}\Pi + \beta^R\Phi_R + \beta^z\Phi^z
	\right]^n_{i,j}
\eeq
for $i=N_R$, $2\le j \le N_z-1$,
\beq
\Delta_t \hat{\Pi}^n_{i,j} &=& \mu^{ave}_t\left[ 
\displaystyle{	\frac{R \Delta_R\Pi + z\Delta^f_z\Pi + \Pi}{\sqrt{R^2 + z^2}} }
	\right]^n_{i,j} \\
\Delta_t \left(\hat{\Phi}_R\right)^n_{i,j} &=& \mu^{ave}_t\left[ 
\displaystyle{	\frac{R \Delta_R\left(\Phi_R\right) + z\Delta^f_z\left(\Phi_R\right) + \left(\Phi_R\right)}{\sqrt{R^2 + z^2}} }
	\right]^n_{i,j} \\
\Delta_t \left(\hat{\Phi}_z\right)^n_{i,j} &=& \mu^{ave}_t\left[ 
\displaystyle{	\frac{R \Delta_R\left(\Phi_z\right) + z\Delta^f_z\left(\Phi_z\right) + \left(\Phi_z\right)}{\sqrt{R^2 + z^2}} }
	\right]^n_{i,j} \\
\Delta_t\hat{\phi}^n_{i,j} &=& 
\mu^{ave}_t\left[
	\frac{1}{ab}\Pi + \beta^R\Phi_R + \beta^z\Phi^z
	\right]^n_{i,j}
\eeq
for $j=1$, $1\le i \le N_R$,
\beq
\Delta_t \hat{\Pi}^n_{i,j} &=& \mu^{ave}_t\left[ 
\displaystyle{	\frac{R \Delta_R\Pi + z\Delta^b_z\Pi + \Pi}{\sqrt{R^2 + z^2}} }
	\right]^n_{i,j} \\
\Delta_t \left(\hat{\Phi}_R\right)^n_{i,j} &=& \mu^{ave}_t\left[ 
\displaystyle{	\frac{R \Delta_R\left(\Phi_R\right) + z\Delta^b_z\left(\Phi_R\right) + \left(\Phi_R\right)}{\sqrt{R^2 + z^2}} }
	\right]^n_{i,j} \\
\Delta_t \left(\hat{\Phi}_z\right)^n_{i,j} &=& \mu^{ave}_t\left[ 
\displaystyle{	\frac{R \Delta_R\left(\Phi_z\right) + z\Delta^b_z\left(\Phi_z\right) + \left(\Phi_z\right)}{\sqrt{R^2 + z^2}} }
	\right]^n_{i,j} \\
\Delta_t\hat{\phi}^n_{i,j} &=& 
\mu^{ave}_t\left[
	\frac{1}{ab}\Pi + \beta^R\Phi_R + \beta^z\Phi^z
	\right]^n_{i,j}
\eeq
for $j={\rm R_z}$, $1\le i \le N_R$,
\beq
\Delta_t \hat{\Pi}^n_{i,j} &=& \mu^{ave}_t\left[ 
\displaystyle{	\frac{R \Delta^f_R\Pi + z\Delta^f_z\Pi + \Pi}{\sqrt{R^2 + z^2}} }
	\right]^n_{i,j} \\
\Delta_t \left(\hat{\Phi}_R\right)^n_{i,j} &=& \mu^{ave}_t\left[ 
\displaystyle{	\frac{R \Delta^f_R\left(\Phi_R\right) + z\Delta^f_z\left(\Phi_R\right) + \left(\Phi_R\right)}{\sqrt{R^2 + z^2}} }
	\right]^n_{i,j} \\
\Delta_t \left(\hat{\Phi}_z\right)^n_{i,j} &=& \mu^{ave}_t\left[ 
\displaystyle{	\frac{R \Delta^f_R\left(\Phi_z\right) + z\Delta^f_z\left(\Phi_z\right) + \left(\Phi_z\right)}{\sqrt{R^2 + z^2}} }
	\right]^n_{i,j} \\
\Delta_t\hat{\phi}^n_{i,j} &=& 
\mu^{ave}_t\left[
	\frac{1}{ab}\Pi + \beta^R\Phi_R + \beta^z\Phi^z
	\right]^n_{i,j}
\eeq
for $i=1$, $j=1$,
\beq
\Delta_t \hat{\Pi}^n_{i,j} &=& \mu^{ave}_t\left[ 
\displaystyle{	\frac{R \Delta^b_R\Pi + z\Delta^f_z\Pi + \Pi}{\sqrt{R^2 + z^2}} }
	\right]^n_{i,j} \\
\Delta_t \left(\hat{\Phi}_R\right)^n_{i,j} &=& \mu^{ave}_t\left[ 
\displaystyle{	\frac{R \Delta^b_R\left(\Phi_R\right) + z\Delta^f_z\left(\Phi_R\right) + \left(\Phi_R\right)}{\sqrt{R^2 + z^2}} }
	\right]^n_{i,j} \\
\Delta_t \left(\hat{\Phi}_z\right)^n_{i,j} &=& \mu^{ave}_t\left[ 
\displaystyle{	\frac{R \Delta^b_R\left(\Phi_z\right) + z\Delta^f_z\left(\Phi_z\right) + \left(\Phi_z\right)}{\sqrt{R^2 + z^2}} }
	\right]^n_{i,j} \\
\Delta_t\hat{\phi}^n_{i,j} &=& 
\mu^{ave}_t\left[
	\frac{1}{ab}\Pi + \beta^R\Phi_R + \beta^z\Phi^z
	\right]^n_{i,j}
\eeq
for $i=N_R$, $j=1$,
\beq
\Delta_t \hat{\Pi}^n_{i,j} &=& \mu^{ave}_t\left[ 
\displaystyle{	\frac{R \Delta^f_R\Pi + z\Delta^b_z\Pi + \Pi}{\sqrt{R^2 + z^2}} }
	\right]^n_{i,j} \\
\Delta_t \left(\hat{\Phi}_R\right)^n_{i,j} &=& \mu^{ave}_t\left[ 
\displaystyle{	\frac{R \Delta^f_R\left(\Phi_R\right) + z\Delta^b_z\left(\Phi_R\right) + \left(\Phi_R\right)}{\sqrt{R^2 + z^2}} }
	\right]^n_{i,j} \\
\Delta_t \left(\hat{\Phi}_z\right)^n_{i,j} &=& \mu^{ave}_t\left[ 
\displaystyle{	\frac{R \Delta^f_R\left(\Phi_z\right) + z\Delta^b_z\left(\Phi_z\right) + \left(\Phi_z\right)}{\sqrt{R^2 + z^2}} }
	\right]^n_{i,j} \\
\Delta_t\hat{\phi}^n_{i,j} &=& 
\mu^{ave}_t\left[
	\frac{1}{ab}\Pi + \beta^R\Phi_R + \beta^z\Phi^z
	\right]^n_{i,j}
\eeq
for $i=1$, $j=N_z$, and finally
\beq
\Delta_t \hat{\Pi}^n_{i,j} &=& \mu^{ave}_t\left[ 
\displaystyle{	\frac{R \Delta^b_R\Pi + z\Delta^b_z\Pi + \Pi}{\sqrt{R^2 + z^2}} }
	\right]^n_{i,j} \\
\Delta_t \left(\hat{\Phi}_R\right)^n_{i,j} &=& \mu^{ave}_t\left[ 
\displaystyle{	\frac{R \Delta^b_R\left(\Phi_R\right) + z\Delta^b_z\left(\Phi_R\right) + \left(\Phi_R\right)}{\sqrt{R^2 + z^2}} }
	\right]^n_{i,j} \\
\Delta_t \left(\hat{\Phi}_z\right)^n_{i,j} &=& \mu^{ave}_t\left[ 
\displaystyle{	\frac{R \Delta^b_R\left(\Phi_z\right) + z\Delta^b_z\left(\Phi_z\right) + \left(\Phi_z\right)}{\sqrt{R^2 + z^2}} }
	\right]^n_{i,j} \\
\Delta_t\hat{\phi}^n_{i,j} &=& 
\mu^{ave}_t\left[
	\frac{1}{ab}\Pi + \beta^R\Phi_R + \beta^z\Phi^z
	\right]^n_{i,j}
\eeq
for $i=N_R$, $j=N_z$.
This condition is derived by transforming the spherically symmetric out-going
boundary condition to axisymmetric coordinates.  This is a reasonable approach
to take since at large distances away from the collision, the emitted 
wave-fronts {\em do}
becomes spherical. 

After each update using the evolution equations, 
dissipation is added independently in each spatial direction
wherever the action of the appropriate dissipation operator is 
well-defined.
Specifically, the fields are updated a second time (second updated quantities
denoted by $\tilde{}$\ ) using $\mu^{\rm diss}_R$ according
to
\begin{eqnarray}
\tilde{\Pi}^{n+1}_{i,j} &=& \hat{\Pi}^{n+1}_{i,j} + \mu^{\rm diss}_R\hat{\Pi}^n_{i,j} \\
\left(\tilde{\Phi}_R\right)^{n+1}_{i,j} &=& \left(\hat{\Phi}_R\right)^{n+1}_{i,j} 
		+ \mu^{\rm diss}_R\left(\hat{\Phi_R}\right)^n_{i,j} \\
\left(\tilde{\Phi}_z\right)^{n+1}_{i,j} &=& \left(\hat{\Phi}_z\right)^{n+1}_{i,j} 
		+ \mu^{\rm diss}_R\left(\hat{\Phi_z}\right)^n_{i,j} \\
\tilde{\phi}^{n+1}_{i,j} &=& \hat{\phi}^{n+1}_{i,j} + \mu^{\rm diss}_R\hat{\phi}^n_{i,j}
\end{eqnarray}
where $3\le i \le N_r-2$ and $1\le j \le  N_z$. 
Then the fields are updated a third time (third updated 
quantities have {\em no} accent) using $\mu^{\rm diss}_z$ according to
\begin{eqnarray}
\Pi^{n+1}_{i,j} &=& \tilde{\Pi}^{n+1}_{i,j} + \mu^{\rm diss}_z \tilde{\Pi}^n_{i,j} \\
\left(\Phi_R\right)^{n+1}_{i,j} &=& \left(\tilde{\Phi}_R\right)^{n+1}_{i,j} 
		+ \mu^{\rm diss}_z\left(\tilde{\Phi}_R\right)^n_{i,j} \\
\left(\Phi_z\right)^{n+1}_{i,j} &=& \tilde{\left(\tilde{\Phi}_z\right)}^{n+1}_{i,j} 
		+ \mu^{\rm diss}_z\left(\tilde{\Phi}_z\right)^n_{i,j} \\
\phi^{n+1}_{i,j} &=& \tilde{\phi}^{n+1}_{i,j} + \mu^{\rm diss}_R\tilde{\phi}^n_{i,j}
\end{eqnarray}
where $3\le j \le N_z-2$ and $1\le i \le N_r$.
The three update steps comprise one iteration.  
Typically, the whole process is repeated until the solutions at the ($n+1$) time step
converge to one part in $10^{10}$. 

\section{Testing the 2-D code}

Although the 1-D MIB code worked very well, it was not obvious {\it a priori}
that the same success would be achievable in a 2-D axisymmetric code.  
However, we find that the code is stable and passes all the desired tests;
it is second-order convergent in general, it conserves energy to 
second-order in particular,
and outgoing radiation is absorbed without reflection, so that there is no
noticeable contamination of the interior solution (at least at 
the level of accuracy at which we work).

Again, since equation (\ref{eq:2Dboxphi}) is a {\it flatspace} wave equation,
obtaining a globally conserved energy is straightforward.  The spacetime
admits a timelike Killing vector, $t^\nu$, and therefore has a 
conserved current, $J_\mu = t^\nu T_{\mu\nu}$. 
A gaussian surface is constructed between two spacelike hypersurfaces
within a cylindrical spatial domain $0\le R \le R_{\rm endcap}$, 
$-z_{\rm endcap}\le z \le z_{\rm endcap}$,
with normals $n_\mu = (\pm1,0,0,0)$ (the timelike normal to the hypersurfaces),
$n_R^\mu = (0,a^{-1},0,0)$ (the normal to the side of the spatial gaussian cylinder),
and 
$n_z^\mu = (0,0,0,\pm b^{-1})$ (the normals to the top and bottom of the
spatial gaussian cylinder).
To obtain the conserved energy at a time, $t_f$, the energy contained within 
the bubble,
\begin{equation}
E_{\rm bubble}(t) = 
2\pi
\hspace{-0.5cm}
\int\limits_{\! \! 0}^{\ \ \ R_{endcap}} 
\hspace{-0.4cm}
\int\limits_{-z_{\rm endcap}}^{\ \ \ z_{\rm endcap}}
\hspace{-0.2cm}
R \left(
\frac{\Pi^2}{2a^2 b^2} + \frac{\Phi_R^2}{2a^2} + \frac{\Phi_z^2}{2b^2} 
+ V\left(\phi\right)
\right) dr\ dz,
\end{equation}
(where the integrand is evaluated at time $t$) is added to the total radiated 
energy, 
$E_{\rm rad}(t) =
E^{R_{\rm endcap}}_{\rm rad} +
E^{Z^+_{\rm endcap}}_{\rm rad} +
E^{Z^-_{\rm endcap}}_{\rm rad}$, where
\begin{equation}
E^{R_{\rm endcap}}_{\rm rad}(t) =
2\pi R
\int\limits_{\! \! 0}^{\ t} 
\hspace{-0.2cm}
\int\limits_{-z_{\rm endcap}}^{\ \ \ z_{\rm endcap}}
\hspace{-0.2cm}
\left(
\frac{\Pi \Phi_R}{ab}
\right)
dz \ dt'
\end{equation}
(where the integrand is evaluated at $R=R_{\rm endcap}$)
and
\begin{equation}
E^{Z^\pm_{\rm endcap}}_{\rm rad}(t) =
2\pi 
\int\limits_{\! \! 0}^{\ t} 
\hspace{-0.4cm}
\int\limits_{\! \! 0}^{\ \ \ R_{\rm endcap}}
\hspace{-0.2cm}
R\left(
\frac{\Pi \Phi_z}{ab}
\right)
dR \ dt'
\end{equation}
(where the integrand is evaluated at $z=\pm z_{\rm endcap}$).
The sum, $E_{\rm total} = E_{\rm bubble} + E_{\rm rad}$,
remains conserved to within  a percent\footnote{
Using $R_{\rm max}=25$ and $z_{\rm max/min}=\pm 25$.}
at $201\times 401$ gridpoints 
(see figure \ref{fig:2Decon} for energy as a function
of time, see figure \ref{fig:2Dcfacen} to see more clearly the energy being 
conserved as per second order convergence).
\begin{figure}
\epsfxsize=14cm
\centerline{\epsffile{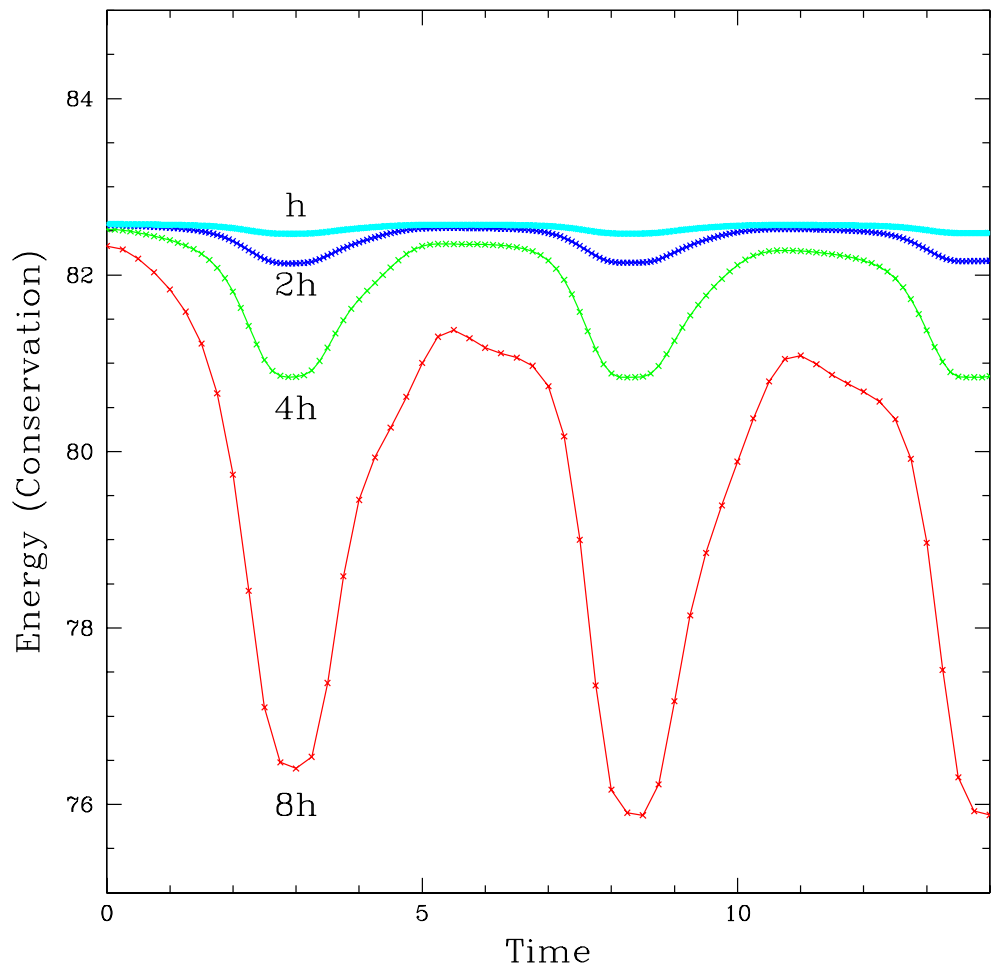}}
\caption[Energy Conservation test of 2D code]
{\small \label{fig:2Decon}
Energy conservation of 2D code at four discretizations, 
$41\times 81$,
$81\times 161$,
$161\times 321$, and
$321\times 641$ gridpoints (8h, 4h, 2h, and h, respectively).
The code conserves energy approximately to one part in 13, 50,
200, and 800, for levels 8h, 4h, 2h, and h, respectively, thus
displaying energy conservation consistent with a second-order 
convergent code.
}
\label{fig:2Decon}
\end{figure} \noindent
\begin{figure}
\epsfxsize=14cm
\centerline{\epsffile{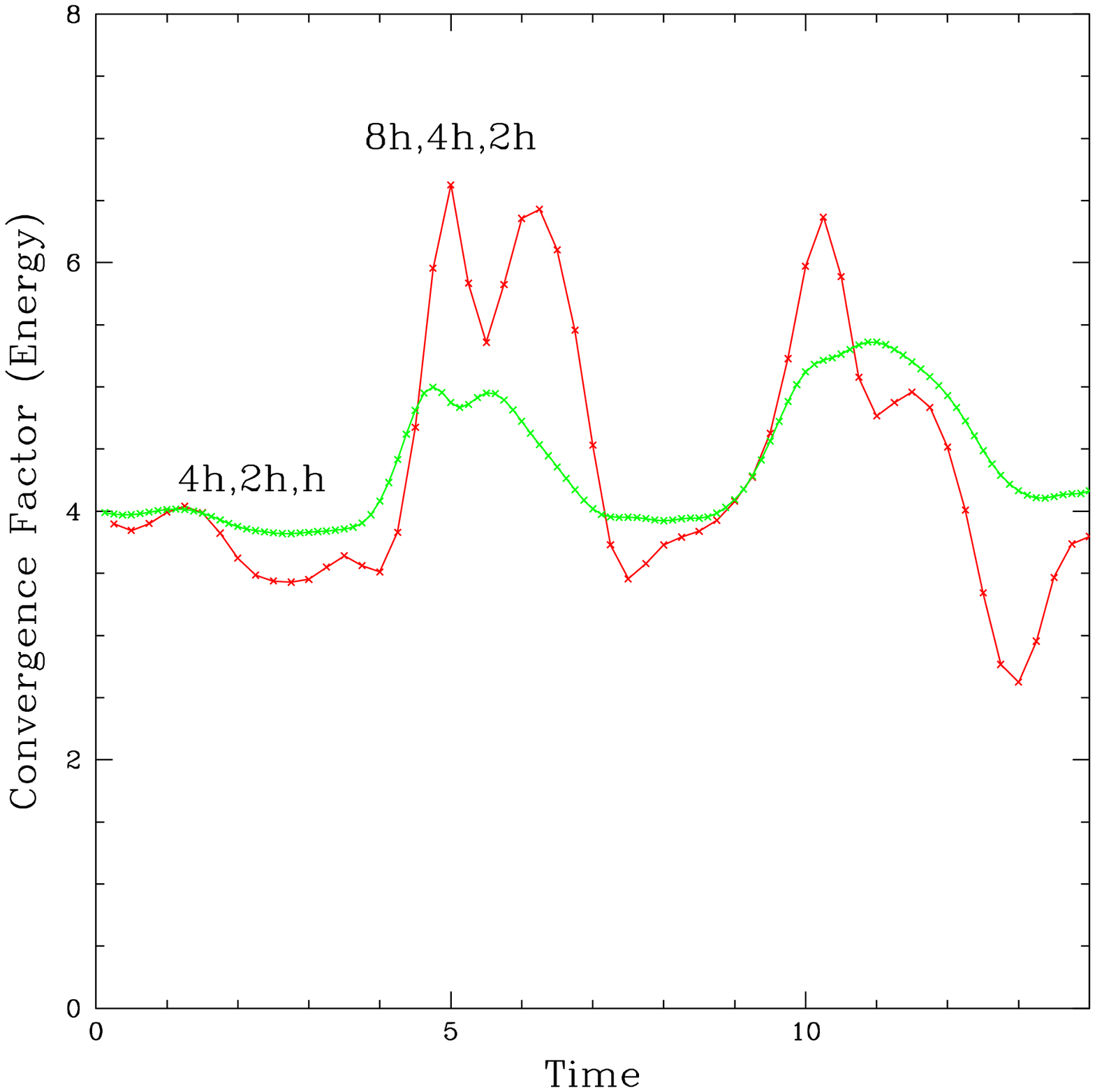}}
\caption[Convergence Factor for the 2D code, the conserved energy.]
{\small \label{fig:2Dcfacen}
Convergence factors,
$\left(E_{4h} - E_{2h}\right)/ \left(E_{2h} - E_{h}\right)$ in red and
$\left(E_{8h} - E_{4h}\right)/ \left(E_{4h} - E_{2h}\right)$ in green,
for the conserved energy.
As with the convergence tests in chapter \ref{chap:1D} (figure 
\ref{fig:convtest}, in particular) a convergence factor of four
indicates second-order convergence.
The code clearly conserves energy to second order.
}
\label{fig:2Dcfacen}
\end{figure} \noindent
In addition to conserving energy (to second order), the code displays second
order convergence
in all the evolved field variables (see figure \ref{fig:2Dcfacphi}).
\begin{figure}
\epsfxsize=14cm
\centerline{\epsffile{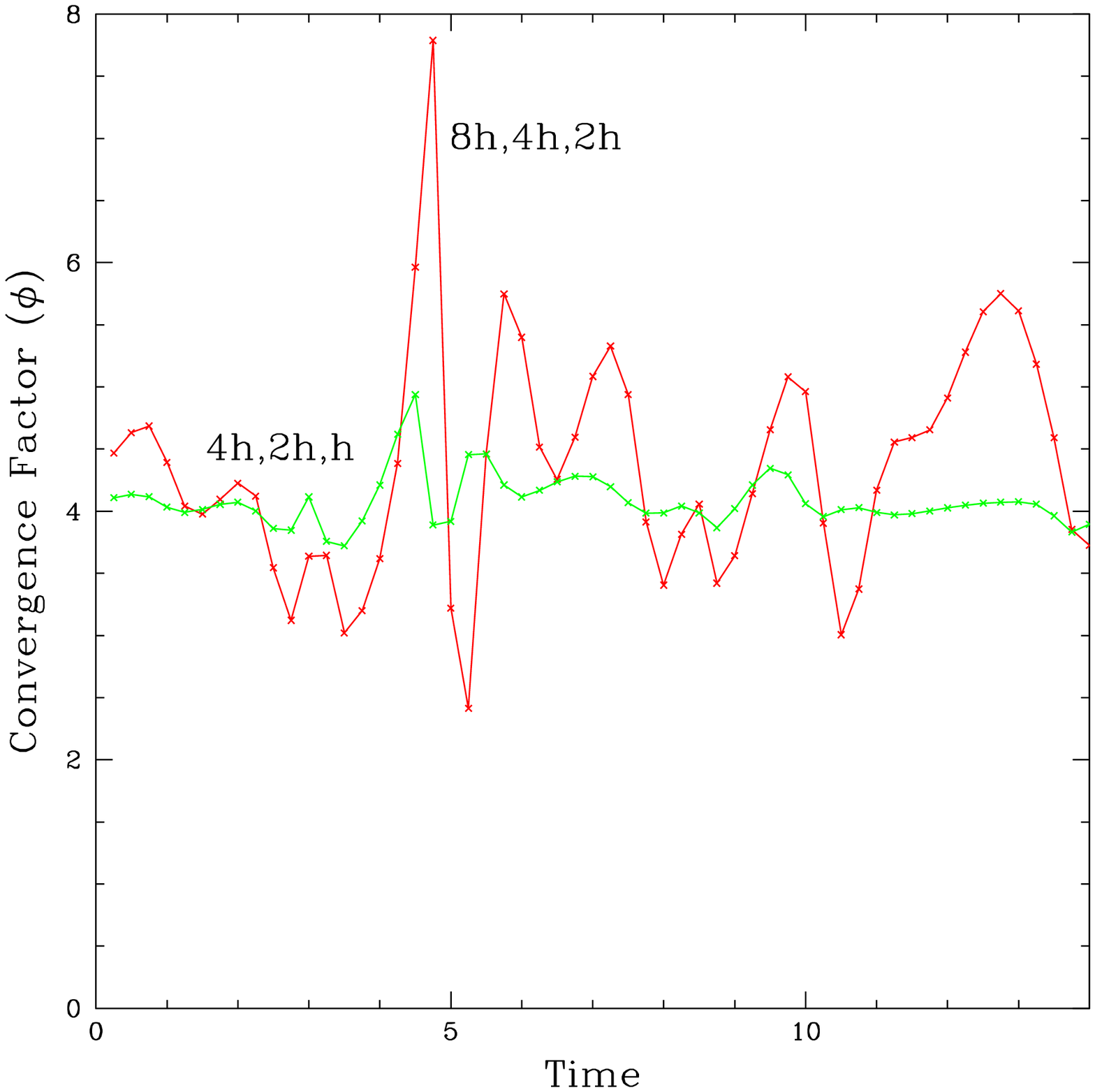}}
\caption[Convergence test of the 2D code for the field $\phi$]
{\small \label{fig:2Dcfacphi}
Convergence of 2D code for the field $\phi$ at four discretizations, 
$41\times 81$,
$81\times 161$,
$161\times 321$, and
$321\times 641$ gridpoints (8$h$, 4$h$, 2$h$, and $h$, respectively).
The convergence factors for levels 8$h$, 4$h$, and 2$h$ and levels 4$h$, 2$h$, and $h$
are shown in red and green, respectively.  The value of roughly four observed 
for each convergence factor indicates that the code is indeed second-order.
}
\label{fig:2Dcfacphi}
\end{figure} \noindent
However, convergent energy conservation and convergent field variables do
not imply that the MIB system is absorbing the outgoing radiation properly.

To verify that there is no significant reflected radiation off the outer boundary 
that contaminates the solution (exactly following the methods described in section
\ref{sec:codetest}) the MIB solution is compared to an ideal large-grid solution by
taking the $\ell_2$-norm of the difference at every point,
\begin{equation}
||A||_F = \left( \frac{1}{MN}\sum_{ij} |a_{ij}|^2\right)^{1/2}
\label{eq:frobnorm}
\end{equation}
where $a_{ij}$ are the components of an $M\times N$ matrix, $A$.
This norm is compared to the $\ell_2$-norm of the 
difference between the large-grid solution and a solution using an out-going boundary 
condition known not to work well (again, the massless outgoing boundary
condition, OBC).  Figure \ref{fig:2Dcontam} shows that after two crossing
times, the OBC solution becomes contaminated and the solution error increases 
dramatically (approximately three orders of magnitude!) while the MIB solution
always remains around or below $10^{-5}$ and shows no dramatic increase in
solution error indicative of contamination. 

\begin{figure}
\epsfxsize=14cm
\centerline{\epsffile{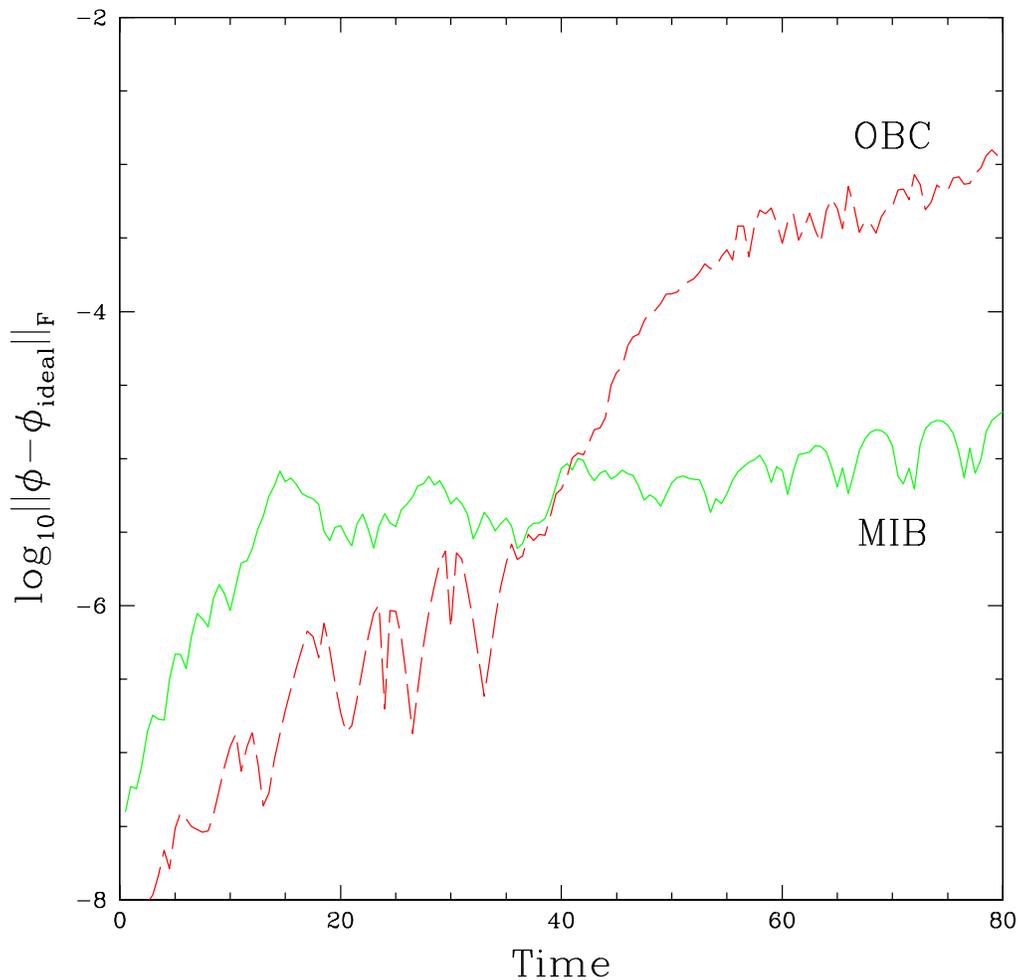}}
\caption[Contamination test for 2D code]
{\small \label{fig:2Dcontam}
Plot comparing the OBC (dashed red) and MIB (solid green) 
solutions to an ``ideal'' solution.
The OBC solution is obtained from using a massless outgoing
boundary condition, the MIB solution is obtained by solving the 
system in axisymmetric MIB coordinates, and the ideal solution is 
obtained by evolving the solution in standard ($R$,$z$,$t$) coordinates on 
a grid large enough to ensure no reflection off the outer boundaries.    
The error estimates are obtained from the $\ell_2$-norm of the difference 
between the trial solutions (OBC or MIB) and the ideal solution, 
$||\phi-\phi_{\rm ideal}||_F$.
Contamination of the OBC solution is observed at two crossing times, $t\approx 40$,
where the error estimate increases almost three orders of magnitude.
}
\label{fig:2Dcontam}
\end{figure} \noindent

Just like with the 1-D code, since the 2-D MIB code conserves energy 
quadratically, has quadratically convergent field variables, 
and is in agreement with the large-grid solution; it is
an acceptable means of solving equations 
(\ref{eq:2Dpidot}),(\ref{eq:2Dphirdot}),(\ref{eq:2Dphizdot}), and (\ref{eq:2Dphidot}),
while being dramatically more computationally efficient than either dynamical grid
methods or large-grid methods.  The computational demand for the MIB system 
grows {\it linearly} with the oscillon lifetime, while for dynamical or large-grid methods,
the computational demand grows as the {\it cube} of the oscillon lifetime\footnote{
Assuming that the outgoing radiation demands the grid to grow in both
the $R$ and $z$ direction.}.

\section{Boosting the Spherically Symmetric Oscillons as Initial Data}

The main (certainly
most fun) reason for creating an axisymmetric code to solve the
KG equation was to investigate what happens when two oscillons
collide.  

The first step in most dynamical calculations is to generate
meaningful initial data.
Since there is no ``closed-form'' oscillon solution, it is actually
non-trivial to generate ``boosted'' oscillon data.  There
are many approximate initial data configurations that give rise to a 
gaussian-shaped bubble that can move at the desired velocity.  
For example, ``boosting''
\be
\label{eq:approx_oscillon_noboost}
\phi(\tilde{R},\tilde{z},\tilde{t}) = 
\displaystyle{
	\phi_0\exp\left(-\frac{\tilde{R}^2}{\sigma_R^2}
			-\frac{\tilde{z}^2}{\sigma_z^2}\right)
		\sin\left(\omega \tilde{t}\right)
	}
\ee
by applying the coordinate transformation for a Lorentz boost along
the z-axis (using $\displaystyle{\gamma=\frac{1}{\sqrt{1-v^2}}}$):
\be
\phi(R,z,t) = 
\displaystyle{
	\phi_0\exp\left(-\frac{R^2}{\sigma_R^2}
			-\frac{\left( \gamma\left( z - v t \right)
				\right)^2}{\sigma_z^2}\right)
		\sin\left(\omega \gamma\left(t - v z \right) \right),
	}
\ee
usually results in an oscillon that moves at roughly the
desired velocity\footnote{Vacuum assumed to be at $\phi=0$.}.  
The problem, however, is that even with the best
possible choices of the free parameters ($\phi_0$, $\omega$,
$\sigma_R$, $\sigma_z$)
the approximate oscillon (\ref{eq:approx_oscillon_noboost})
is {\it not} a solution to the (Lorentz invariant) Klein-Gordon 
equation.  Therefore, trying to generate initial data in this 
manner often results in dramatically different global behavior
depending on the ``boost velocity'', $v$ 
(ie. behavior that is not Lorentz invariant and can be as dramatic as leaving
the universe in different vacuum states!).
The goal is to be able to compare oscillons being collided at
different velocities while knowing exactly how each oscillon 
behaves in its own rest-frame.

We choose to numerically evolve the
oscillon in its rest frame (which can be done efficiently in {\it spherical}
symmetry) and then  {\it numerically} boost the solution.
The single boosted oscillon then, of course, retains all of its
Lorentz invariant properties regardless of boost velocity.
The collision initial data is constructed from the superposition 
of two boosted oscillons (one at some $z<0$ boosted in the $z^+$ direction, 
and the other at some $z>0$ boosted in the $z^-$ direction).  

The equations for the transformation between the boosted (tilde) and rest
(non-tilde) frames are just the equations for a Lorentz boost along
the z-axis:
\begin{equation}
\begin{array}{rclcrcl}
\tilde{t} &=& \gamma\left( t - v z \right) & \hspace{1cm}&
t &=& \gamma\left( \tilde{t} + v \tilde{z} \right) \\
\tilde{z} &=& \gamma\left( z - v t \right) & &
z &=& \gamma\left( \tilde{z} + v \tilde{t} \right)\\
\tilde{R} &=& R & & 
R &=& \tilde{R} \\ 
\tilde{\theta} &=& \theta & & 
\theta &=& \tilde{\theta} 
\label{boosteq}
\end{array}
\end{equation}
where again, 
$\displaystyle{\gamma = \frac{1}{\sqrt{1-v^2}}}$.
Since $\phi$ is a scalar field, it transforms trivially as
\begin{equation}
\tilde{\phi}(\tilde{t},\tilde{R},\tilde{\theta},\tilde{z})=
\phi\left((\gamma\left( t - v z \right),R,\theta,\gamma\left( z - v t \right)\right),
\label{phitrans}
\end{equation}
while the other field variables (being derivatives of the scalar field) transform
as forms 
$\partial_{\mu'}\phi = 
\displaystyle{\frac{\partial x^\mu}{\partial x^{\mu'}}\partial_{\mu}\phi}$ which
gives
\begin{equation}
\begin{array}{rcl}
\tilde{\Pi} &=&  \displaystyle{\gamma\Pi + \gamma v \Phi_z} \\
&=&  \displaystyle{\gamma\Pi + \frac{\gamma v z}{\sqrt{R^2 + z^2}} \Phi_r}.\\
\label{pitrans}
\end{array}
\end{equation}
where $\Phi_r=\partial_r \phi$ and is obtained from the spherically 
symmetric evolution.
(Remember that \hbox{$r \equiv \displaystyle{\sqrt{x^2 + y^2 + z^2}}$} and 
\hbox{$R\equiv \displaystyle{\sqrt{x^2 + y^2}}$}.)

Equations (\ref{boosteq}) imply that to obtain data for $\tilde{t}=0$,
$\tilde{R}_{\rm min}\le \tilde{R}\le \tilde{R}_{\rm max}$, and
$\tilde{z}_{\rm min}\le \tilde{z} \le \tilde{z}_{\rm max}$, 
the rest frame solution
must be known over the spacetime domain 
$\gamma v\tilde{z}_{\rm min}\le t\le \gamma v\tilde{z}_{\rm max}$,
$\tilde{R}_{\rm min}\le R \le \tilde{R}_{\rm max}$, and
$\gamma \tilde{z}_{\rm min}\le z \le \gamma \tilde{z}_{\rm max}$.
Time symmetric initial data was used and the $z$ domain was
chosen such that $\tilde{z}_{\rm min}=-\tilde{z}_{\rm max}$
which allowed the time domain for the
spherical evolution to be restricted to 
$0\le t\le \gamma \,v\,\tilde{z}_{\rm max}$ (see figure \ref{fig:boostslice}
for a schematic relating a ``new'' $t=0$ slice to the ``old'' spacetime 
($\tilde{t}$,$\tilde{z}$) 
domain, for a typical $R={\rm constant}$ slice).
\begin{figure}
\epsfxsize=14cm
\centerline{\epsffile{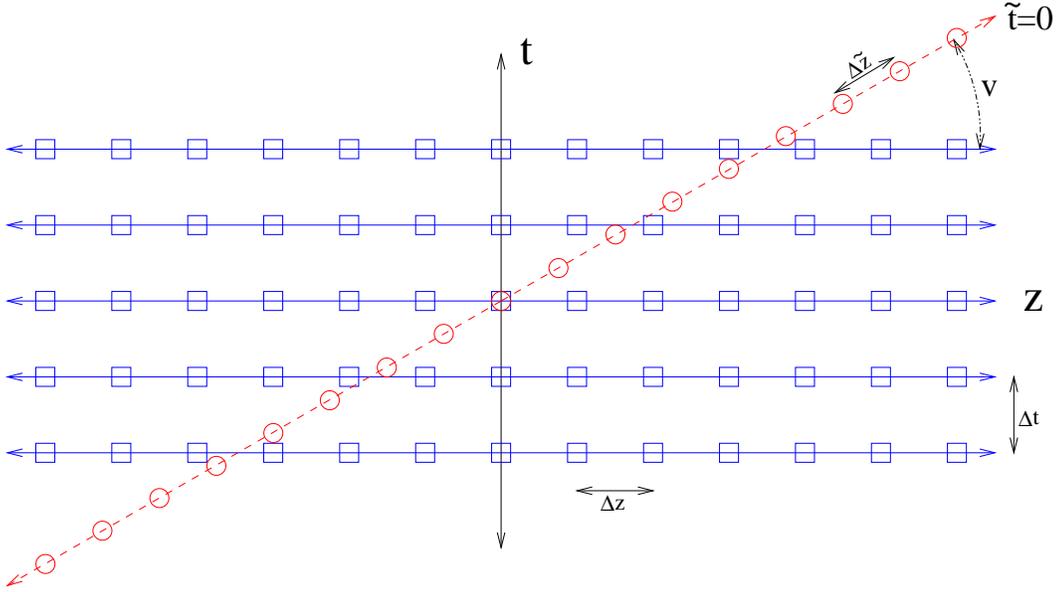}}
\caption[Schematic of boosted initial data generation]
{\small \label{fig:boostslice}
A schematic (for a sample constant-$R$ slice)
displaying how the ($R$,$\tilde{z}$,$\tilde{t}_0$) boosted data is obtained 
by interpolating from rest-frame ($R$, $z$,$t$) dynamic data.
Data points that do not match up exactly between the tilde and non-tilde frames
are linearly interpolated in both space and time (note that figure
is only a schematic and not drawn to scale, $\Delta z$ and $\Delta t$ were always taken 
to be very small compared to $\Delta \tilde{z}$).
The blue squares are gridpoints lying on $t$-constant hypersurfaces 
in the rest-frame. 
Red circles are gridpoints lying on the $\tilde{t}=0$ boosted-frame
hypersurface (the initial data being obtained).
To obtain boosted data for $\tilde{z}_{min}\le \tilde{z}\le \tilde{z}_{max}$ at
$\tilde{t}=0$, the rest-frame radius must be known over the domain
$\gamma v\tilde{z}_{min}\le t \le \gamma v\tilde{z}_{max}$ and
$\gamma \tilde{z}_{min}\le z\le \gamma \tilde{z}_{max}$.
Since $R$ and $\tilde{R}$ are orthogonal to the boost, the field
remains unaffected in that direction.
}
\label{fig:boostslice}
\end{figure}

A boosted oscillon can then be obtained by time evolving 
the desired (rest frame) oscillon throughout the necessary spacetime
domain and transforming the field variables using equations 
(\ref{phitrans},\ref{pitrans}).  Since very few of the 
($t$, $R$, $z$) gridpoints match exactly any gridpoint from the 
spherically symmetric ($t$,$r$) grid, linear interpolation in both space
and time was used
(the 1-D grid that was interpolated from is at a {\it much} 
finer resolution than the 2-D grid). 
A schematic of an  constant-$R$ slice can be seen in
figure \ref{fig:boostslice}.
Finally, {\it collision initial data} is obtained by forming
the superposition of two boosted oscillons (obviously, boosted
{\it at} one another.).

\subsection{Testing the Numerically Boosted Initial Data}

The ``numerical boosting'' of the spherically symmetric oscillon 
data was tested by monitoring the Lorentz invariant behavior
of expanding bubble formation.
For a spherically symmetric  gaussian pulse, instantaneously at rest,
\begin{eqnarray}
\phi(r,0) &=& \phi_F + \left( \phi_T-\phi_F\right) \exp\left(-\frac{r^2}{\sigma_r^2}\right)
\label{eq:atrestID}\\
\vbox{\vspace{0.75cm}} \dot{\phi}(r,0) &=& 0                                    
\label{eq:gaussianID}
\end{eqnarray}
so  $\phi(\infty,t)=\phi_F$ and $\phi(0,t)=\phi_T$,
there will be a critical ``bubble radius'', $\sigma_r^*$, 
for which time evolution of $\sigma_r > \sigma_r^*$
will lead to an expanding bubble 
that converts false vacuum to true vacuum everywhere. For 
$\sigma_r < \sigma_r^*$ the time evolution will eventually 
result in dispersal of the field (possibly after an oscillon 
stage).  Since $\sigma_r^*$ is defined to be the {\it rest-frame}
critical radius for expanding bubble formation, this number should not change and
the boosting of the critical field configuration should always lie on the 
threshold of expanding bubble formation.
\begin{figure}
\epsfxsize=14cm
\centerline{\epsffile{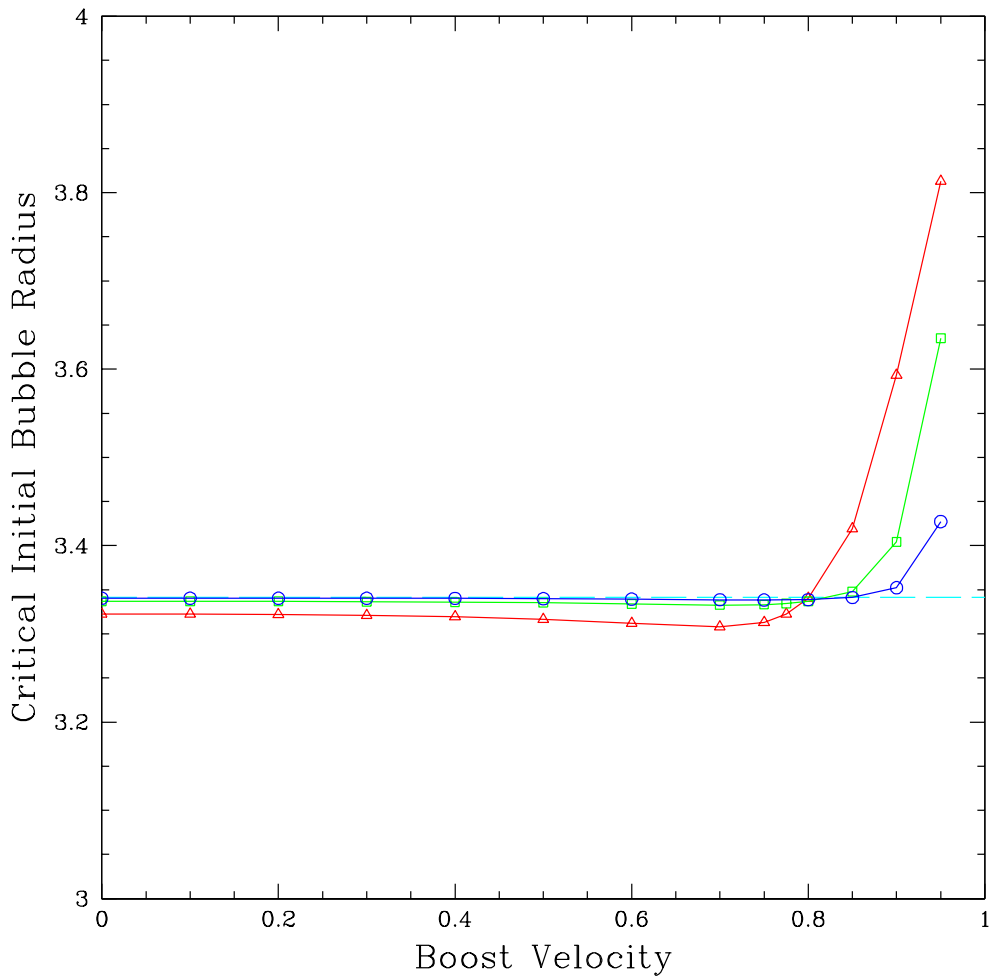}}
\caption[Critical initial bubble radius versus boost velocity ]
{
Plot of the Critical Initial Bubble Radius versus Boost Velocity 
for three discretizations ($81\times 161$ in red triangle, 
$161\times 321$ in green squares, and $321\times 641$ in blue circles).
Since the initial bubble radius parameter in the code is the 
{\it rest-frame} initial radius, the Lorentz invariant behavior 
of contraction or expansion (which governs the threshold value of 
$r_0$) should {\it not} change as a function
of boost velocity.  The ``ideal'' result is the (dashed cyan) horizontal line at
$r_0\approx 3.34$ (determined spherically symmetric collapse at $N_r=2049$).
The deviation at large boost velocity is due
to the increasing effects of length contraction of the bubble.
}
\label{fig:boostconv1}
\end{figure} 
Figure (\ref{fig:boostconv1}) 
shows the critical bubble radius as a function 
of boost velocity for three resolutions 
($81\times 161$, $161\times 321$, $321\times 641$);
the critical bubble radius remains
constant to within one-quarter percent at $161\times 321$ for $v\leq 0.8$,
and to within one percent at $v=0.85$.  The three resolutions show convergent 
properties (ie. the graph of $\sigma_r^*$ vs. boost velocity approaches
a constant as $h\rightarrow 0$)
and suggest that the deviation from the expected value of 
$\sigma_r^*$ at high boost velocity is due to the steep gradients 
which primarily result from the length contraction of the bubble.

\section{Collision and Endstate Detection}

In the spherically symmetric ADWP collapse, there were two possible endstates:
expanding bubble formation {\it or} bubble collapse (usually resulting in the 
formation of an oscillon).
However, in axisymmetric collision simulations, there are four possible 
endstates that we consider\footnote{Of course, there could be more ways
of classifying them into more endstates, but we choose four.}:
expanding bubble formation, annihilation, soliton-like transmission,
and coalescence.

\begin{figure}[h]
\vbox{\vspace{-2cm}}
\centerline{\vbox{
                \hbox{\hspace{1cm}}
                \hbox{\epsfxsize =10cm\epsffile{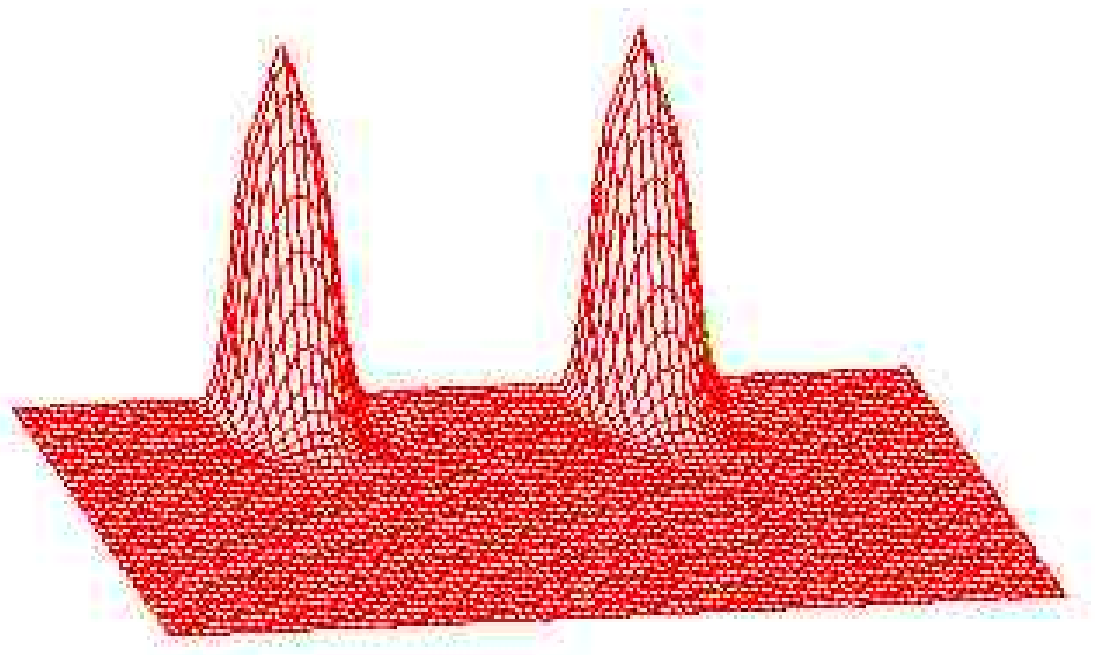}}
                \hbox{\hspace{2.5cm}(a)}
                \hbox{\epsfxsize =10cm\epsffile{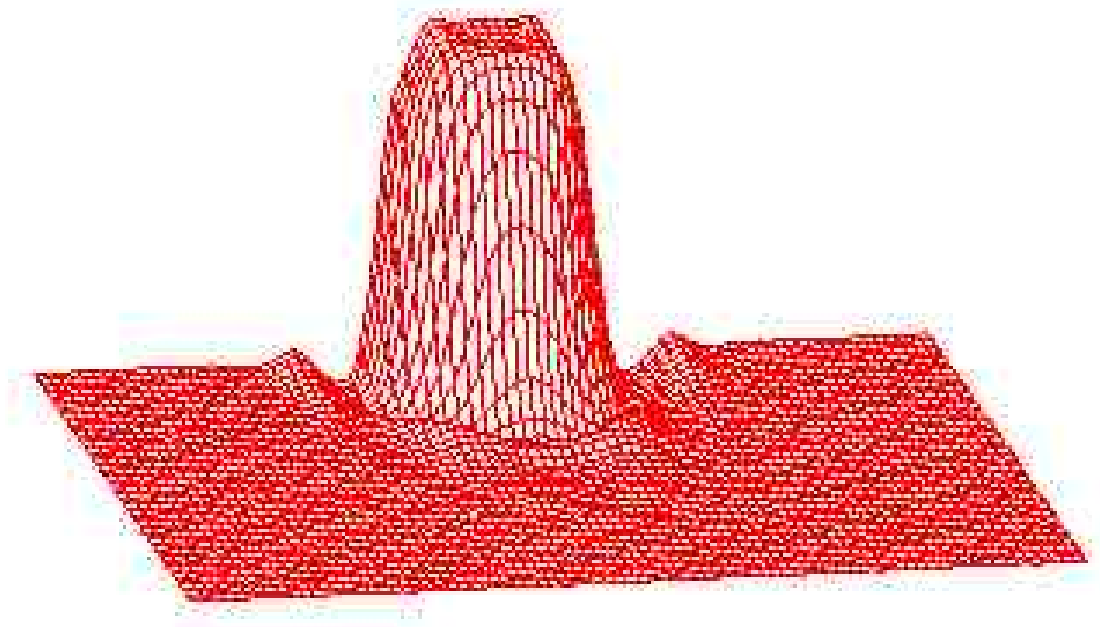}}
                \hbox{\hspace{2.5cm}(c)}
                  }
            \hbox{\hspace{-2cm}}
            \vbox{
                \hbox{\epsfxsize =10cm\epsffile{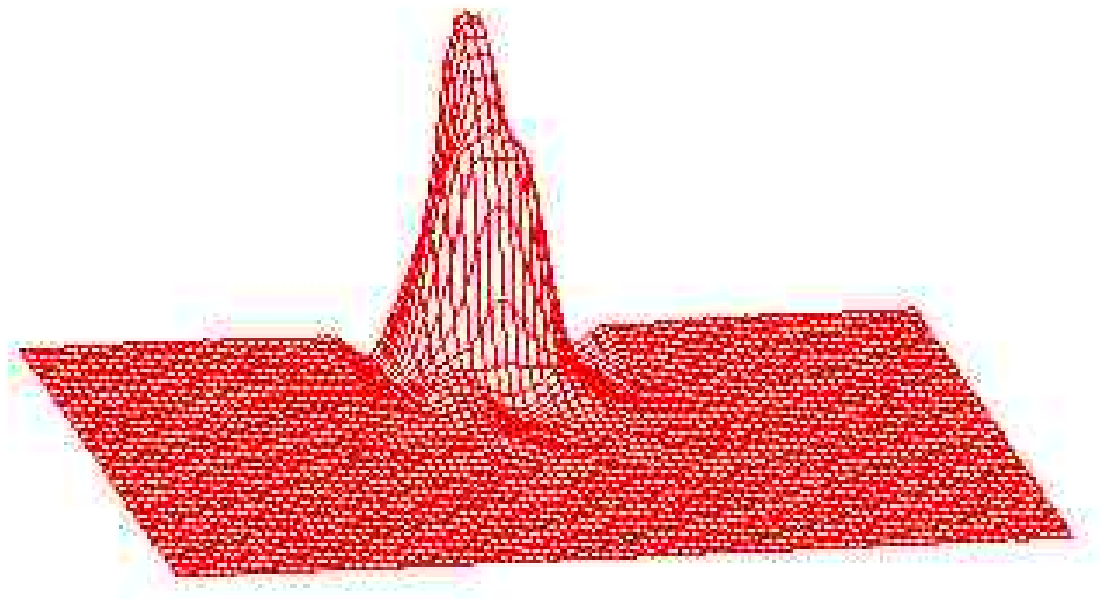}}
                \hbox{\hspace{2.5cm}(b)}
                \hbox{\epsfxsize =10cm\epsffile{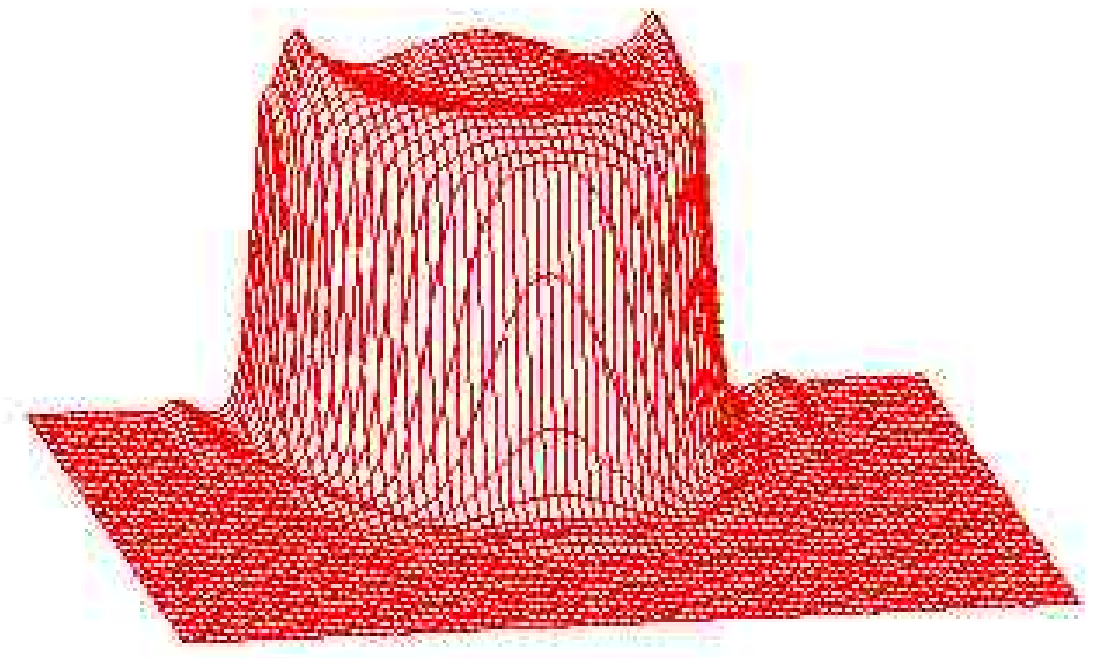}}
                \hbox{\hspace{2.5cm}(d)}
                  }
           }
\caption[Time series showing expanding bubble formation]{\small \label{fig:tseries_tvac} 
A times series showing the collision of two oscillons forming an 
expanding true vacuum bubble. 
The initial frame (a) shows the two oscillons boosted toward one another; 
the fields interpolate between the vacua so that their peaks are at the 
true vacuum while away from the oscillons is the false vacuum.
Frames (b) and (c) show the nonlinear interactions during the collision.
Frame (d) shows the beginning of an expanding bubble; the steep field 
gradient is the bubble wall.  Inside the bubble wall is the true vacuum and
the bubble wall will continue to move outward (eventually reaching 
the speed of light).
}
\end{figure}

{\it Expanding bubble formation} occurs when the whole spacetime is converted
from the (unstable) false vacuum to the (stable) true vacuum.  
This can occur trivially when the bubbles that are being boosted 
at one another each have initial rest-frame radii above the critical 
radius for expanding bubble formation (for $\delta=0.1$, $\sigma^*_r \approx 3.3$); 
this is the case where each bubble would lead to an expanding bubble independently. 
However, a more interesting case involves the collision of  two oscillons, each with a rest-frame
radius {\it less than} the critical radius for expanding bubble formation.
In this case, each oscillon independently would {\it not} lead 
to an expanding bubble, but together they overcome the potential barrier of 
$V_A(\phi)$ and a {\it false-to-true} phase transition ensues 
(see figure \ref{fig:TFcollisioncartoon} for heuristic explanation).
\begin{figure}
\epsfxsize=8cm
\centerline{\epsffile{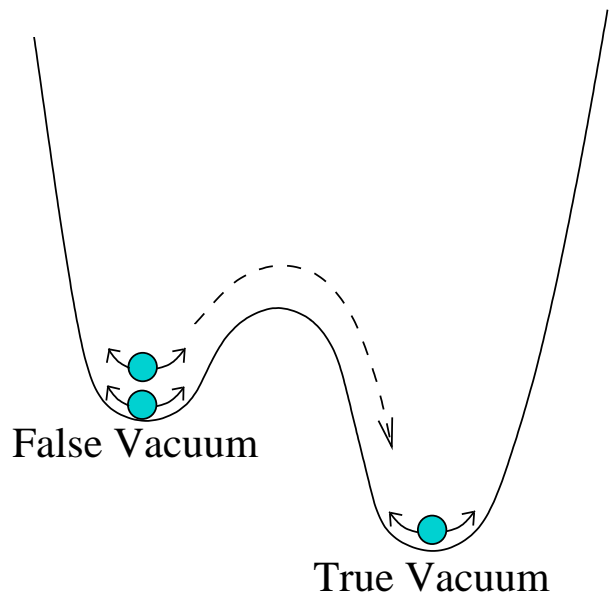}}
\caption[Colliding bubbles lead to phase transition]
{\small \label{fig:TFcollisioncartoon}
Heuristic representation of what happens {\it energetically}  when two 
oscillons collide to form an expanding true vacuum bubble. 
Individually, the two bubble configurations 
(represented by the two balls oscillating about the false vacuum)
are far from one another and do not individually have 
enough energy to overcome the potential barrier; if left alone, 
they would eventually disperse, leaving the spacetime in the false vacuum. 
However, when the two oscillons collide, their combined energy is enough overcome 
the potential barrier and form an expanding bubble that
converts the spacetime to true vacuum.  A time series of an actual
collision of this type can be seen in figure \ref{fig:tseries_tvac}.
}
\label{fig:TFcollisioncartoon}
\end{figure} \noindent
The code detects bubble formation by computing the 
area (${\int dR\ dz}$, 
since we work in two spatial dimensions) that is in
the true-vacuum and stopping the evolution after some empirically determined
threshold is reached. In practice, it is quite easy to see 
where the ``surface tension'' in the bubble wall can no longer compete with 
the field's ``attraction'' to the true-vacuum\footnote{We are also guided
by critical radius measured from the 1-D ADWP results.}; 
the detection threshold is then taken to be safely above this point.  
A time series of snapshots of an actual 2-D collision that results in 
the formation of an expanding bubble can be seen in figure
\ref{fig:tseries_tvac}.

The other three endstates being considered do {\it not} form 
expanding bubbles and therefore leave the spacetime in the false 
vacuum.  Two of these three endstates, annihilation and
soliton-like transmission are forms of dispersal, while 
coalescing oscillons remain within the computational domain.
To help distinguish between these endstates, three second
moments are considered\footnote{Note that these are clearly not {\it physical}
moments, but they work well to distinguish between the various endstates of the
oscillon-oscillon collisions.}:
\begin{eqnarray}
M^2_R(t) &=& \displaystyle{\frac{\displaystyle{\int R^2 \tilde{\rho} \ dR \ dz}}{\displaystyle{\int \tilde{\rho} \ dR \ dz}}} \\
M^2_z(t) &=& \frac{\displaystyle{\int z^2 \tilde{\rho} \ dR \ dz}}{\displaystyle{\int \tilde{\rho} \ dR \ dz}} \\
M^2_r(t) &=& \frac{\displaystyle{\int \left(R^2 + z^2\right) \tilde{\rho} \ dR \ dz}}{\displaystyle{\int \tilde{\rho} \ dR \ dz}}
\end{eqnarray}
where 
\begin{equation}
\tilde{\rho}(R,z,t) = \displaystyle{
\frac{1}{2}\left( 
 \frac{\Pi^2}{a^2 b^2}
+\frac{\Phi_R^2}{a^2}
+\frac{\Phi_z^2}{b^2}
\right).
}
\end{equation}
$\tilde{\rho}$ is the energy density due to the time derivative and gradients 
of the field, neglecting the potential terms.  
Including the potential term would lead to large contributions to the moments from 
{\it outside} the bubble, whereas the goal here is to know the the location of the surface 
of the bubbles.
For all of the collisions discussed here, the initial data are 
symmetric about the $z=0$ plane, and hence the collision also occurs
at $z=0$.  

{\it Annihilation} occurs when two oscillons collide and interact
in such a way that the field is no longer localized and all the radiation 
disperses.  
The code determines this to be the endstate when both $M^2_R$ and $M^2_z$
rise above an empirically chosen threshold.  This was typically taken
to be around $100$ which implies most of the matter is around 
$r=\sqrt{R^2 + z^2}=10$), which, in turn, is indicative of dispersal 
for oscillons
with typical radii around $r\approx 3$.  
A time series of a collision
resulting in annihilation can be seen in figure \ref{fig:tseries_ann}.
\begin{figure}[h]
\vbox{\vspace{-2cm}}
\centerline{\vbox{
                \hbox{\hspace{1cm}}
                \hbox{\epsfxsize =10cm\epsffile{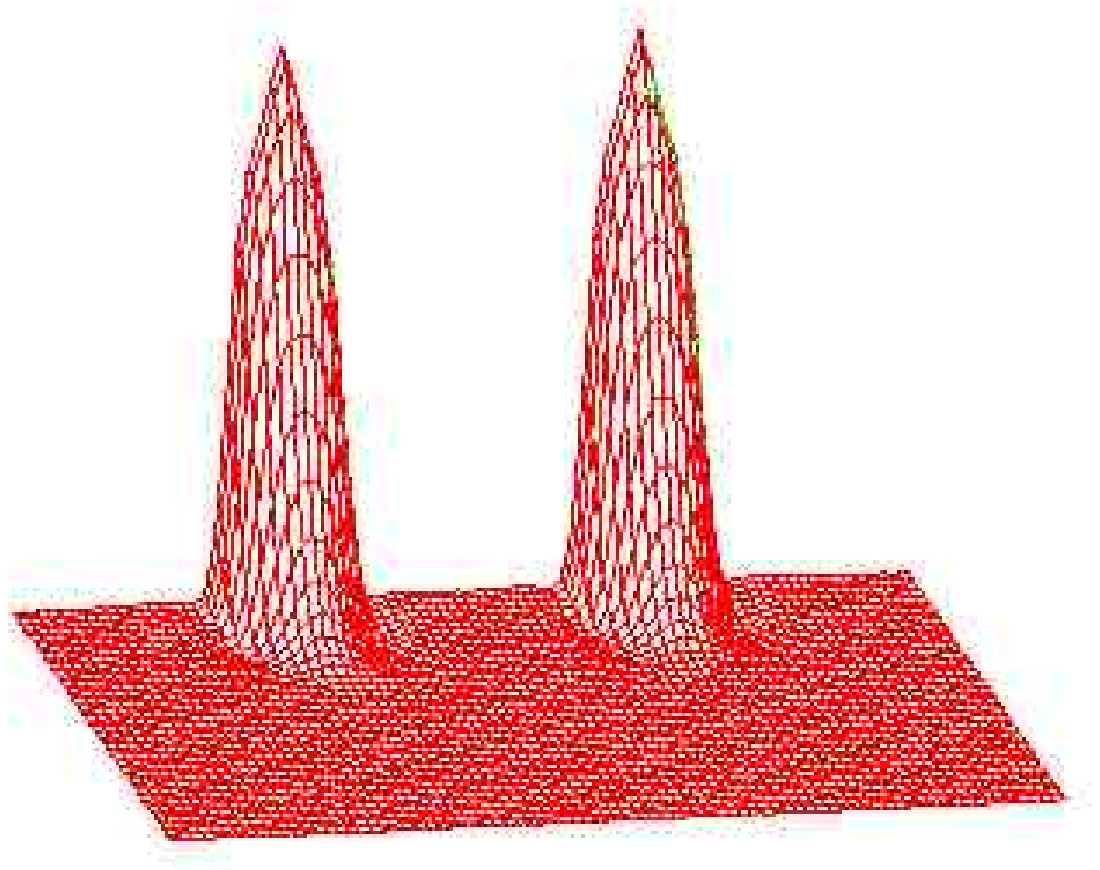}}
                \hbox{\hspace{2.5cm}(a)}
                \hbox{\epsfxsize =10cm\epsffile{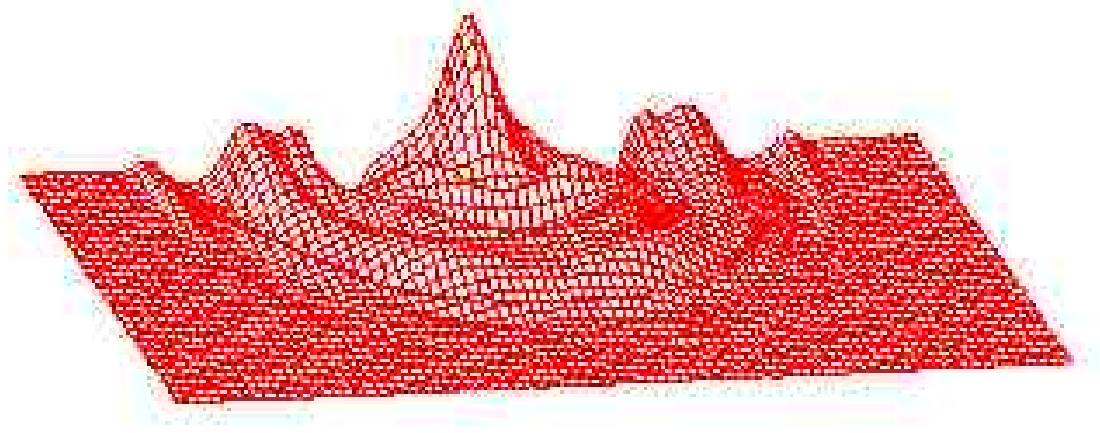}}
                \hbox{\hspace{2.5cm}(c)}
                  }
            \hbox{\hspace{-2cm}}
            \vbox{
                \hbox{\epsfxsize =10cm\epsffile{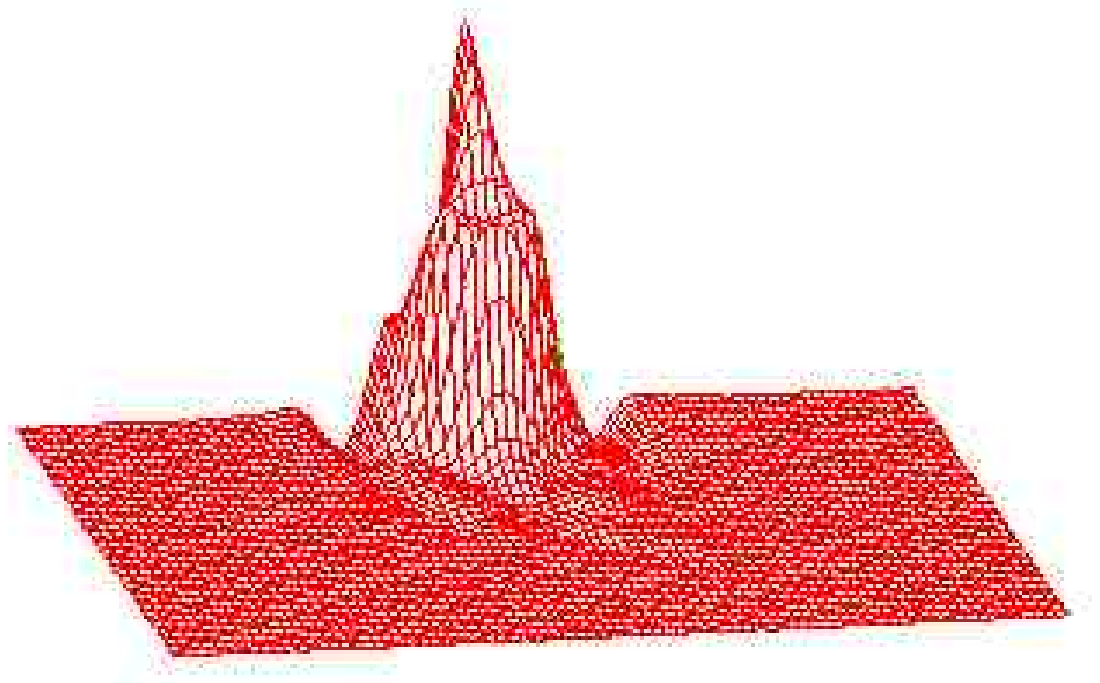}}
                \hbox{\hspace{2.5cm}(b)}
                \hbox{\epsfxsize =10cm\epsffile{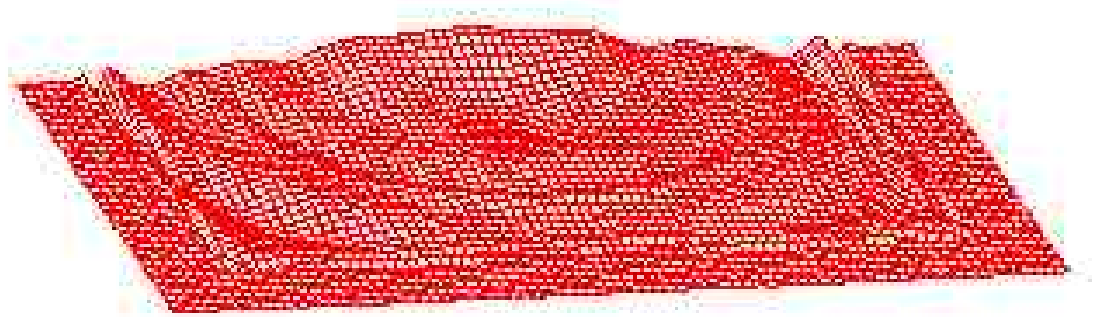}}
                \hbox{\hspace{2.5cm}(d)}
                  }
           }
\caption[Time series showing bubble annihilation]{\small \label{fig:tseries_ann} 
A times series showing the collision of two oscillons that annihilate each
other.
The initial frame (a) shows the two oscillons boosted toward one another; 
the fields interpolate between the vacua so that their peaks are at the 
true vacuum while away from the oscillons is the false vacuum.
Frames (b) and (c) show the nonlinear interactions during the collision.
Frame (d) shows the endstate consisting of almost entirely 
of outgoing radiation.  The spacetime is left in the false vacuum.
}
\label{fig:tseries_ann} 
\end{figure}

{\it Soliton-like Transmission} occurs when two oscillons collide, interact, and 
then pass through (or reflect off) one another while each oscillon stays
localized.   
\begin{figure}[ht]
\vspace{-1cm}
\centerline{\vbox{
                \hbox{\hspace{1cm}}
		\hbox{\epsfxsize =7cm\epsffile{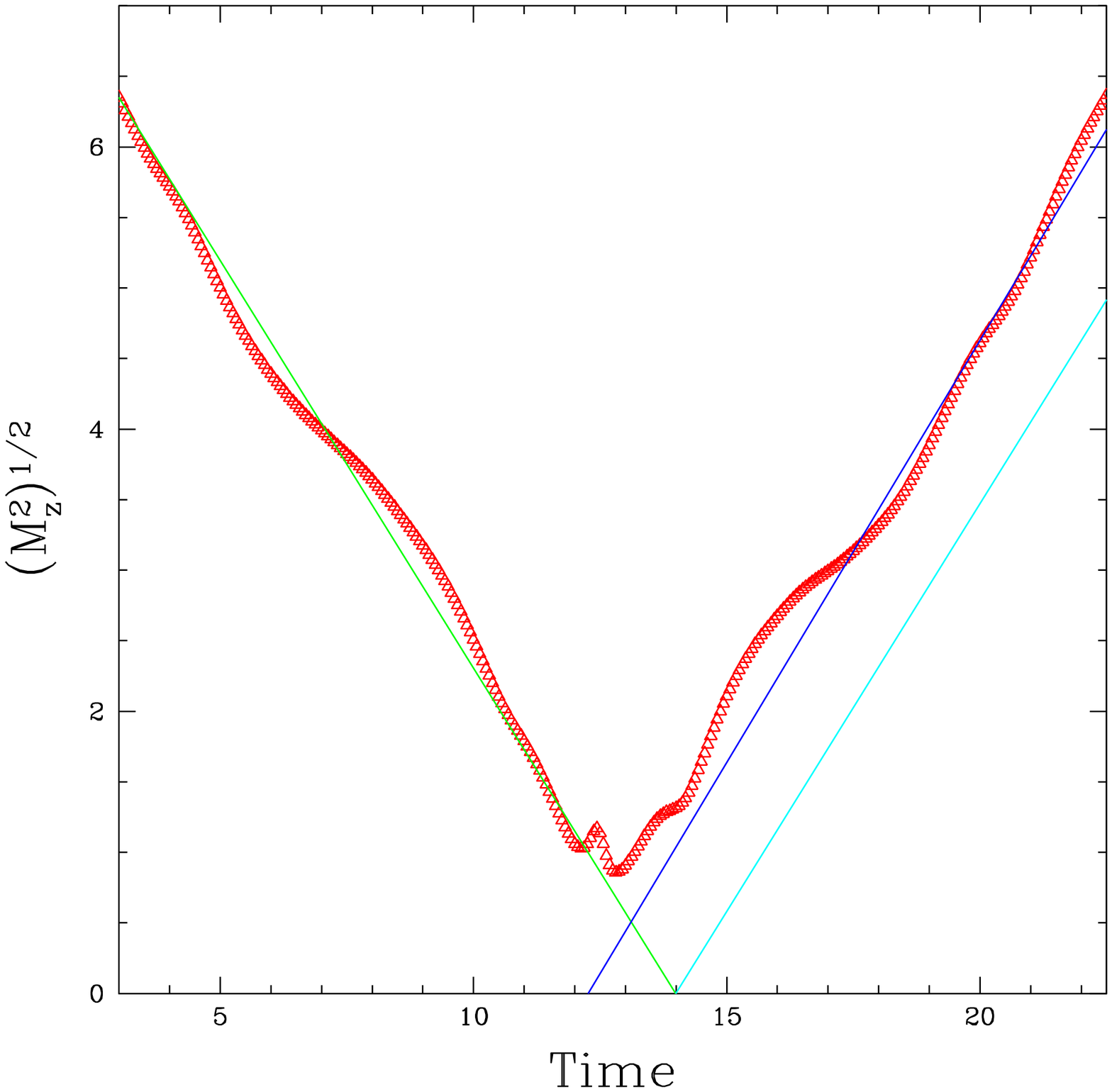}}
                \hbox{\hspace{2cm}(a) $v_b=0.6$, $\sigma_0=2.15$}
                \hbox{\epsfxsize =7cm\epsffile{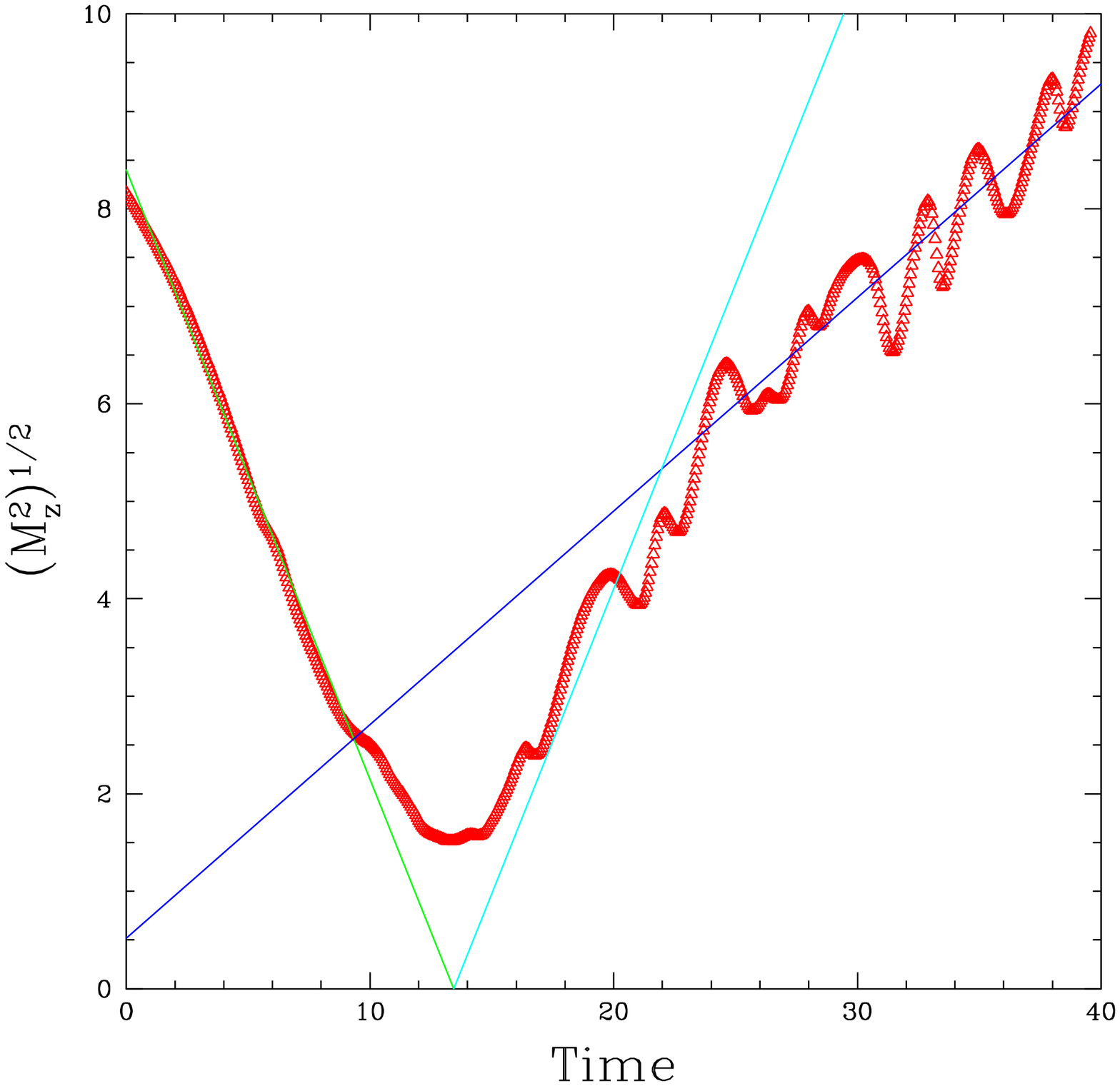}}
                \hbox{\hspace{2cm}(c) $v_b=0.6$, $\sigma_0=2.675$}
                  }
            \hbox{\hspace{0.1cm}}
            \vbox{
		\hbox{\epsfxsize =7cm\epsffile{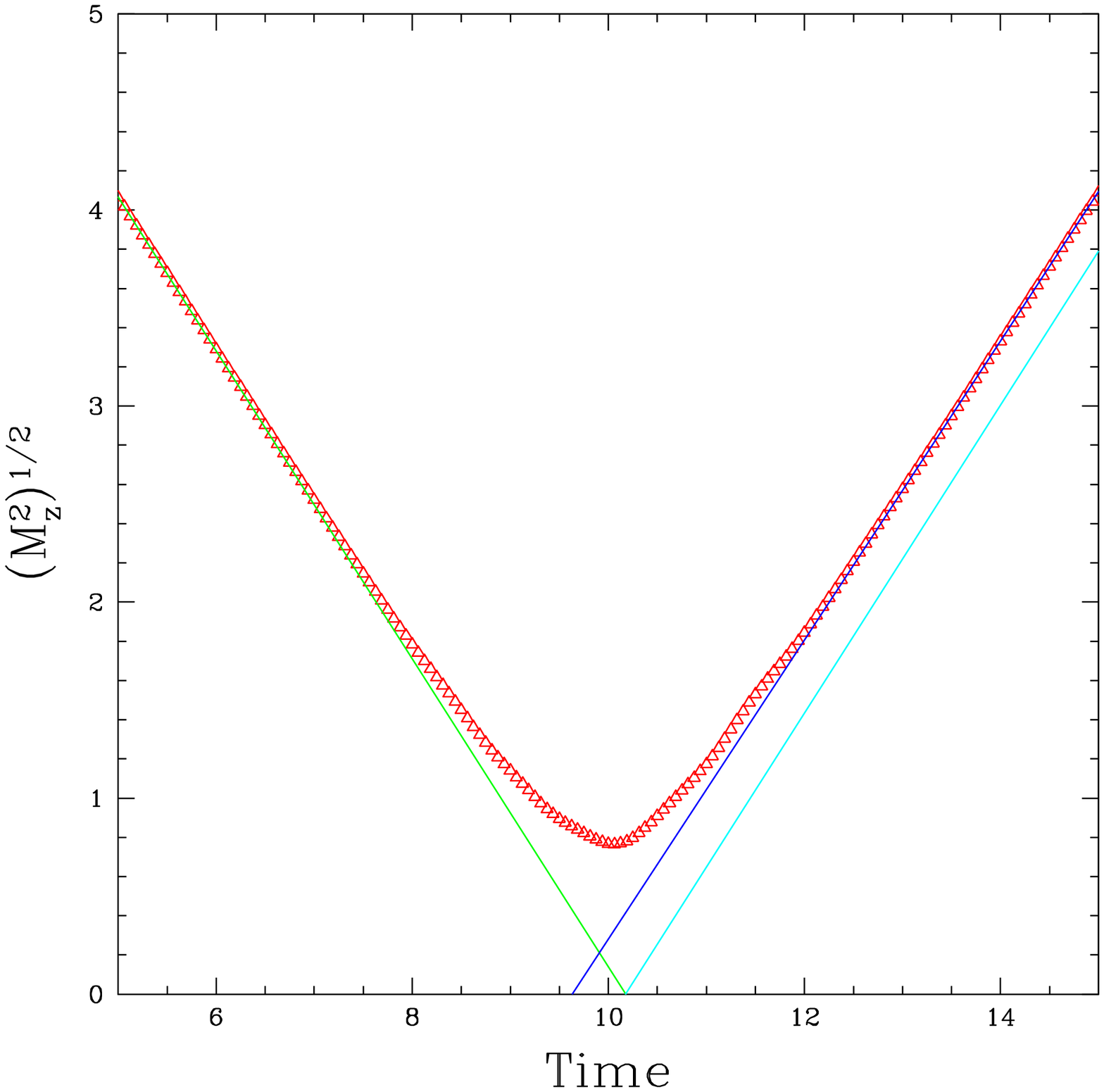}}
                \hbox{\hspace{2cm}(b) $v_b=0.85$, $\sigma_0=2.0$}
		\hbox{\epsfxsize =7cm\epsffile{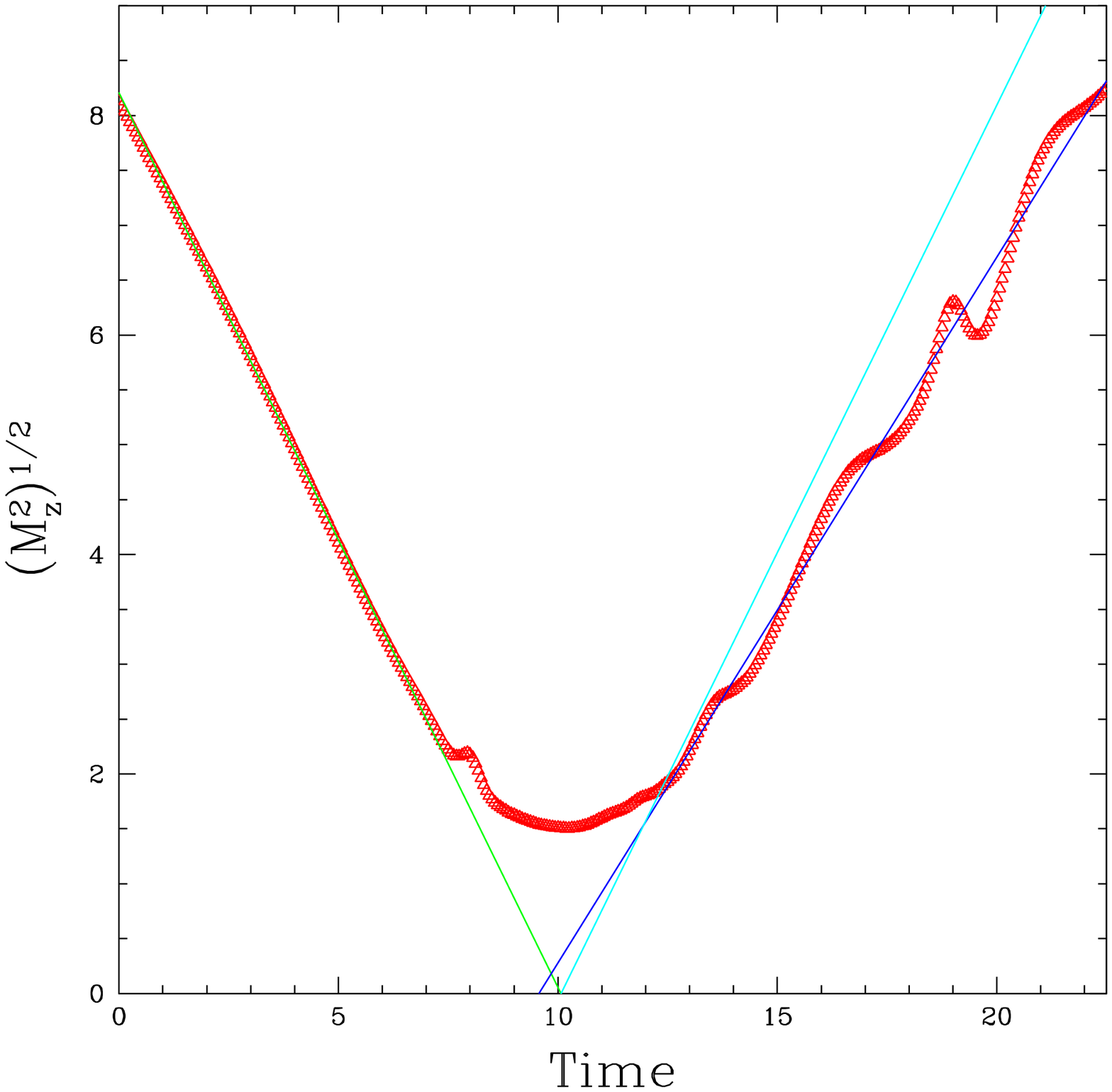}}
                \hbox{\hspace{2cm}(d) $v_b=0.85$, $\sigma_0=3.325$}
                  }
           }
\caption[$\Delta v$ and $\Delta \theta$ for soliton interaction]{\small \label{fig:sollag}
Plots of the location, $(M^2_z)^{1/2}$, of the oscillon located 
in the ``upper'' ($z>0$) half of the computational domain as a function of time. 
In either plot, the green line is the best fit for the ingoing position as
a function of time, the blue line is the best fit for the outgoing position, 
and the cyan line is the path that the oscillon 
{\it would have taken} had it not interacted (nonlinearly) with the other oscillon 
(direct reflection across $z=0$).
During the transmission the velocities (slopes) can increase {\it or} decrease
while the outgoing path was always seen to be shifted backward in time (x-intercept
always moved left).
Table \ref{tbl:sollag} shows the measured time lags and changes in velocity
for these four evolutions.
}
\label{fig:sollag}
\end{figure}
Again, this is easily determined by monitoring the second moments.  Since the oscillons
pass through one another, the $z$ moment will get large as they leave the computational 
domain.  However, since each oscillon remains localized, the $R$ moment will stay
small (on the order of the typical oscillon radius of $r\approx 3$).
A time series of a collision
resulting in soliton-like transmission can be seen in figure \ref{fig:tseries_sol}.
\begin{figure}
\vbox{\vspace{-2cm}}
\centerline{\vbox{
                \hbox{\hspace{1cm}}
                \hbox{\epsfxsize =10cm\epsffile{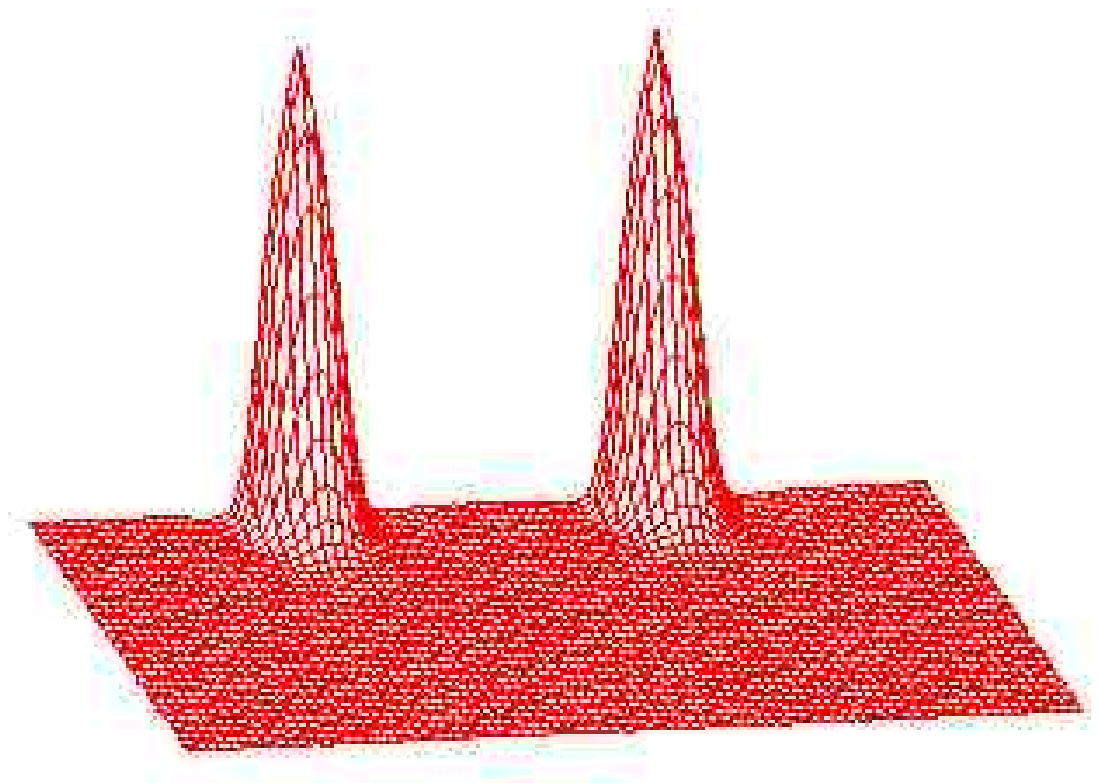}}
                \hbox{\hspace{2.5cm}(a)}
                \hbox{\epsfxsize =10cm\epsffile{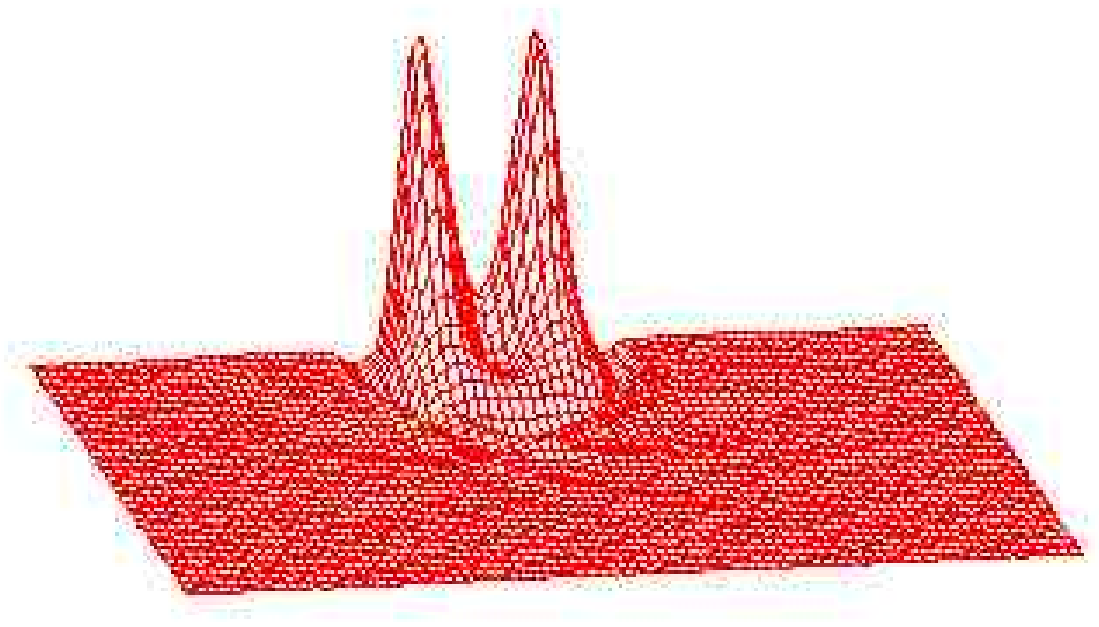}}
                \hbox{\hspace{2.5cm}(c)}
                  }
            \hbox{\hspace{-2cm}}
            \vbox{
                \hbox{\epsfxsize =10cm\epsffile{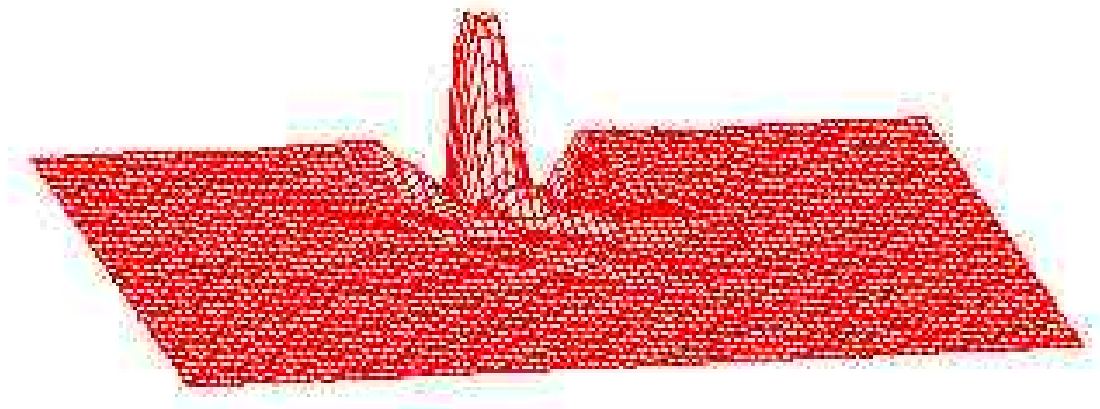}}
                \hbox{\hspace{2.5cm}(b)}
                \hbox{\epsfxsize =10cm\epsffile{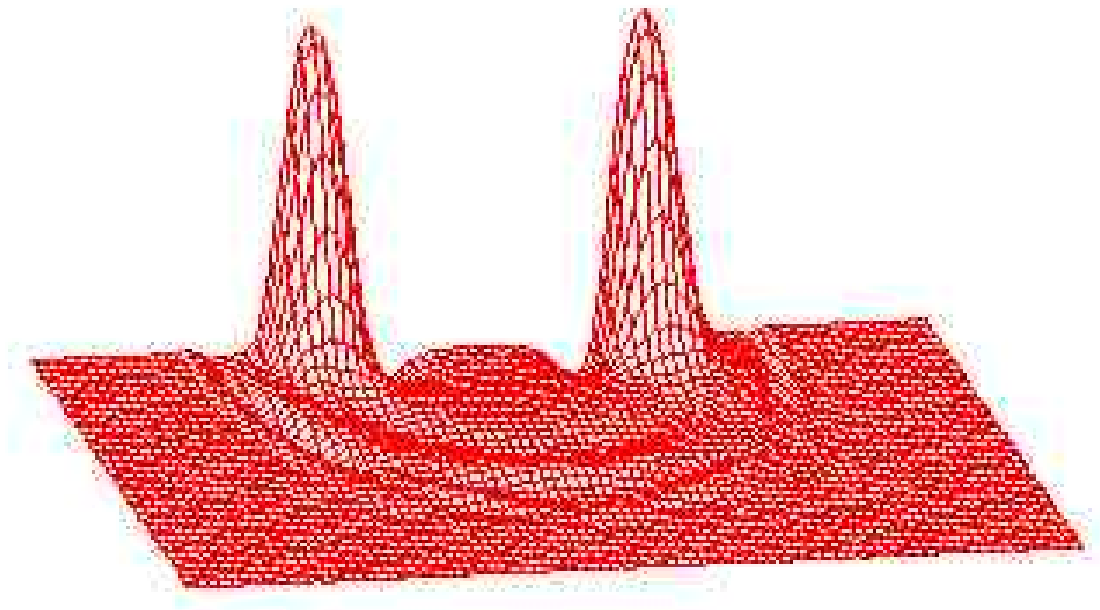}}
                \hbox{\hspace{2.5cm}(d)}
                  }
           }
\caption[Time series showing soiton-like transmission]{\small  \label{fig:tseries_sol}
A times series showing the collision of two oscillons that pass through 
(or reflect off) one another.
The initial frame (a) shows the two oscillons boosted toward one another; 
the fields interpolate between the vacua so that their peaks are at the 
true vacuum while away from the oscillons is the false vacuum.
Frames (b) and (c) show the nonlinear interactions during the collision.
Frame (d) shows the endstate consisting of two oscillons moving outward
along the $z$-axis. The spacetime is left in the false vacuum.
}
 \label{fig:tseries_sol}
\end{figure}
\begin{table}
\centerline{
\begin{tabular}{cc|cc}
\hline
 $v_b$ & $\sigma_0$ & $\Delta |v|$ & $\Delta T$ \\ 
\hline
0.6 & 2.15   & +0.02 & -1.7  \\
0.6 & 2.675  & -0.41 & -15.8 \\
0.85 & 2.00  & -0.02 & -0.6  \\
0.85 & 3.325 & -0.17 & -0.5  \\
\hline
\end{tabular}
}
\caption[Table of $\Delta v$ and $\Delta T$ for sample oscillon transmission]{
\small \label{tbl:sollag}
Table of change in the oscillon velocity (magnitude), 
$\Delta |v|$, and time ``lag'', $\Delta T$, for four sample soliton-like collisions.
Oscillons (and solitons) that interact {\it nonlinearly} 
are usually characterized by a time shift or change in velocity 
arising from coupling between modes. 
If the oscillons interacted linearly, $\Delta |v|$ and $\Delta T$ would be zero.
The oscillons explored here 
almost always slow down or stay at {\it roughly} the same speed and
the time lag (measured by extrapolating back to $z=0$) almost always is negative.
This is contrary to what is usually seen in (1+1) dimensional soliton interactions
where  $\Delta T > 0$.
}
\label{tbl:sollag}
\end{table}

Lastly, {\it Coalescence} occurs when the two oscillons collide, interact, 
radiate away some of their energy, and survive to form a single 
oscillon.  
\begin{figure}[h]
\vbox{\vspace{-2cm}}
\centerline{\vbox{
                \hbox{\hspace{1cm}}
                \hbox{\epsfxsize =10cm\epsffile{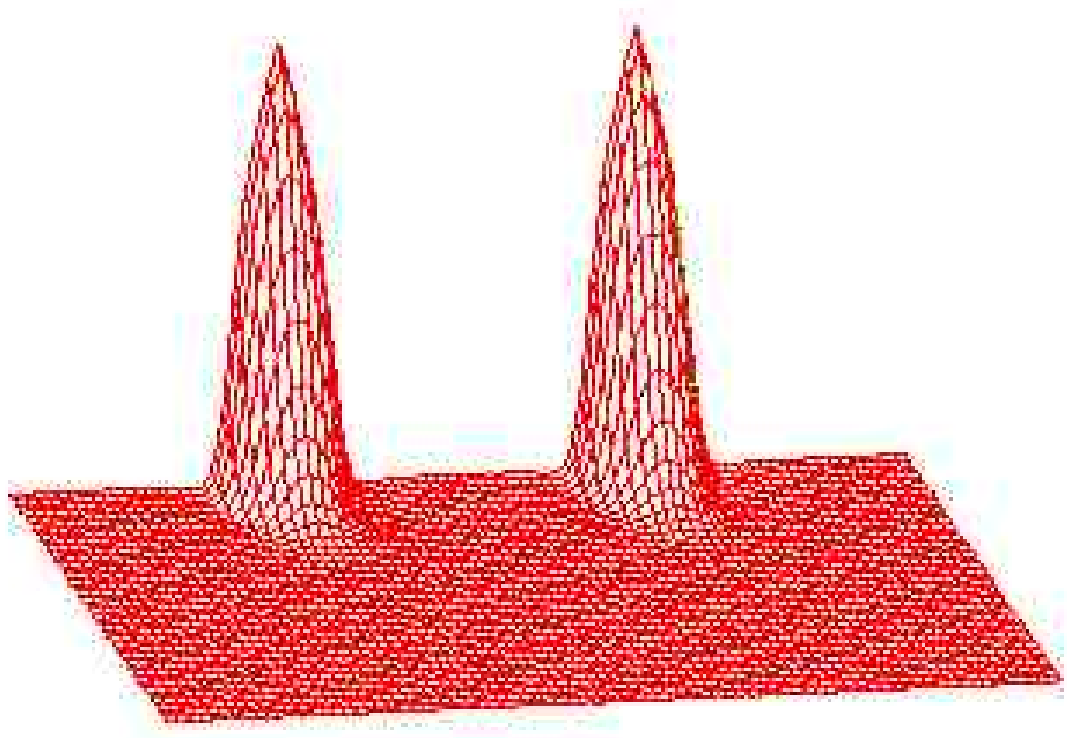}}
                \hbox{\hspace{2.5cm}(a)}
                \hbox{\epsfxsize =10cm\epsffile{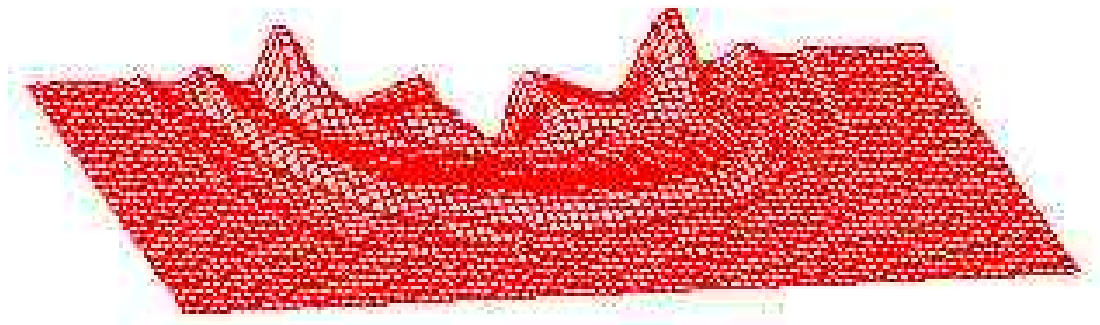}}
                \hbox{\hspace{2.5cm}(c)}
                  }
            \hbox{\hspace{-2cm}}
            \vbox{
                \hbox{\epsfxsize =10cm\epsffile{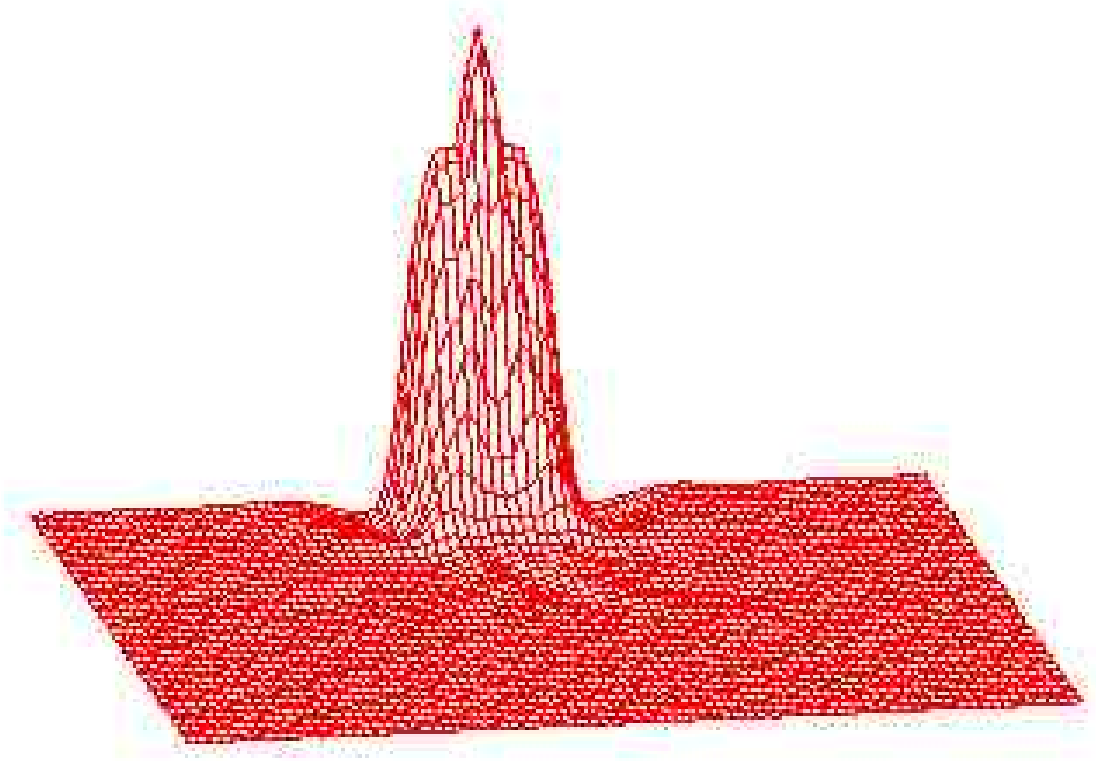}}
                \hbox{\hspace{2.5cm}(b)}
                \hbox{\epsfxsize =10cm\epsffile{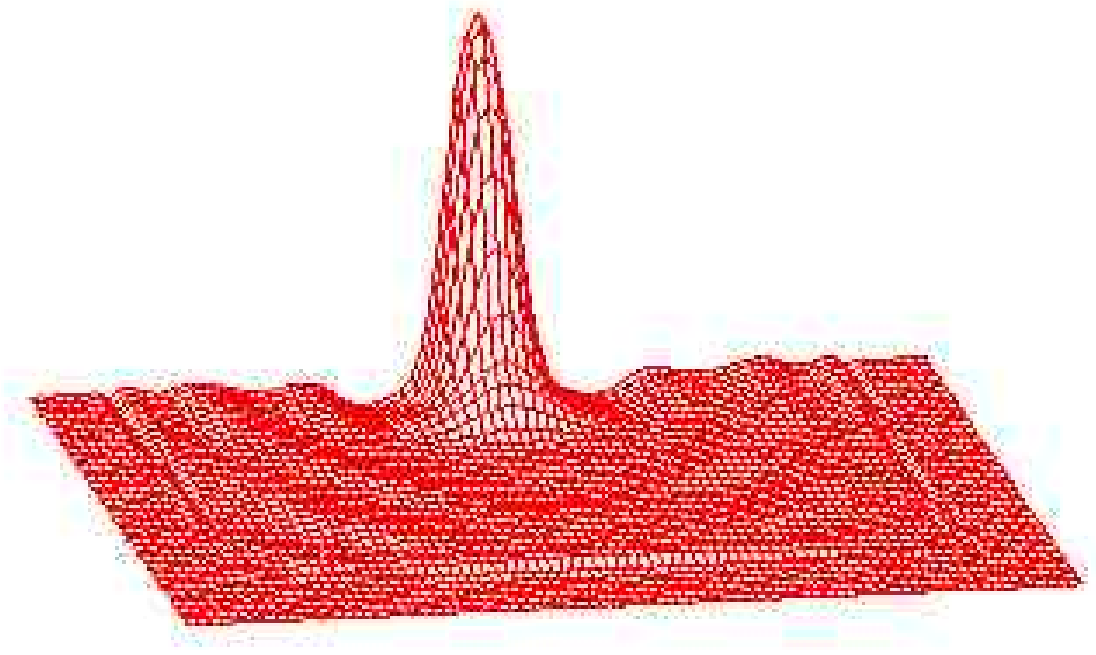}}
                \hbox{\hspace{2.5cm}(d)}
                  }
           }
\caption[Time series showing soliton coalescence]{\label{fig:tseries_coal} \small 
A times series showing the collision of two oscillons that coalesce.
The initial frame (a) shows the two oscillons boosted toward one another; 
the fields interpolate between the vacua so that their peaks are at the 
true vacuum while away from the oscillons is the false vacuum.
Frames (b) and (c) show the nonlinear interactions during the collision,
where roughly half the total energy of the system is converted into
radiation.
Frame (d) shows the endstate consisting of one remaining oscillon,
oscillating on top of the false vacuum.
}
\end{figure}
Coalescence is the default endstate for the code's
detection algorithm since it is what happens if an expanding bubble does not
form or the field does not disperse in the form of annihilation or soliton-like 
transmission.
Unfortunately, simulations with this endstate are the most computationally 
intensive because it is hard to know how long to wait before deciding
that the solution will not disperse or form an expanding bubble 
(particularly on the threshold of expanding bubble formation).
A time series of an oscillon-oscillon collision that results in 
coalescence can be seen in figure \ref{fig:tseries_coal}.
The endstate classification logic is summed up in Table \ref{tab:endclass}.
\begin{table}
\centerline{
\begin{tabular}{|c|c|c|c|}
\hline
	& ${\rm \left(Area\right)_{True\  Vacuum}}$ & $M^2_R$	& $M^2_z$ \\
\hline
Expanding Bubble& large			& -- 		& -- \\ 
Annihilation 	&  small		& large 	& large \\
Soliton-like 	&  small		& small 	& large \\
Coalescence 	&  small		& small 	& small \\
\hline
\end{tabular}
}
\caption[Logic table for endstate classification]{
\small \label{tab:endclass}
Logic table for endstate classification.  ${\rm \left(Area\right)_{True\  Vacuum}}$ is the 
``area'' ($\int dz\ dR$) of space that is in the true vacuum.  $M^2_R$ and $M^2_z$
are the second $R$ and $z$ moments, respectively, of the energy density-like 
function, $\tilde{\rho}$.
The expanding bubbles are easily separated from the other endstates; once the area
of space at the true vacuum reaches a given threshold, the collision is flagged
as forming an expanding bubble.
Annihilation and soliton-like transmission both occur when the field disperses 
($M^2_z$ becomes large)
and are differentiated by the mass moment in the radial direction;
if $M^2_R$ is small, the field is localized indicative of transmission, whereas
if $M^2_R$ is large, the field has clearly dispersed.
}
\label{tab:endclass}
\end{table}

\section{Parameter Space Survey \label{sec:pspace}}

\begin{figure}
\epsfxsize=14cm
\vbox{\vspace{-2cm}}
\centerline{\epsffile{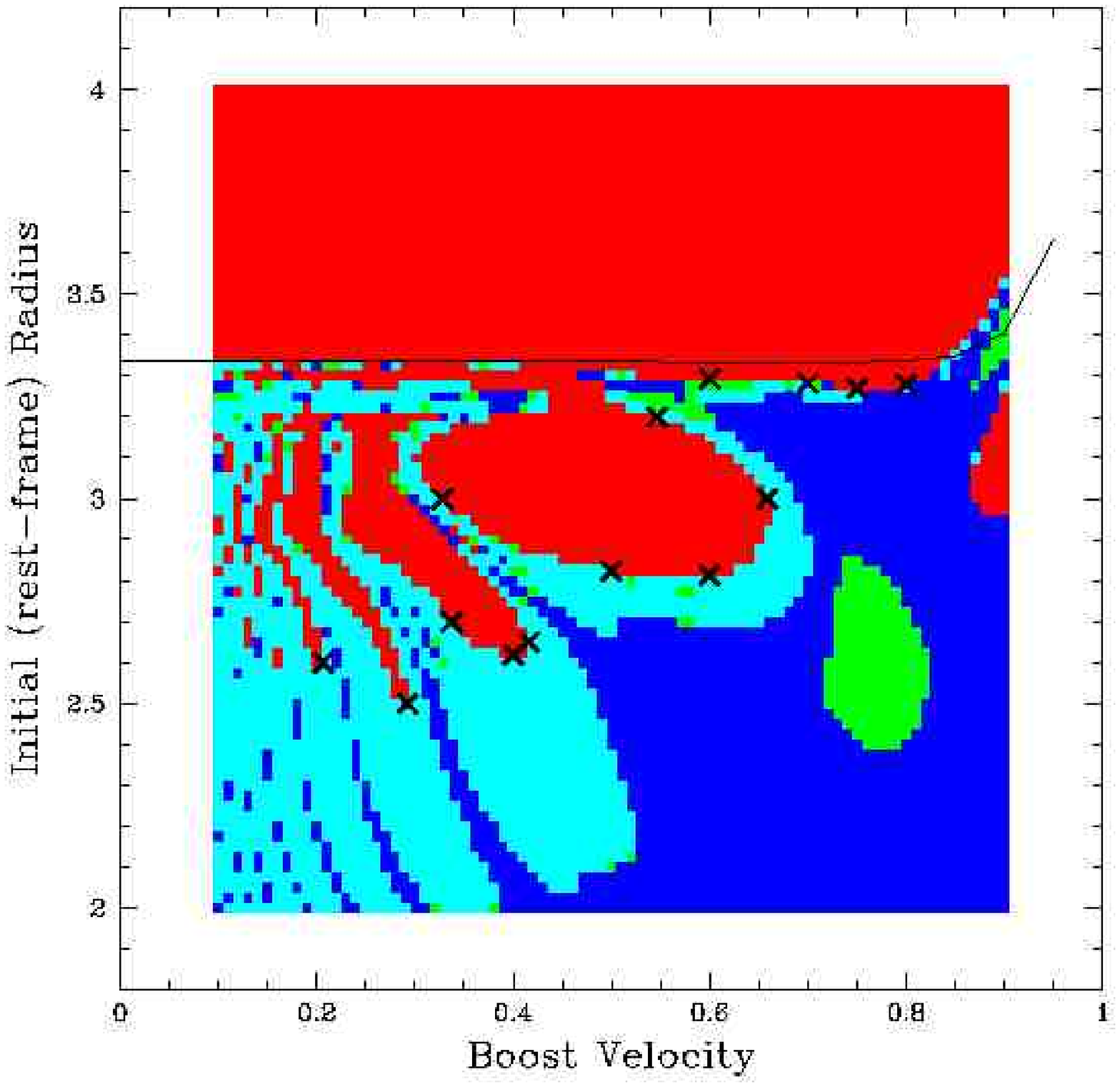}}
\caption[$\sigma_0$-$v_b$ parameter space survey]
{
Two dimensional, $\sigma_0$-$v_b$, 
parameter space survey of axisymmetric bubble collisions (colored 
according to the endstate of the system);
the vertical axis measures the initial rest-frame radius of the bubble,  $\sigma_0$,
while the horizontal axis measures the velocity that each oscillon is boosted at,
$v_b$.
The red region corresponds to expanding vacuum bubble formation, the 
cyan region to coalescence, the blue region to soliton-like transmission, 
and the green to annihilation.
The black curve is the line representing the threshold for expanding bubble
formation for individual boosted bubbles (bubbles above this line would form 
expanding vacuum bubbles even if {\it not} colliding).
The black $\times$'s correspond to points on the threshold of expanding bubble formation
for which critical phenomena was explored (see table \ref{tab:2dcrit}).
Each collision was performed on a 161x321 grid.
}
\label{fig:2Dpspace1}
\end{figure}

In the spherically symmetric ADWP evolutions, a natural parameter to adjust in
the exploration of the model was the initial width of the collapsing bubble,
$\sigma_r$ of equation (\ref{eq:atrestID}).
This one dimensional parameter space is extended here by adding the boost velocity,
$v_b$, as a second parameter.  
Although the initial data is still highly symmetric (the colliding 
oscillons are the same size and oscillate in phase), 
there is still interesting structure to the parameter space. 
In figure \ref{fig:2Dpspace1} we see that for a large area of parameter space, 
there are collisions where bubbles that would otherwise have (eventually) dispersed 
actually combine to form an expanding true vacuum bubble.  

 \noindent

\begin{figure}[h]
\vbox{\vspace{-1.5cm}}
\centerline{\vbox{
                \hbox{\hspace{1cm}}
                \hbox{\epsfxsize =7.0cm\epsffile{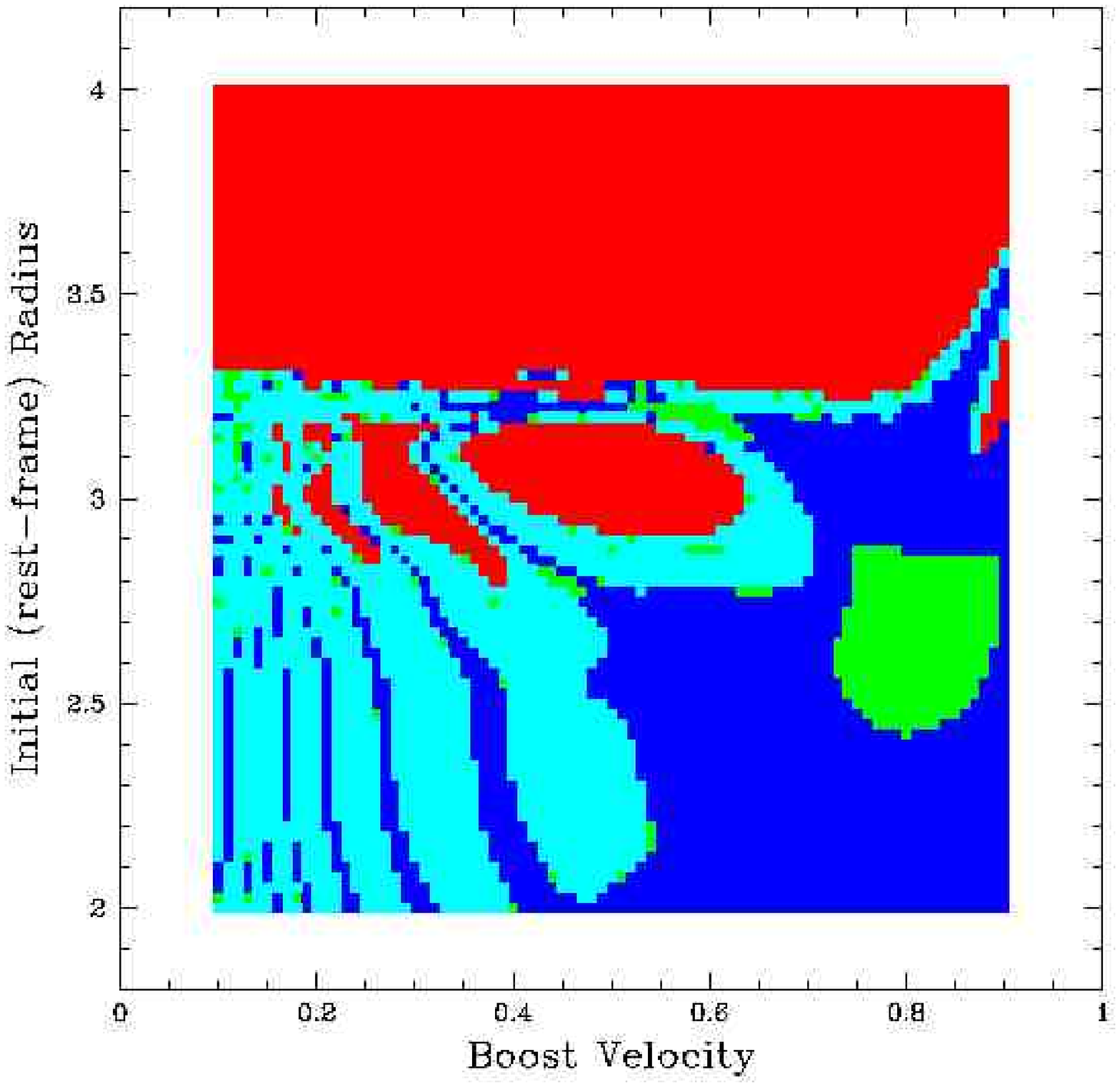}}
                \hbox{\hspace{2.5cm}(a) $81\times 161$, $h=0.25$}
		\hbox{\vspace{0.5cm}}
                \hbox{\epsfxsize =7.0cm\epsffile{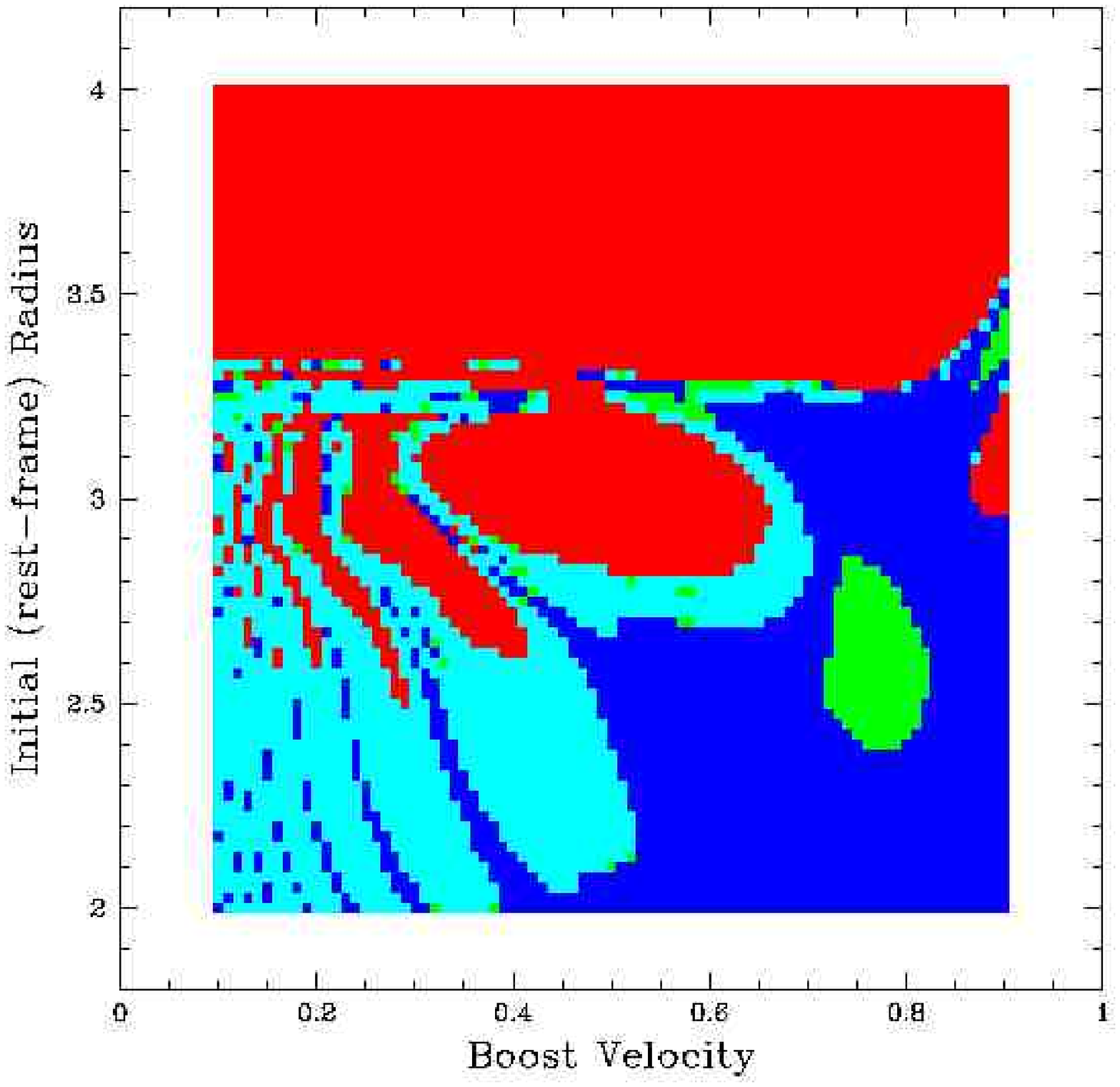}}
                \hbox{\hspace{2.5cm}(c) $161\times 321$, $h=0.125$}
                  }
            \hbox{\hspace{0cm}}
            \vbox{
                \hbox{\epsfxsize =7.0cm\epsffile{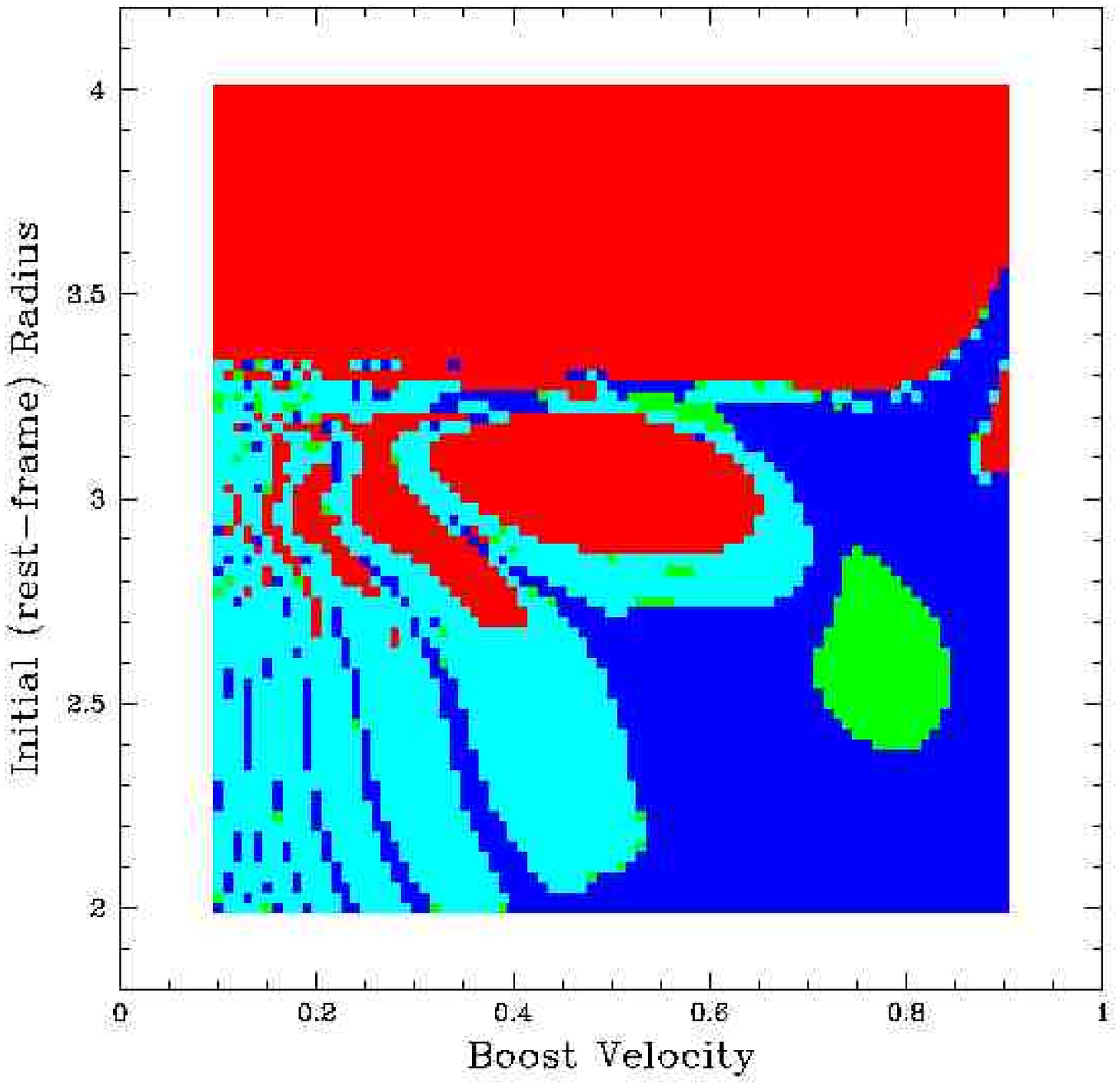}}
                \hbox{\hspace{2.5cm}(b) $114\times 227$, $h\approx 0.177$}
		\hbox{\vspace{0.5cm}}
                \hbox{\epsfxsize =7.0cm\epsffile{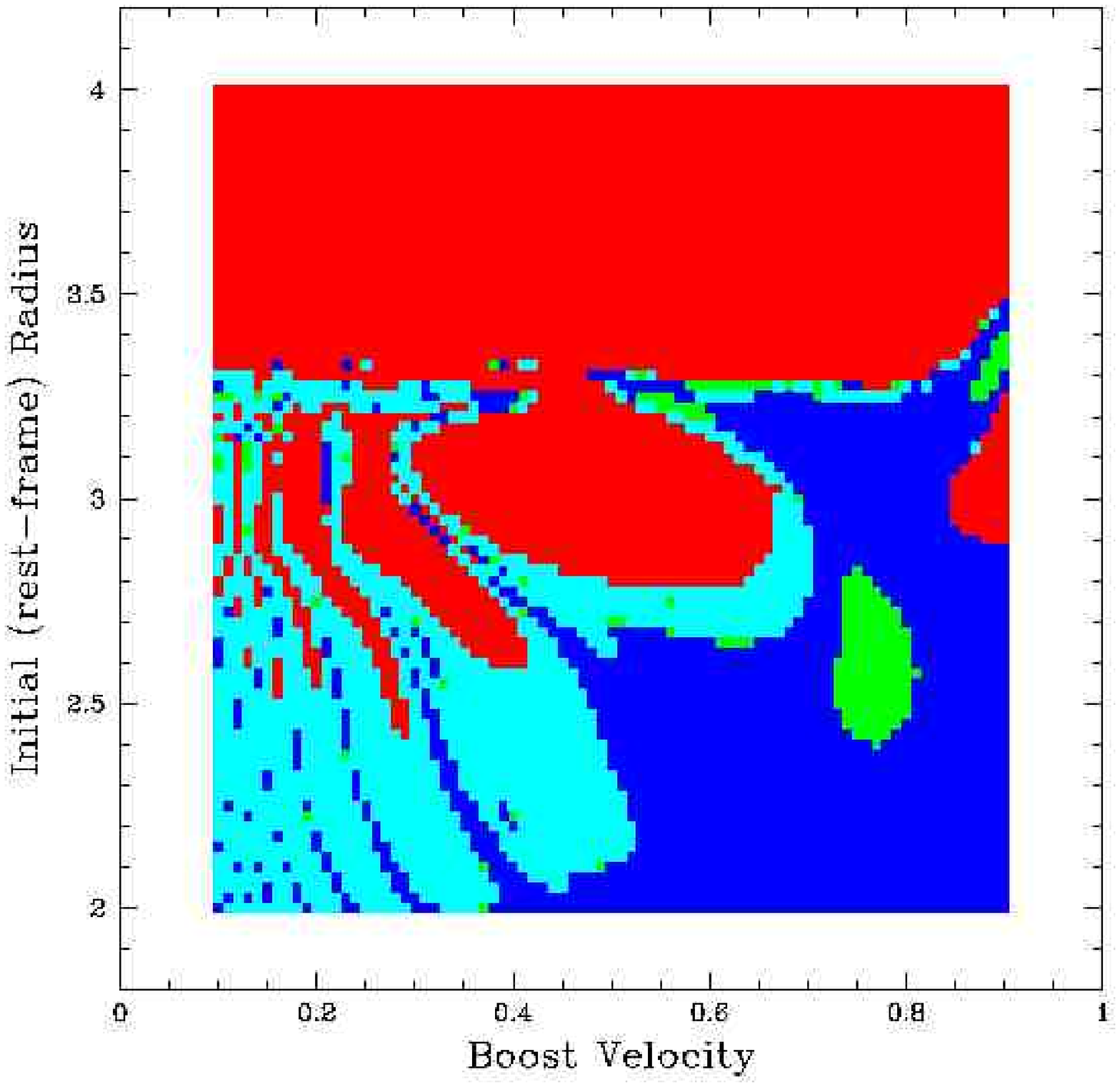}}
                \hbox{\hspace{2.5cm}(d) $227\times 453$, $h\approx 0.088$}
                  }
           }
\caption[Convergence of features on parameter space Surveys]{
\label{fig:pspace_conv} \small 
$\sigma_0$-$v_b$ parameter space surveys at four different
discretizations ($\Delta r = \Delta z = h$).
Although no particular convergence factor was measured, the shapes of the 
colored regions appear to be converging to those of figure (d), at least
well enough to support the {\it existence} of both the large oval-shaped 
red region centered around $v_B\approx 0.5$, $\sigma_0\approx 3$ and the 
side-lobes that are present for smaller boost velocities (to the left
of the main oval).
}
\label{fig:pspace_conv}
\end{figure}

\section{Threshold of Expanding Bubble Formation in 2D Collisions \\
\hbox{({\it 2D Critical Phenomena})}}

At the end of chapter \ref{chap:1D}, a time scaling law was presented 
for solutions on the threshold of expanding bubble formation for 
spherically symmetric bubble collapse.  
With the 2D parameter space surveys presented
in section \ref{sec:pspace} (for collisions), there also is a boundary
separating expanding bubble formation from solutions that leave
the space in the false vacuum.  The natural question is then:
{\it Does a similar scaling law exist, and if so, does it have the 
same critical exponent?}

The threshold of expanding bubble formation was explored first by 
fixing the boost velocity, $v_b$, and then bisecting between two values of 
the initial rest frame radius, $\sigma_r$ (one that forms an expanding vacuum 
bubble and one that does not).  Next, the threshold was explored along a different
direction in parameter space by fixing $\sigma_r$ and bisecting between two 
values of $v_b$.  Time scaling relations were observed for each bisection 
and the critical exponents were observed to be quite similar 
($\gamma \approx 2.2$, see table \ref{tab:2dcrit}).  The locations 
in parameter space of these critical points can be seen by the 
crosses in figure \ref{fig:2Dpspace1}.
The similarity between the critical exponents for the {\it axisymmetric}
collision of two gaussian bubbles and the {\it spherically symmetric} 
collapse of a single gaussian bubble 
suggests that the dominant unstable mode 
observed with the axisymmetric critical solution might be spherically 
symmetric.

\begin{figure}
\epsfxsize=14cm
\centerline{\epsffile{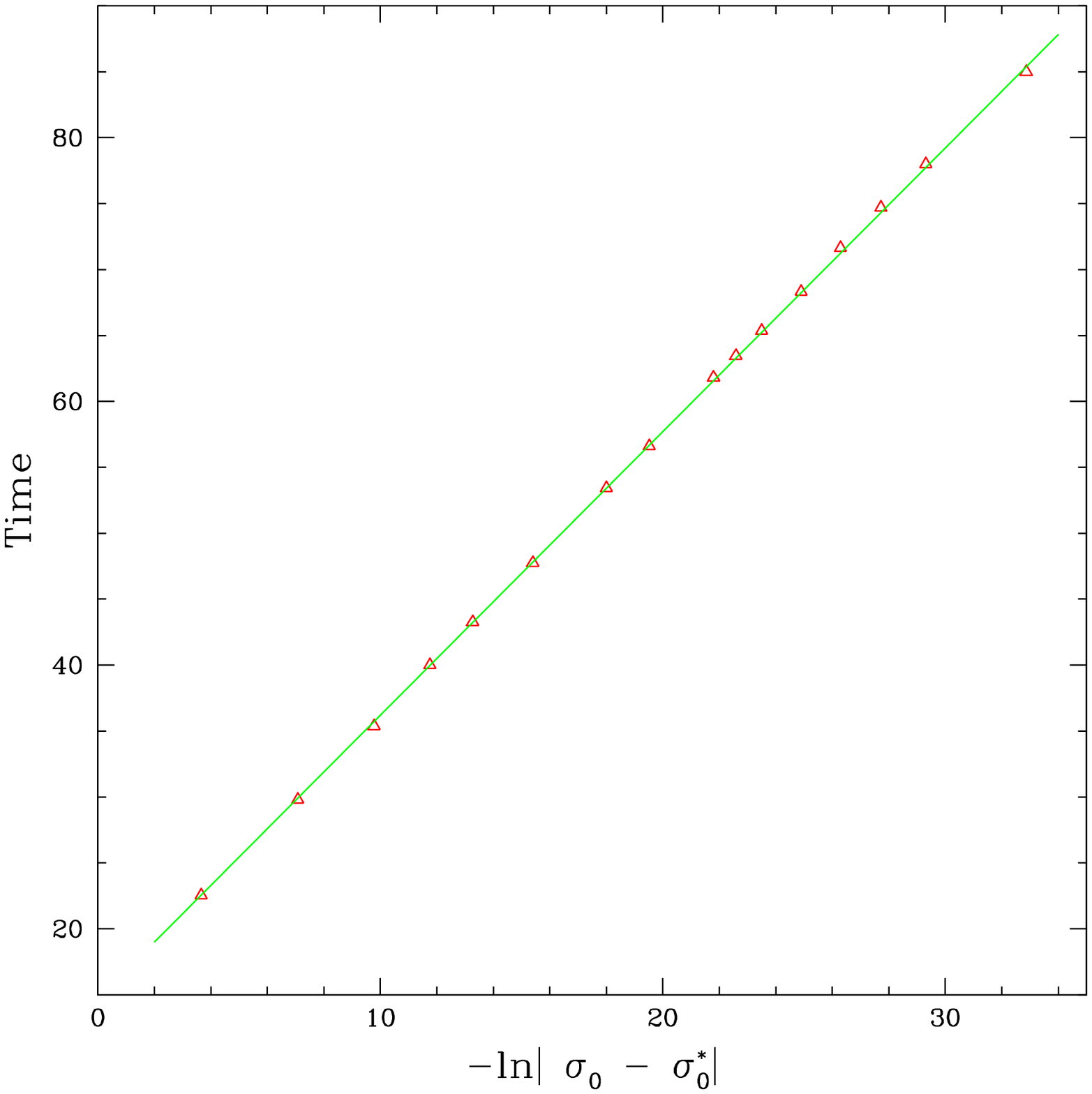}}
\caption[Time scaling on the threshold of expanding bubble formation]
{\small \label{fig:2dtscale}
Sample Time versus $-\ln|\sigma_0-\sigma^*_0|$ plot for the threshold
of expanding bubble formation.
The points plotted are for $\sigma_0 > \sigma^*_0$ and therefore all produce 
expanding bubbles.  
The scaling exponent (slope) was measured to be  $\gamma\approx 2.15$.
All of the threshold parameters explored produced similar scaling
exponents and are tabulated in figure \ref{tab:2dcrit}.
}
\label{fig:2dtscale}
\end{figure} \noindent

\begin{table}
\centerline{
\begin{tabular}{|c|c|c|c|c|}
\hline
$\sigma^*_r$ & $\Delta\sigma$ &  $v^*_b$ & $\Delta v^*_b$ & $\gamma$ \\
\hline
2.620314246044920 	& $\sim 10^{-14}$  & 0.4  & fixed & 2.18 \\
2.824162086503015 	& $\sim 10^{-14}$  & 0.5  & fixed & 2.15 \\
2.813209312722570 	& $\sim 10^{-14}$  & 0.6  & fixed & 2.19 \\
3.293020631075245	& $\sim 10^{-14}$  & 0.6  & fixed & 2.15 \\
3.281889627766725	& $\sim 10^{-14}$  & 0.7  & fixed & 2.15 \\
3.270433308449535 	& $\sim 10^{-14}$  & 0.75 & fixed & 2.15 \\
3.278874383866785	& $\sim 10^{-14}$  & 0.8  & fixed & 2.17 \\
2.5 		& fixed & 0.292882455348505 & $\sim 10^{-14}$ & 2.18 \\
2.6 		& fixed & 0.206552206744452 & $\sim 10^{-14}$ & 2.18 \\
2.65		& fixed & 0.415868755640645 & $\sim 10^{-14}$ & 2.17 \\ 
2.7 		& fixed & 0.337408336301504 & $\sim 10^{-14}$ & 2.17 \\
3.0 		& fixed & 0.658702864946936 & $\sim 10^{-14}$ & 2.20 \\
3.0 		& fixed & 0.327480589060684 & $\sim 10^{-14}$ & 2.17 \\
3.2 		& fixed & 0.546131322091751 & $\sim 10^{-14}$ & 2.19 \\
\hline
\end{tabular}
}\caption[2D collision time scaling critical exponents]
{\small \label{tab:2dcrit} 
Table of critical exponents and threshold parameter values for a two-dimensional
parameter space survey of axisymmetric collisions.
The entire survey can be seen in figure \ref{fig:2Dpspace1}, where the 
black x's denote the threshold values in this table.
The points were chosen to span the space as much as possible and the bisection
search was performed in both directions of the parameter space; nevertheless
the critical exponents are remarkably similar, suggesting a universal scaling law
(see figure \ref{fig:2dtscale}).
}
\end{table}


\chapter{Trial Function Approach to Critical Solutions \label{chap:trial}}  

This chapter discusses both exact and 
approximate {\it critical} (non-radiative) oscillon solutions 
obtained through a trial function approach.
Differential equations (\ref{modestruct0}) and (\ref{modestructn}) 
are rederived using a very general trial function ansatz.
By assuming more constrained gaussian and hyperbolic secant trial 
functions, approximate critical oscillon solutions are
also obtained.

\section{Trial Functions \& Variational Approach \label{sec:trial1}}

In section \ref{subsec:modestruct}, the ordinary differential equations
(\ref{modestruct0}) and (\ref{modestructn}) were derived by inserting
the ansatz 
\be
\phi(r,t) = -1 + \phi_0(r) + \sum_{n=1}^{\infty}\phi_n(r) \cos\left(n\omega t\right)
\label{eq:ansatz2}
\ee
into the equation of motion (\ref{eq:boxphi}) and matching $\cos(n \omega t)$ 
terms.  
The same set of differential equations can be obtained by inserting
(\ref{eq:ansatz2}) into the action
\begin{equation}
\begin{array}{rcl}
S &=& \displaystyle{
\int d^4x \ {\mathcal L} 
}\\
  &=& 
\displaystyle{
4\pi\int r^2\left( 
\frac{1}{2}\left(\dot{\phi}\right)^2
-\frac{1}{2}\left( \phi' \right)^2
-\frac{1}{4}\left(\phi^2 -1 \right)^2
\right)
dr \ dt \
}
\end{array}
\label{eq:simpleaction}
\end{equation}
and integrating over time to obtain a dimensionally reduced 
action\footnote{Of course, the second line in (\ref{eq:simpleaction}) is also 
dimensionally reduced by integrating over $d\Omega$, which is trivial 
in spherical symmetry.  The more interesting case arises when imposing the 
(discrete) periodic boundary condition in time.}. The reduced action is
then varied with respect to the individual modes, $\phi_0$ and $\phi_n$, to obtain the 
(ordinary differential) equations of motion.

Ansatz (\ref{eq:ansatz},\ref{eq:ansatz2}) is actually obtained by starting 
with the most general possible ansatz 
\be
\phi(r,t) = -1 + \phi_0(r) + \sum_{n=1}^{\infty}A_n(r) 
\sin\left(n\omega t + \theta_n\right)
\label{eq:ansatz_gen}
\ee
and imposing boundary conditions that match the phases of the each mode
and that demand that the maximum amplitude of each mode is equal to $A_n(0)$.
The reduced action obtained by integrating (\ref{eq:simpleaction}) and (\ref{eq:ansatz2}) 
over time is
\begin{equation}
\begin{array}{rclcl}
S&=& \displaystyle{ 
4\pi \int^\infty_{r=0} dr \int^T_{t=0} \ {\mathcal L}{ (r,t)} }\hbox{\hspace{-1.5cm}}&&
\\
&=& \vbox{\vspace{1cm}}\displaystyle{ 
\frac{4\pi^2}{\omega} \int^\infty_{r=0} dr 
} & \hspace{-2cm} \left\{\vbox{\vspace{0.75cm}} \right. 
\hbox{\hspace{-2cm}}
&
\displaystyle{ 
\frac{1}{2}\omega^2 \sum^\infty_{n=1} n^2 \phi_n^2 - \left(\phi_0'\right)^2 -
\frac{1}{2} \sum^\infty_{n=1}\left(\phi_n'\right)^2 
}\\
&& && \displaystyle{ 
-\frac{1}{2}\phi_0^2\left(\phi_0 -2\right)^2 
- \frac{1}{4}\left( 6 \phi_0^2 -12\phi_0 +4\right)\sum^\infty_{n=1}\left(\phi_n\right)^2
} \\
&&&& \displaystyle{
-\frac{1}{2}\left(\phi_0-1\right)\sum^\infty_{n,p,q=1}\phi_n\phi_p\phi_q
 		\left(\delta_{n,\pm p \pm q}\right)
} \\
&&&& \displaystyle{
\left.
-\frac{1}{16}\sum^\infty_{n,m,p,q=1}\phi_n\phi_m\phi_p\phi_q
 		\left(\delta_{n,\pm m \pm p \pm q}\right)
\right\} } \\
&=& \displaystyle{ 
\frac{4\pi^2}{\omega}\int^\infty_{r=0} dr \ {\mathcal L}(r) }\hbox{\hspace{-1.5cm}}&&\\
&&&&
\end{array}
\end{equation}
where again\footnote{
We also made judicious use of 
$\displaystyle{\int_{t=0}^{2\pi/\omega} \hbox{\hspace{-0.5cm}}\cos(m \omega t) 
\cos(n \omega t) = \left(\frac{\pi}{\omega}\right) \delta_{m,n}}$, for
$m,n \geq 1$.  This is why the ansatz was split up into the $\phi_0$ and
$\phi_n$ modes, since the above integral is $2\pi/\omega$ 
for $m=n=0$, not $\pi/\omega\delta_{m,n}$ as for $m,n \geq 1$. 
} the pluses and minuses sum over every permutation (ie. 
$\delta_{n,\pm p \pm q}=\left(
 \delta_{n, + p + q}
+\delta_{n, + p - q}
+\delta_{n, - p + q}
+\delta_{n, - p - q} 
\right)$).
Unfortunately, ansatz (\ref{eq:ansatz2}) is so general that the spatial integration 
cannot be done in closed-form.
However, one can still vary the (partially) reduced action with 
respect to $\phi_0$ and $\phi_n$ 
to obtain the usual Euler-Lagrange equations,
\begin{eqnarray}
\displaystyle{
\partial_\mu \left( \frac{ \partial {\mathcal L}(r) }{\partial \left(\partial_\mu \phi_0\right)}\right)
- \frac{\partial {\mathcal L}(r)}{\partial \phi_0}}
&=&0 \\
\displaystyle{
\partial_\mu \left( \frac{ \partial {\mathcal L}(r) }{\partial \left(\partial_\mu \phi_n\right)}\right)
- \frac{\partial {\mathcal L}(r)}{\partial \phi_n}}
&=&0,
\end{eqnarray}
to obtain a set of {\it differential} equations which (again) are
\begin{equation}
\begin{array}{rcl}
\frac{1}{r^2} \left( r^2 \phi_0' \right)'
&=& \phi_0\left( \phi_0 - 1\right)\left(\phi_0-2\right) \\
&&+ \frac{3}{2} \left( \phi_0 -1\right) \sum\limits_n \left( \phi_n\right)^2\\
&&+ \frac{1}{4}\sum\limits_{n,p,q}\phi_n\phi_p\phi_q\left(
\delta_{n, \pm p \pm q} \right) ,
\end{array}
\label{modestruct0_2}
\end{equation}
\begin{equation}
\begin{array}{rcl}
\frac{1}{r^2}\left( r^2 \phi_n' \right)' 
&=& \left( 3 \left(\phi_0-1\right)^2 -
\left( n^2 \omega^2 +1\right)\right)\phi_n\\
&&+\frac{3}{2} \left(\phi_0-1\right) \sum\limits_{p,q}\phi_p\phi_q 
\left(\delta_{n,\pm p \pm q}\right) \\
&&+\frac{1}{4}\sum\limits_{m,p,q}\phi_m\phi_n\phi_q\left(
\delta_{n,\pm m \pm p \pm q}\right),
\end{array}
\label{modestructn_2}
\end{equation}
respectively.  
The solution to equations (\ref{modestruct0_2}) and (\ref{modestructn_2}) for $\omega=1.38$ can be seen in 
Figure \ref{fig:modstruct}.  
The solution matches extremely well to the Fourier decomposed dynamic data.  Again, this
is not surprising because the only assumption made restricting the ansatz 
that could prevent solutions to 
(\ref{modestruct0_2}) and (\ref{modestructn_2}) from being solutions to the true PDE equations 
of motion was that the solution is periodic, a property which is observed empirically
for the resonant oscillons.

\section{Approximate Oscillons From Constrained Ansatz}

Although the resonant non-radiative solutions found in section
\ref{sec:trial1} match the dynamic data extremely well, they are
still obtained from solving ordinary differential equations.
With a well chosen and more constrained trial function, both the time
and space integrations in the action can be evaluated in closed-form.
Variational methods are then applied to obtain a set of {\it algebraic} 
equations that can be solved to obtain the best possible values for the 
free parameters of the trial function.
The ansatz is taken to be
\begin{equation}
\phi(r,t) = -1 + A\ f\left(\frac{r}{\sigma}\right)\sin\left(\omega t + \theta\right),
\label{eq:generaltrialf}
\end{equation}
where $\sigma$ is related to the width of the field configuration and
$A$ is the amplitude of the oscillations.
Equation (\ref{eq:generaltrialf}) is subjected to the following boundary conditions:
\begin{eqnarray}
\phi(r,0) &=& -1 \\
\phi(r,\pi/(2\omega)) &=& -1 + A  f\left(\frac{r}{\sigma}\right),
\label{eq:fBCs}
\end{eqnarray}
which results in a choice of phase, $\theta=0$, and sets $A$ equal to the 
{\it maximum} amplitude of the oscillating mode.
This yields 
\begin{equation}
\phi(r,t) = -1 + A\ f\left(\frac{r}{\sigma}\right)\sin\left(\omega t\right)
\label{eq:specifictrialf}
\end{equation}
for the general form of the trial function.  Equation (\ref{eq:specifictrialf})
is inserted into the action and can be written in terms of 
definite integrals.
\begin{equation}
\begin{array}{rcl}
S &=& 
\displaystyle{
  4\pi \int^\infty_0 dr \int^{\pi/(2\omega)}_0 dt \ r^2
\left\{
\frac{1}{2}\dot{\phi}^2
-\frac{1}{2}\phi'^2 
-\frac{1}{2}\left( \phi^2 - 1\right)^2 
\right\} 
} \\
&=&
\displaystyle{
\vbox{\vspace{0.75cm}}
\pi\left\{
2A^2\sigma^3\omega c_2 g_1 -\omega^{-1}
\left(
2A^2\sigma   c_1 g_2 +
4A^2\sigma^3 c_2 g_2 -
4A^3\sigma^3 c_3 g_3 +
 A^4\sigma^3 c_4 g_4 
\right)
\right\} 
}
\label{eq:redaction}
\end{array}
\end{equation}
where the $c_i$ are dimensionless constants 
\begin{equation}
\begin{array}{rclcrcl}
c_1 &\equiv& \displaystyle{\int^\infty_0 du\ u^2 
\left(\partial_u f(u)\right)^2}  & \hbox{\hspace{1cm}}&
c_2 &\equiv& \displaystyle{\int^\infty_0 du\ u^2 f(u)^2} \\
\vbox{\vspace{0.75cm}}
c_3 &\equiv& \displaystyle{\int^\infty_0 du\ u^2 f(u)^3}   & &  
c_4 &\equiv& \displaystyle{\int^\infty_0 du\ u^2 f(u)^4}
\label{eq:spaceintegrals}
\end{array}
\end{equation}
obtained by taking $u\equiv r/\sigma$ 
and the $g_i$ are dimensionless constants 
\begin{equation}
\begin{array}{rcccl}
g_1 &\equiv& \displaystyle{\int^{\pi/2}_0 d\tau\  
\cos^2(\tau)} &=&
\displaystyle{\frac{\pi}{4}}\\
\vbox{\vspace{0.75cm}}
g_2 &\equiv& \displaystyle{\int^{\pi/2}_0 d\tau\  \sin^2(\tau)} &=&
\displaystyle{\frac{\pi}{4}}\\
\vbox{\vspace{0.75cm}}
g_3 &\equiv& \displaystyle{\int^{\pi/2}_0 d\tau\  \sin^3(\tau)} &=&
\displaystyle{\frac{2\pi}{3}}\\
\vbox{\vspace{0.75cm}}
g_4 &\equiv& \displaystyle{\int^{\pi/2}_0 d\tau\  \sin^4(\tau)} &=&
\displaystyle{\frac{3\pi}{16}}
\end{array}
\end{equation}
obtained by taking $\tau\equiv \omega t$.
The reduced action (\ref{eq:redaction}) will be the same
even if $\sin(\omega t)$ in equation (\ref{eq:specifictrialf})
is replaced with an arbitrary function, $h(\omega t)$.
However, the constants, $g_1$, $g_2$, $g_3$, and $g_4$, 
will have to be recomputed according to
\begin{equation}
\begin{array}{rclcrcl}
g_1 &\equiv& \displaystyle{\int^{\tau_1}_{\tau_0} d\tau\ 
\left(\partial_\tau h(\tau)\right)^2}  & \hbox{\hspace{1cm}}&
g_2 &\equiv& \displaystyle{\int^{\tau_1}_{\tau_0} d\tau\ h(\tau)^2} \\
\vbox{\vspace{0.75cm}}
g_3 &\equiv& \displaystyle{\int^{\tau_1}_{\tau_0} d\tau\ h(\tau)^3}   & &  
g_4 &\equiv& \displaystyle{\int^{\tau_1}_{\tau_0} d\tau\ h(\tau)^4}
\label{eq:spaceintegrals}
\end{array}
\end{equation}
and the limits of integration need to be chosen appropriately.

Now that the action is completely reduced, 
varying with respect to the free parameters yields a system
of {\it algebraic} equations,
\begin{eqnarray}
\displaystyle{\frac{\partial S}{\partial \sigma}}&=&0 \label{eq:dSdsig}\\
\displaystyle{\frac{\partial S}{\partial A}}&=&0\label{eq:dSdA},
\end{eqnarray}
that can be solved for $A$ and $\sigma^2$,
\begin{eqnarray}
A_\pm &=& \displaystyle{
\frac{
6c_3g_3\pm \sqrt{ 36c_3^2g_3^2 -16c_2c_4g_4\left(2g_2-g_1\omega^2\right)}
}{2c_4g_4}
} \label{eq:trialA}\\
\vbox{\vspace{0.75cm}}
\sigma^2_\pm &=& \displaystyle{
\frac{-2c_1g_2}{
3\left(c_4g_4A_\pm^2 -4c_3g_3A_\pm +2c_2\left(2g_2-g_1\omega^2\right)\right)}
}.\label{eq:trialsig2}
\end{eqnarray}

The reader may be wondering why, since the action is also a function of $\omega$,
that the equation $\displaystyle{\frac{\partial S}{\partial \omega}=0}$ was not
added to the system of equations, (\ref{eq:dSdsig}) and (\ref{eq:dSdA}).
Unfortunately, the trial function method does not work well to determine 
$\omega$.  
Looking at typical plots of the action\footnote{$f(r/\sigma)$ is taken
to be a gaussian for these figures, but the property holds for 
the $\sech$ and $\sech^2$ functions also discussed below.} 
as a function of the amplitude and width 
(figures \ref{fig:AOmegaS} and \ref{fig:SigOmegaS}, respectively),
it is apparent that there are no (non-trivial) points where  
$\displaystyle{\frac{\partial S}{\partial \omega}=0}$.
Forcing this equation to be solved in conjunction with 
(\ref{eq:dSdsig}) and (\ref{eq:dSdA}) results in  unphysical complex solutions.
In the following subsections, we choose three forms for $f(r/\sigma)$ and 
look at the amplitude and widths as a function of $\omega$.

\begin{figure}
\epsfxsize=12cm
\centerline{\epsffile{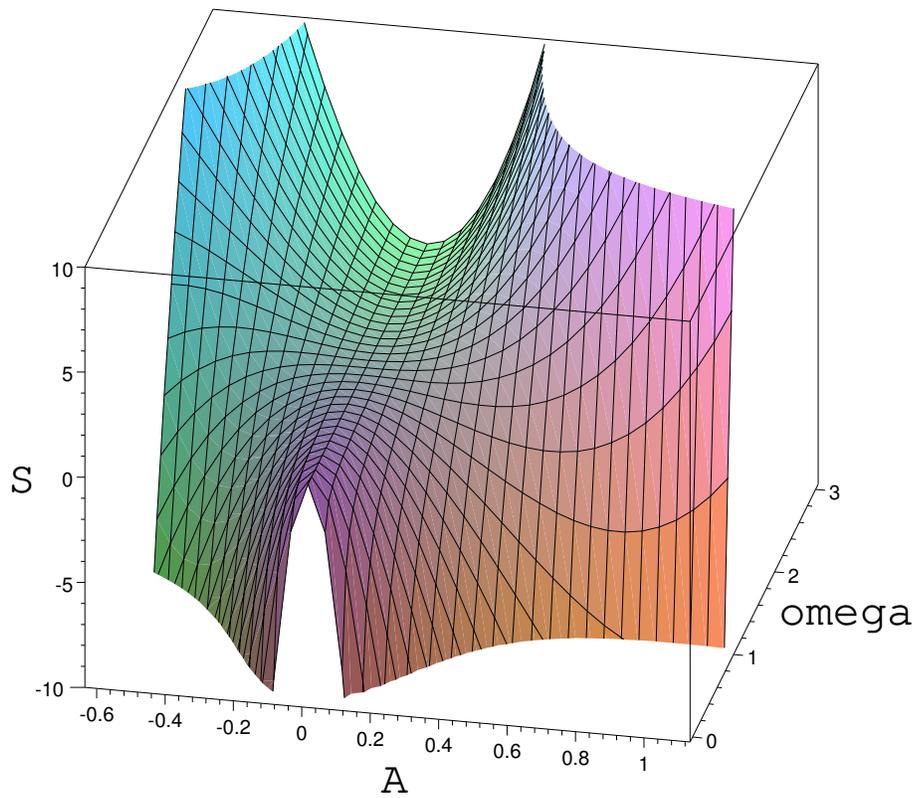}}
\caption[Reduced action as function of $A$ and $\omega$]
{\small \label{fig:AOmegaS}
Plot of the reduced action as a function of the oscillon amplitude,
$A$, and the frequency, $\omega$ (while setting $\sigma=3.6$).
Looking along $\omega$-constant slices, it is clear that there
are many points that satisfy 
$\displaystyle{\frac{\partial S}{\partial A}=0}$, whereas
when looking at $A$-constant slices, there are no points that 
satisfy $\displaystyle{\frac{\partial S}{\partial \omega}=0}$
(except for the trivial $A=0$ solution).
}
\label{fig:AOmegaS}
\end{figure} \noindent

\begin{figure}
\epsfxsize=12cm
\centerline{\epsffile{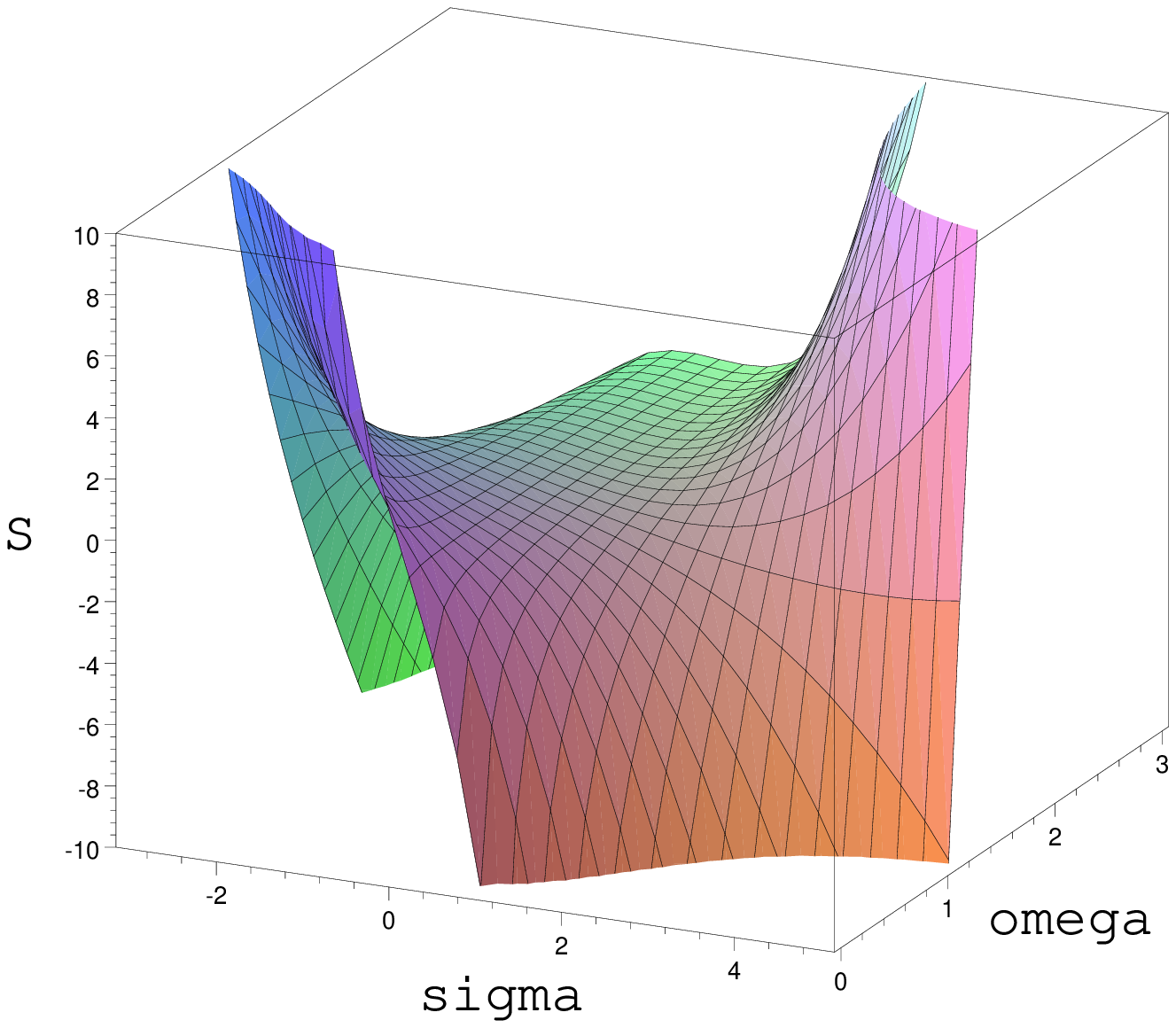}}
\caption[Reduced action as function of $\sigma$ and $\omega$]
{\small \label{fig:SigOmegaS}
Plot of the reduced action as a function of the oscillon width,
$\sigma$, and the frequency, $\omega$ (while setting $A=0.62$).
Looking along $\omega$-constant slices, it is clear that there
are many points that satisfy 
$\displaystyle{\frac{\partial S}{\partial \sigma}=0}$, whereas
when looking at $\sigma$-constant slices, there are no points that 
satisfy $\displaystyle{\frac{\partial S}{\partial \omega}=0}$
(except for the trivial $\sigma=0$ solution).
}
\label{fig:SigOmegaS}
\end{figure} \noindent

\subsection{Gaussian Trial Functions\label{sec:gautri}}

In this subsection the field configuration is assumed to have a gaussian
shape (as assumed in \cite{copeland:1995}),
\begin{equation}
\displaystyle{
f\left(\frac{r}{\sigma}\right) = \exp\left(-\frac{r^2}{\sigma^2}\right)
}.
\label{eq:gau}
\end{equation}
Now that $f(r/\sigma)$ has been chosen, the integrals (\ref{eq:spaceintegrals}) 
can now be performed:
\begin{equation}
\begin{array}{rclcrcl}
\vbox{\vspace{0.75cm}}
c_1 &=& \displaystyle{\frac{3}{8}\sqrt{\frac{\pi}{2}}} & \hbox{\hspace{1cm}} &
c_2 &=& \displaystyle{\frac{1}{8}\sqrt{\frac{\pi}{2}}}
\\
\vbox{\vspace{0.75cm}}
c_3 &=& \displaystyle{\frac{1}{12}\sqrt{\frac{\pi}{3}}}  & &
c_4 &=& \displaystyle{\frac{1}{32}\sqrt{\pi}}.
\label{eq:GAUintegrals}
\end{array}
\end{equation}
Using equations (\ref{eq:trialA}) and (\ref{eq:trialsig2}), 
$A_\pm$ and $\sigma^2_\pm$ can be plotted as a function of $\omega$
(figure \ref{fig:gauAsig}).
Unfortunately, for $\omega<0.437$ the discriminant in equation 
(\ref{eq:trialA}) is negative and gives complex solutions.
Furthermore, where $A_-$ is real $\sigma^2_-$ is negative, 
resulting in an imaginary oscillon width!
However, there is a part of the ``+'' root ($0.62<\omega<1.41$) 
that has positive $\sigma^2_+$ and can therefore support physical 
solutions.  Table \ref{tbl:gauparams} compares sample solutions
(located near the x's in figure \ref{fig:gauAsig}) 
to estimates obtained from the dynamic solution.
\begin{table}
\centerline{
\begin{tabular}{|c||c||c|c|}
\hline
 	     & Dynamic  &         &       \\ 
Parameter    & Solution & $A_+=A_D$& $\sigma_+=\sigma_D$  \\
\hline 
$A$          & 0.62  & 0.62   & 0.39   \\
$\sigma$     & 3.6   & 2.9    & 3.6    \\
$\omega$     & 1.4   & 1.26   & 1.32   \\
\hline
\end{tabular}
}
\caption[Estimated and determined values of $A$, $\sigma$, and $\omega$ (gaussian)]{
\small\label{tbl:gauparams}
Table of free parameter values ($A$, $\sigma$, and $\omega$) 
determined using a gaussian trial function (third and fourth columns)
compared to the  
values estimated from the spherically symmetric 
dynamics (second column); the ${\rm D}$ subscript indicates
the value was obtained from the dynamic solution.
The third column contains results obtained using the 
trial function approach  
where $A_+$ equals the amplitude of the $n=1$ mode 
dynamic solution; 
the fourth column contains results where 
$\sigma_+$ equals the width of the $n=1$ mode 
dynamic solution. 
The data from the third and fourth columns occur around the x's in
figure \ref{fig:gauAsig}.
}
\label{tbl:gauparams}
\end{table}

\begin{figure}
\epsfxsize=12cm
\centerline{\epsffile{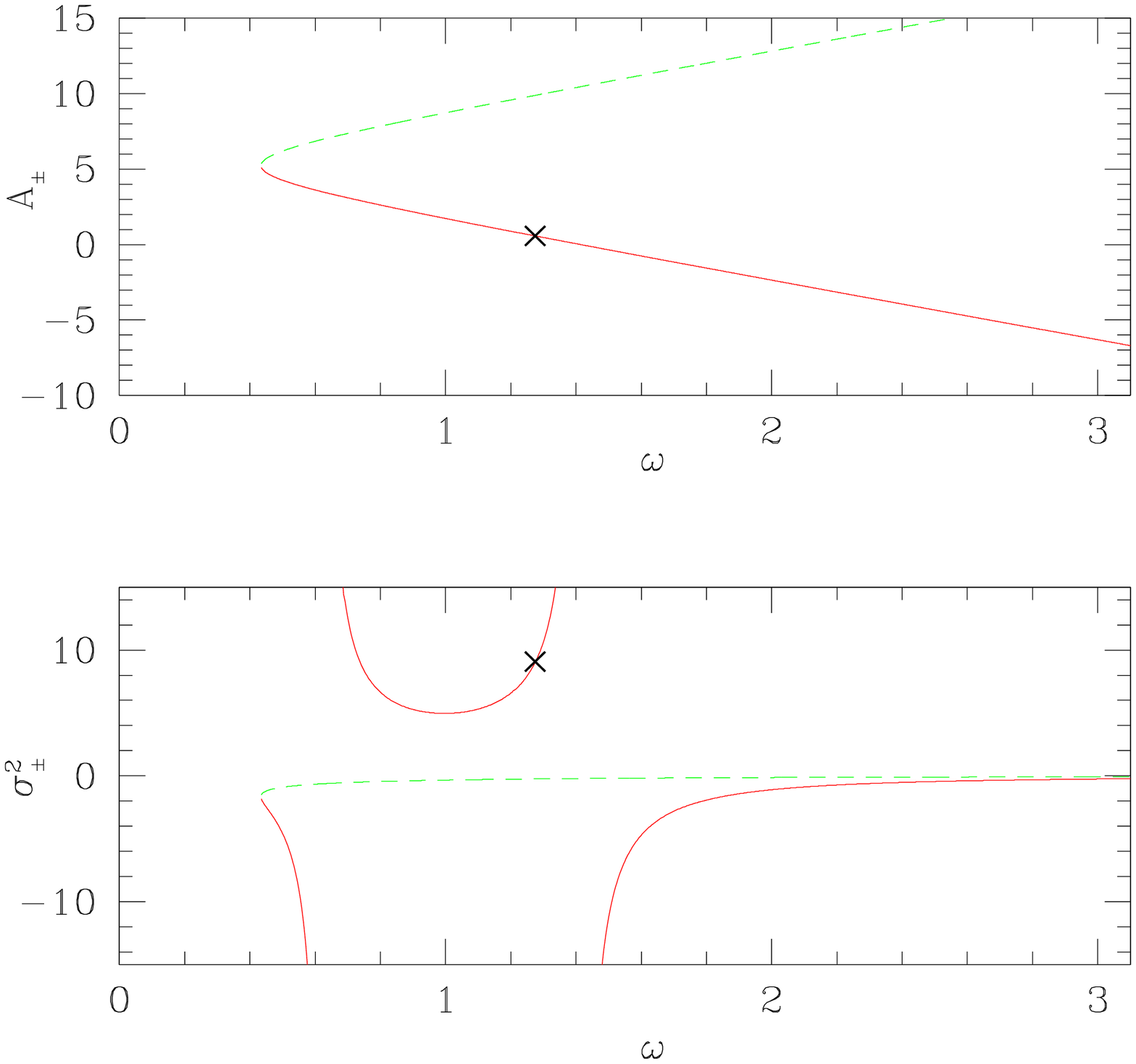}}
\caption[$A_\pm$ and $\sigma^2_\pm$ as a function of $\omega$ for gaussian $f(r/\sigma)$ ]
{\small \label{fig:gauAsig}
Plot of $A_\pm$ (top) and $\sigma^2_\pm$ (bottom) as a function of $\omega$ 
for gaussian $f(r/\sigma)$.
The ``+'' roots are plotted in dashed green, while the ``--'' roots 
are plotted in solid red.
The solution is not plotted for $\omega < 0.437$ as the discriminant in 
equation (\ref{eq:trialA}) is negative and only real $A$ are of interest.
The ``+'' roots are all unphysical since  $\sigma^2_+$ is less than zero everywhere,
which means that solutions within this region have an imaginary width.
The solution for $\sigma^2_-$ is also unphysical {\it outside} the region 
$0.62 < \omega <1.41$ for the same reason.
Fortunately, there are solutions that are both physical (real roots) 
and similar to those seen in the dynamic simulations (near the x's).
}
\label{fig:gauAsig}
\end{figure} \noindent


\subsection{Hyperbolic Secant Trial Functions\label{sec:sechtri}}

In this subsection the field configuration is assumed to have a hyperbolic
secant shape,
\begin{equation}
\displaystyle{
f\left(\frac{r}{\sigma}\right) = {\rm sech}\left(\frac{r}{\sigma}\right)
}.
\label{eq:sech}
\end{equation}
With the new choice of $f(r/\sigma)$, the integrals (\ref{eq:spaceintegrals}) 
are now reevaluated:
\begin{equation}
\begin{array}{rclcrcl}
\vbox{\vspace{0.75cm}}
c_1 &=& \displaystyle{\frac{1}{36}\pi^2 +\frac{1}{3} } & \hbox{\hspace{1cm}} &
c_2 &=& \displaystyle{\frac{1}{12}\pi^2 }
\\
\vbox{\vspace{0.75cm}}
c_3 &=& 0.3670959657 & &
c_4 &=& \displaystyle{\frac{1}{18}\pi^2 - \frac{1}{3}}.
\label{eq:SECHintegrals}
\end{array}
\end{equation}
Using equations (\ref{eq:trialA}) and (\ref{eq:trialsig2}), the new
$A_\pm$ and $\sigma^2_\pm$ are plotted as a function of $\omega$
(figure \ref{fig:sechAsig}).
Complex solutions exist for $\omega<0.595$ where (as with the 
gaussian trial function) the discriminant in (\ref{eq:trialA})
is negative.
The ${\rm sech}$ function also yields unphysical solutions for all of the
``--'' roots since $\sigma^2_-$ is negative everywhere.
Again, there is a part of the ``+'' solution ($0.73<\omega<1.41$) 
that has positive $\sigma^2_+$ and can therefore support physical 
solutions.  Table \ref{tbl:sechparams} compares sample solutions
(located near the x's in figure \ref{fig:sechAsig}) to estimates obtained 
from the dynamic solution.
\begin{table}
\centerline{
\begin{tabular}{|c||c||c|c|}
\hline
             & Dynamic  &         &       \\ 
Parameter    & Solution & $A_+=A_D$& $\sigma_+=\sigma_D$  \\
\hline 
$A$             & 0.62  & 0.62 	  & 0.72   \\
$\sigma$        & 2.2   & 2.5     & 2.2    \\
$\omega$        & 1.4   & 1.29    & 1.27   \\
\hline
\end{tabular}
}
\caption[Estimated and determined values of $A$, $\sigma$, and $\omega$ (sech)]{
\small\label{tbl:sechparams}
Table of free parameter values ($A$, $\sigma$, and $\omega$) 
determined using a ${\rm sech}$ trial function (third and fourth columns)
compared to the  
values estimated from the spherically symmetric 
dynamics (second column); the ${\rm D}$ subscript indicates
the value was obtained from the dynamic solution.
The third column contains results obtained using the 
trial function approach  
where $A_+$ equals the amplitude of the $n=1$ mode 
dynamic solution; 
the fourth column contains results where 
$\sigma_+$ equals the width of the $n=1$ mode 
dynamic solution. 
The data from the third and fourth columns occur around the x's in
figure \ref{fig:sechAsig}.
}
\label{tbl:sechparams}
\end{table}

\begin{figure}
\epsfxsize=12cm
\centerline{\epsffile{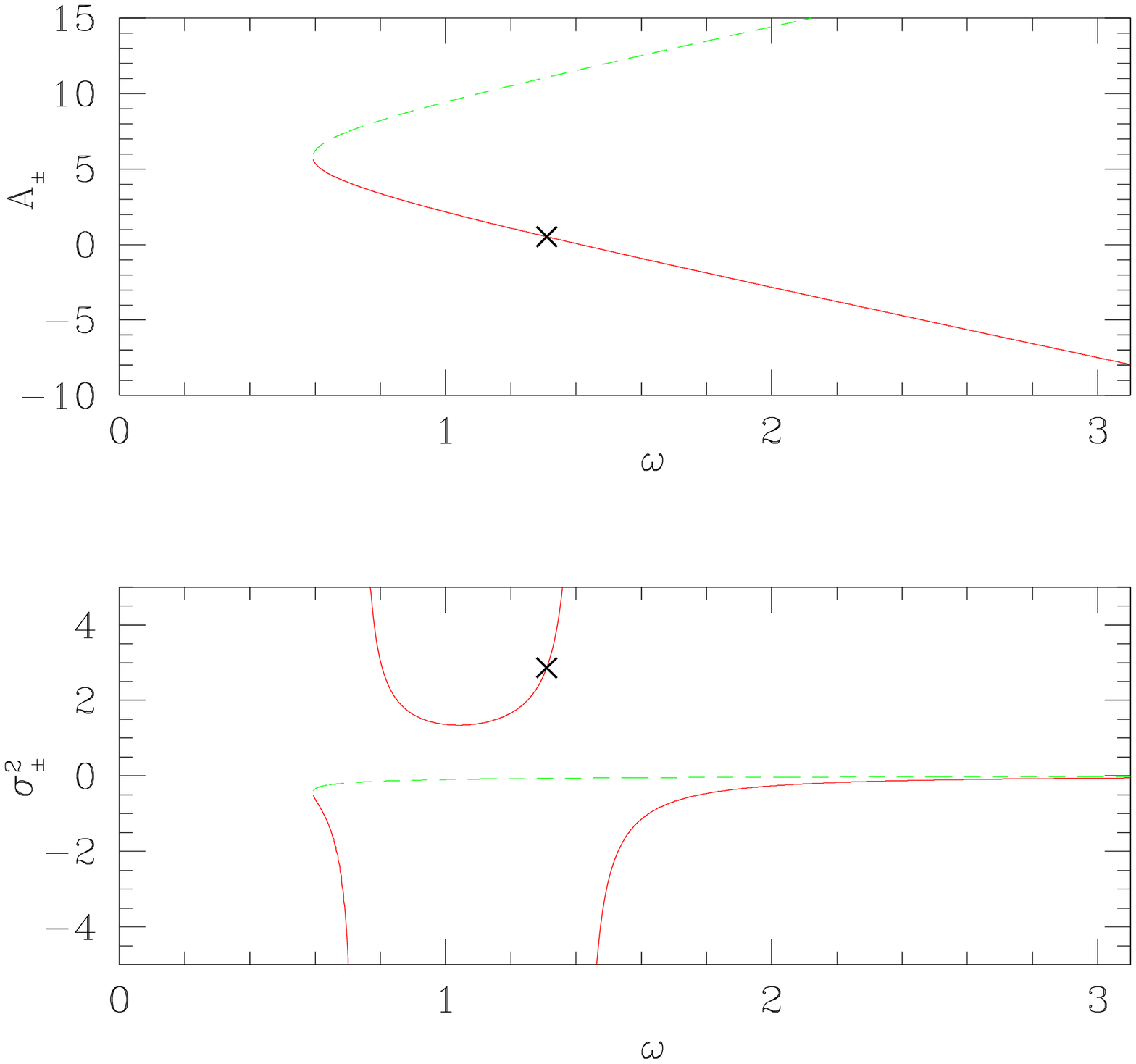}}
\caption[$A_\pm$ and $\sigma^2_\pm$ as a function of $\omega$ for sech $f(r/\sigma)$ ]
{\small \label{fig:sechAsig}
Plot of $A_\pm$ (top) and $\sigma^2_\pm$ (bottom) as a function of $\omega$
for hyperbolic secant $f(r/\sigma)$.
The ``+'' roots are plotted in dashed green, while the ``--'' roots 
are plotted in solid red.
The solution is not plotted for $\omega < 0.595$ as the discriminant in 
equation (\ref{eq:trialA}) is negative and only real $A$ are of interest.
The ``+'' roots are all unphysical since  $\sigma^2_+$ is less than zero everywhere,
which means that solutions within this region have an imaginary width.
The solution for $\sigma^2_-$ is also unphysical {\it outside} the region 
$0.73 < \omega <1.41$ for the same reason.
Fortunately, there are solutions that are both physical (real roots) 
and similar to those seen in the dynamic simulations (near the x's).
}
\label{fig:sechAsig}
\end{figure} \noindent

\subsection{Hyperbolic Secant (Squared) Trial Functions\label{sec:sechtri}}

Lastly, in this subsection the field configuration is assumed to have a hyperbolic
secant squared shape,
\begin{equation}
\displaystyle{
f\left(\frac{r}{\sigma}\right) = {\rm sech}^2\left(\frac{r}{\sigma}\right)
}.
\label{eq:sech2}
\end{equation}
With the new choice of $f(r/\sigma)$, the integrals (\ref{eq:spaceintegrals}) 
are now reevaluated:
\begin{equation}
\begin{array}{rclcrcl}
\vbox{\vspace{0.75cm}}
c_1 &=& \displaystyle{\frac{2}{45}\pi^2 } & \hbox{\hspace{1cm}} &
c_2 &=& \displaystyle{\frac{1}{18}\pi^2 -\frac{1}{3} }
\\
\vbox{\vspace{0.75cm}}
c_3 &=& 0.1053157513 & &
c_4 &=& \displaystyle{\frac{4}{105}\pi^2 - \frac{14}{45}}.
\label{eq:SECH2integrals}
\end{array}
\end{equation}
Using equations (\ref{eq:trialA}) and (\ref{eq:trialsig2}), the new
$A_\pm$ and $\sigma^2_\pm$ are plotted as a function of $\omega$
(figure \ref{fig:sech2Asig}).
Complex solutions exist for $\omega<0.530$ where 
the discriminant in (\ref{eq:trialA}) is negative.
The ${\rm sech}^2$ function also yields unphysical solutions for all of the
``--'' roots since $\sigma^2_-$ is negative everywhere.
Again, there is a part of the ``+'' solution ($0.69<\omega<1.41$) 
that has positive $\sigma^2_+$ and can therefore support physical 
solutions.  Table \ref{tbl:sech2params} compares sample solutions
(located near the x's in figure \ref{fig:sech2Asig}) to estimates obtained 
from the dynamic solution.

\begin{table}
\centerline{
\begin{tabular}{|c||c||c|c|}
\hline
             & Dynamic  &         &       \\ 
Parameter    & Solution & $A_+=A_D$& $\sigma_+=\sigma_D$  \\
\hline 
$A$             & 0.62  & 0.62 	  & 0.36   \\
$\sigma$        & 3.4   & 2.5     & 3.4    \\
$\omega$        & 1.4   & 1.28    & 1.34   \\
\hline
\end{tabular}
}
\caption[Estimated and determined values of $A$, $\sigma$, and $\omega$ (sech squared)]{
\small\label{tbl:sech2params}
Table of free parameter values ($A$, $\sigma$, and $\omega$) 
determined using a ${\rm sech}^2$ trial function (third and fourth columns)
compared to the  
values estimated from the spherically symmetric 
dynamics (second column); the ${\rm D}$ subscript indicates
the value was obtained from the dynamic solution.
The third column contains results obtained using the 
trial function approach  
where $A_+$ equals the amplitude of the $n=1$ mode 
dynamic solution; 
the fourth column contains results where 
$\sigma_+$ equals the width of the $n=1$ mode 
dynamic solution. 
The data from the third and fourth columns occur around the x's in
figure \ref{fig:sech2Asig}.
}
\label{tbl:sech2params}
\end{table}

\begin{figure}
\epsfxsize=12cm
\centerline{\epsffile{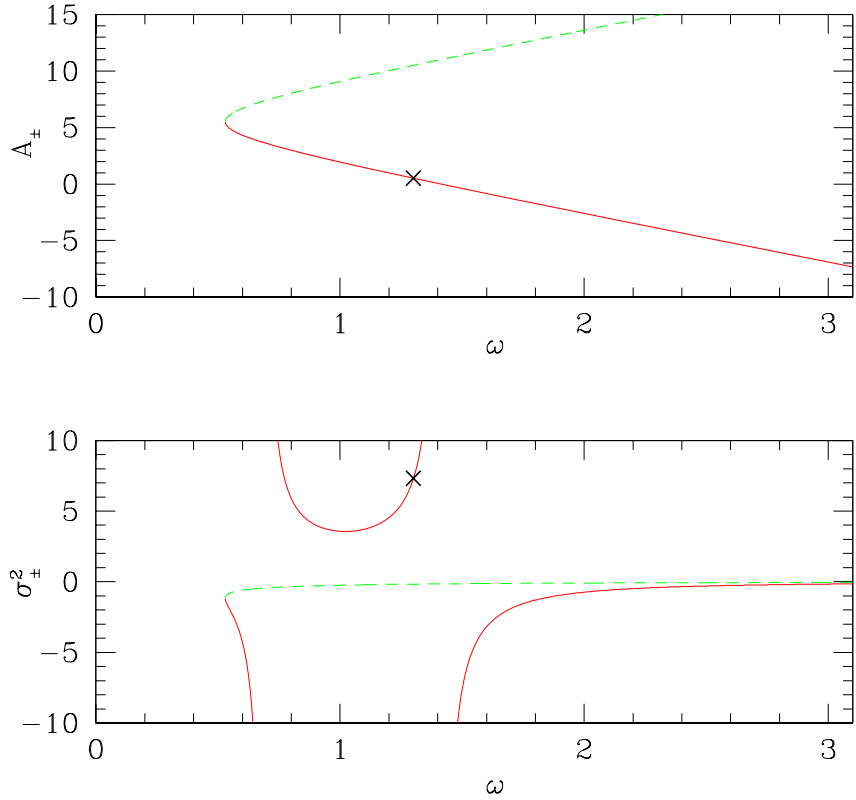}}
\caption[$A_\pm$ and $\sigma^2_\pm$ as a function of $\omega$ for ${\rm sech^2}$ 
$f(r/\sigma)$ ]
{\small \label{fig:sech2Asig}
Plot of $A_\pm$ (top) and $\sigma^2_\pm$ (bottom) as a function of $\omega$
for hyperbolic secant squared $f(r/\sigma)$.
The ``+'' roots are plotted in dashed green, while the ``--'' roots 
are plotted in solid red.
The solution is not plotted for $\omega < 0.530$ as the discriminant in 
equation (\ref{eq:trialA}) is negative and only real $A$ are of interest.
The ``+'' roots are all unphysical since  $\sigma^2_+$ is less than zero everywhere,
which means that solutions within this region have an imaginary width.
The solution for $\sigma^2_-$ is also unphysical {\it outside} the region 
$0.69 < \omega <1.41$ for the same reason.
Fortunately, there are solutions that are both physical (real roots) 
and similar to those seen in the dynamic simulations (near the x's).
}
\label{fig:sech2Asig}
\end{figure} \noindent


\chapter{Conclusions and Future Work}  

Despite its simple appearance and long history, the real scalar field keeps
revealing useful and fascinating physics to those who continue to explore it.
Scalar field theory (both classical and quantum) is being used throughout the frontiers of 
physics; although this work stresses its application to cosmology, the phenomenology 
discussed is widely applicable to the general theory of phase transitions.

Although oscillons had been discovered in 1977 
\cite{bogolyugskii:1977},\cite{bogolyugskii:1976},
and analyzed in slightly more detail in 1995 \cite{copeland:1995}, 
the limitations in computation and numerical methods available to both 
research teams
prevented thorough exploration\footnote{Of course, limited 
computational resources was more of a factor for Bogolyugskii than 
Copeland {\it et al.}, but the numerical methods that both were employing 
were inefficient.}.
This work presents a new technique to absorb outgoing radiation 
in a stable and non-reflecting manner that works well for 
massive fields.
This numerical method allows for the equations of motion
to be evolved on a (relatively) small static lattice, making evolutions
that previously required computational resources proportional to 
the {\it square} of the oscillon lifetime need only computational resources 
{\it directly proportional} to the oscillon lifetime. 
This efficiency is achieved with the combination of solving the 
system of equations in MIB coordinates while including higher order
dissipation.  
In MIB cooridinates outgoing (and ingoing) characteristic velocities go to 
zero near the outer edge of the computational domain so
the outgoing  waves are compressed to nearly the Nyquist limit
on the lattice where they are then quenched by dissipation.
The technique proves to be stable and free of contamination in both one and 
two dimensional problems.

Following the more recent work of Copeland {\it et al.} 
\cite{copeland:1995}, but with the aid of this new numerical 
method, this work reveals many new properties of oscillons
(discussed in chapter \ref{chap:1D}).
In spherical symmetry and using the symmetric double-well potential, 
the method was integral to the discovery of a new resonant phenomenon.
Given the long lifetimes of oscillons (particularly
resonant oscillons) the MIB coordinate evolutions are as much
as one hundred times more computationally efficient than those
previously employed.
The resonances display a time scaling law and reveal 
(on the threshold of successive shape mode modulations) 
a new type of non-radiative oscillon solution.
This solution's existence and form is verified by solving a system 
of ordinary differential equations obtained from a periodic 
non-radiative ansatz.
The method is also used to explore the threshold of expanding
bubble formation for bubble collapse with an asymmetric double-well
potential.  A time scaling law is shown to exist with 
a scaling exponent, $\gamma\approx 2.1$.

Chapter \ref{chap:2D} discusses the use of MIB coordinates in two 
dimensional axisymmetric bubble {\it collisions}.  Evolutions
are observed that result in expanding true-vacuum bubble formation, 
annihilation, coalescence, or soliton-like transmission.
By looking at a two dimensional parameter space, (collision)
boost velocity versus bubble width, the threshold of 
expanding bubble formation is again explored.
The threshold reveals a time scaling law with a scaling
exponent ($\gamma\approx 2.2$) similar to the spherically symmetric collapse
($\gamma\approx 2.1$).
This scaling law appears to be universal in that the scaling 
exponents observed are independent both of the values of the 
critical parameters ($\sigma^*_r$ and $v^*_b$) 
{\it and} which parameter is varied to explore the threshold
(ie. $T = \gamma\ln|p-p^*|$ for $p$ equal to $\sigma$ or $v_B$).
The similarity between the scaling parameters of the spherically
symmetric collapse and the axisymmetric collisions suggests
that the dominant unstable mode (responsible for the critical
phenomena observed) in the axisymmetric collisions might be spherically
symmetric.
In addition to more critical phenomena, the parameter space
surveys reveal that there are many {\it collision-induced}
expanding bubbles.   In other words, there are many bubble configurations
that would otherwise have formed oscillons and eventually dispersed,
but when colliding with other bubbles combine to form an expanding 
vacuum bubble.  
In a system where bubble
collisions might be occurring, nucleation rates for expanding bubbles
will be higher than those originally believed to arise solely 
from thermal (or quantum) fluctuations.

Chapter \ref{chap:trial} discusses oscillon solutions obtained using 
trial function methods.
Judicious choices for trial functions allow the complete integration
of the action; resulting in a system of algebraic equations instead of 
differential equations.
Although the solutions obtained are not solutions to the 
true equations of motion, the predicted values for the amplitude,
frequency, and width of the resonant oscillons are in reasonable
agreement with observation.
Furthermore, since the non-radiative resonant solution discussed in 
chapter \ref{chap:1D} was not known before this work, the mere existence 
of the non-radiative oscillon solution obtained in such a straightforward 
manner was exciting.

Although there is certainly more to be discovered within the nonlinear
Klein-Gordon equation, it would be interesting to see if similar physics
can be found in other models.
It would be interesting to see how the stability of the oscillon solutions
are affected by the coupling to gravity (GR).
The extra gravitational attraction might make oscillons more stable, however,
the extra nonlinearities introduced might have entirely the opposite effect.
It would also be interesting to see if similar phenomena are observable in 
the charged scalar field, both alone and coupled to electromagnetism 
(the charged scalar field with a SDWP coupled to electromagnetism is a simple 
model of superconductivity).
Lastly, the numerical methods used here might be applicable to the solution
of other nonlinear partial differential equations.  

\appendix
\chapter{Will Cosmological Oscillons Form Black Holes?\label{app:A}}

One of the most interesting applications of oscillons
is to cosmological phase transitions resulting from 
spontaneously broken gauge symmetries. 
But whenever one discusses the dynamics of massive objects in a 
cosmological context, two questions naturally arise:  
{\it How large are these objects?} and 
{\it How dense are these objects?}. 
If the object is {\it cosmologically} large, then gravitational 
effects might need to be included since the expansion of the universe
is measurable on the length-scale of the object being studied.
If the object is very dense, then there will be gravitational 
self-interaction that will likely change the dynamics of the system.
This appendix answers these two questions and shows that oscillons
{\bf do not} need to be modeled with general relativistic effects
included.

The first thing to remember in this calculation is that 
(in contradiction to most relativist's conventions), this
thesis uses typical high energy physics units where 
$\hbar = c = 1$, which implies that energy can be 
treated as the fundamental dimension:
\begin{equation}
[{\rm Energy}] = [{\rm Mass}] = [{\rm Length}]^{-1} = [{\rm Time}]^{-1} 
\end{equation}
Two useful conversion factors are $1 GeV = 1.7827\times 10^{-24}g$ 
and $1 GeV^{-1} = 1.9733\times 10^{-14}cm$.

This thesis almost exclusively uses dimensionless units, obtained
by taking
$r=\rho\, m_H$, $t=\tau\, m_H$, and 
$\displaystyle{\phi = \frac{\sqrt{\lambda}}{m_H}\, \psi}$,
where $m_H$ is the mass in the model (the Higgs mass)
and $\rho$, $t$, and $\psi$ are the dimensionful length, time, and
scalar field, respectively\footnote{Also remember $\lambda$ is 
always dimensionless.}.
Converting back to dimensionful quantities, the typical 
oscillon radius and mass/energy are $r\approx 2.5m_{H}^{-1}$ and 
$\displaystyle{M\approx 43 \frac{m_{H}}{\lambda}}$, respectively.
Assuming the Higgs mass is known in GeV, 
converting the oscillon radius radius to $cm$ is performed by
\begin{equation}
\begin{array}{rcl}
r &=& 2.5 m_H^{-1}\\
  &=& \displaystyle{
\left(\frac{2.5}{m_H}\right)
\left(\frac{1.97\times 10^{-14} GeV\, cm}{1}\right)
}\\
  &=& \displaystyle{
\left(\frac{4.93\times 10^{-14} GeV\, cm}{m_H}\right)
},
\end{array}
\end{equation}
so that for an ``electroweak'' oscillon where the Higgs mass is
$m_H \approx 100 GeV$, $r\approx 4.93\times 10^{-16} cm$,
or for a GUT oscillon with $m_H \approx 10^{15} GeV$,
$r\approx 4.93\times 10^{-29} cm$.
It is clear that these are not cosmological in their size, but what about
their densities?

Since these ``particles'' are {\it so} small, it might be inappropriate to even 
consider their classical gravitational Schwarzchild radius, but for an 
oscillon with $M=43 m_{H}$ (taking $\lambda=1$),
the electroweak oscillons have $M\approx 8\times 10^{-21} g$ 
while a GUT oscillon would have $M\approx 8\times 10^{-8}g$.
These oscillons then would have Schwarzchild radii of
$\displaystyle{r_S = 1.13 \times 10^{-50}cm\, \left(\frac{m_H}{GeV}\right)}$,
or $r_S \approx 10^{-48} cm$ and $r_S \approx 10^{-35} cm$, respectively.

So, although we do not know exactly how to describe all the forces
of nature acting together at GUT energy scales, 
the point to be made here is that {\it if} the nonlinear 
Klein-Gordon action is {\it minimally coupled} to gravity, oscillons do not have 
a radius comparable to their Schwarzchild radius until the Higgs mass
approaches $m_H\approx 10^{18} GeV$.

\begin{table}
\centerline{
\begin{tabular}{c|c}
Oscillons & Formula\\
Attribute & (Dependence on $m_H$) \\
\hline
Radius &
\vbox{\vspace{0.75cm}} 
$\displaystyle{r = 4.9\times 10^{-14} cm  \left(\frac{GeV}{m_H}\right)}$\\
Lifetime &
\vbox{\vspace{0.75cm}} 
$\displaystyle{T = 3.3\times 10^{-21} s  \left(\frac{GeV}{m_H}\right)}$\\
Mass   &
\vbox{\vspace{0.75cm}} 
$\displaystyle{M = 7.7\times 10^{-23} g  \left(\frac{m_H}{GeV}\right)}$\\
\vbox{\vspace{0.75cm}} 
Schwarzschild Radius   &
$\displaystyle{r_S = 1.1 \times 10^{-50}cm\, \left(\frac{m_H}{GeV}\right)}$\\
\hline
\end{tabular}
}
\caption[Table of typical oscillon attributes]{
\small \label{tbl:attribues}
Table displaying how to determine typical oscillon attributes them 
as a function of the Higgs mass.
The formulae were derived using typical values (in the dimensionless coordinates 
used throughout this thesis) of $r=2.5$, $T=5000$, and $M=43$, for the oscillon
radius, lifetime, and mass, respectively.
}
\label{tbl:attributes}
\end{table}


\begin{thesisauthorvita}             
Ethan Philip Honda was born in Chicago, IL on December 16, 1974,
the son of Sheila and Edward Honda.
After completing high school at the Illinois Math and Science Academy,
Aurora, IL, in 1992 he entered Loyola University Chicago as a freshman.
He transfered to MIT during the fall of 1993 and received a S.B. in 
physics in 1996.
In the fall of 1996, he entered the graduate school at the University of
Texas at Austin, where he began studying numerical relativity under
the supervision of Matthew Choptuik.
As an adjunct faculty member from 1997 to 2000, he lectured introductory physics classes 
at Austin Community College.
While in graduate school he was engaged to Patricia Eileen Stafford, who he
will be marrying in the spring of 2001.
After graduating he will be working as a postdoctoral fellow
with the Signal Physics Group
at the Applied Research Laboratories in Austin. 
\end{thesisauthorvita}               

\begin{thebibliography}{..}          
\def\NKB{{\it Nucl. Phys. B\ }}


\bibitem{anninos:1991} P.~Anninos, S.~Oliveira, and R.A.~Matzner,
\PRD {\bf 44}, 1147 (1991).

\bibitem{arnowitt:1962} R.~Arnowitt, S.~Deser, and C.W.~Misner,
      ``The dynamics of general relativity'', {\it Gravitation:
       An introduction to current research}, edited by L.~Witten (John Wiley,
       New York, 1962).

\bibitem{balakrishna:1997} J.~Balakrishna, E.~Seidel, and W-M.~Suen,
\PRD {\bf D58}, 104004 (1998).

\bibitem{bishop:1996} N.T.~Bishop, R.~Gomez,
P.R.~Holvorcem, R.A.~Matzner, P.~Papadopoulos, and J.~Winicour,
\PRL {\bf 76}, 4303-4306 (1996). 


\bibitem{bishop2:1996} N.T.~Bishop, R.~Gomez,L.~Lehner, M.~Maharaj,
and J.~Winicour,
\PRD {\bf 60}, 024005 (1999).

\bibitem{bogolyugskii:1976} I.L.~Bogolyugskii, and V.G.~Makhan'kov,
{\em JETP Letters} {\bf 24}, 12 (1976).

\bibitem{bogolyugskii:1977} I.L.~Bogolyugskii, and V.G.~Makhan'kov,
{\em JETP Letters} {\bf 25}, 107 (1977).


\bibitem{brady:1997}  P.R.~Brady, C.M.~Chambers, and 
S.M.C.V.~Goncalves,
\PRD {\bf 56}, 6057--6061 (1997).

\bibitem{brady2:1997} P.R.~Brady and M.J.~Cai,
``Critical phenomena in gravitational collapse'', 
{\it Proceedings of 8th Marcel Grossman Conference}  (World Scientific, Singapore).

\bibitem{campbell:1983} D.K.~Campbell, J.F.~Schonfeld, 
and Charles A. Wingate, {\em Physica} {\bf 9D}, 1-32 (1983). 

\bibitem{chambers:1997} C.M.~Chambers, P.R.~Brady, and 
S.M.C.V.~Goncalves,
``A Critical Look At Massive Scalar Field Collapse''
{\it Proceedings of 8th Marcel Grossman Conference}  (World Scientific, Singapore).

\bibitem{choptuik:1986}  M.W.~Choptuik, 
Ph.D.~Dissertation, The University of British Columbia, 1986.

\bibitem{choptuik:1991} M.W.~Choptuik, 
\PRD {\bf 44}, 3124--3135 (1991).

\bibitem{choptuik:1993} M.W.~Choptuik, 
      \PRL {\bf 70}, 9--12 (1993).

\bibitem{choptuik:1996} M.W.Choptuik, T.~Chmaj, and P.~Bizon,
\PRL {\bf 77}, 424 (1996).

\bibitem{choptuik:1998a} M.W.~Choptuik, ``The 3+1 Einstein equations'', 
        unpublished (1998).

\bibitem{choptuik:1998b} M.W.~Choptuik,
``Finite Difference Methods'',
lecture notes, (1998).

\bibitem{choptuik:1998} M.W.~Choptuik, 
{\tt gr-gc/9803075} (1998).

\bibitem{choptuik:1999} M.W.~Choptuik, 
``Lectures for Taller de Verano 1999 de FENOMEC''
lecture notes, (1999).

\bibitem{choptuik:2000} M.W.~Choptuik, 
personal communication (2000).

\bibitem{coleman:1977} S.~Coleman, 
\PRD {\bf 15}, 2929--2936 (1977).

\bibitem{coleman:1985} S.~Coleman, 
{\it Aspects of symmetry: selected Erice lectures of 
Sidney Coleman}
(Cambridge University Press, Cambridge, 1985).

\bibitem{copeland:1989} E.J.~Copeland, E.W.~Kolb, and K.~Lee,
\NKB {\bf B319}, 501--510 (1989).

\bibitem{copeland:1995} E.J.~Copeland, M.~Gleiser, and H.R.~M\"uller,
\PRD {\bf 52}, 1920--1932 (1995).

\bibitem{courant:1962} R.~Courant and d.~Hilbert,
{\it Methods of Mathematical Physics Volume II}
(Wiley and Sons, New York, 1962).

\bibitem{derrick:1964} G.H.~Derrick,
\JMP {\bf 5}, 1252--1254 (1964).

\bibitem{fetter:1982} A.L.~Fetter and J.D.Walecka,
{\it Quantum Theory of Many-Particle Systems} 
(McGraw-Hill, New York, 1971).

\bibitem{frauendiener:1998a} J.~Frauendiener,
\PRD {\bf 58}, 064002 (1998).

\bibitem{frauendiener:1998b} J.~Frauendiener,
\PRD {\bf 58}, 064003 (1998).

\bibitem{frauendiener:2000}  J.~Frauendiener,
\CQG {\bf 17}, 373-387 (2000).

\bibitem{friedberg:1976} R.~Friedberg, T.D.~Lee, and A.~Sirlin,
\PRD {\bf 13}, 2739--2761 (1976). 

\bibitem{friedberg:1987} R.~Friedberg, T.D.~Lee, and Y.~Pang,
\PRD {\bf 35}, 3658--3677 (1987).

\bibitem{frieman:1989} J.~Frieman, A.V.~Olinto, M.~Gleiser, and
C.~Alcock,
\PRD {\bf 40}, 3241--3251 (1989).

\bibitem{friedrich:1998} H.~Friedrich, 
``15th International Conference on General Relativity
and Gravitation (GR15) Proceedings'' (1997).

\bibitem{gelmini:1994} G.~Gelmini and M.~Gleiser,
\NKB {\bf B419} 129--146 (1994). 

\bibitem{ginzburg:1950} V.N.~Ginzburg and L.D.~Landau,
{\it Zh. Exsp. Teor. Fiz 20} {\bf 20}, 1064 (1950).


\bibitem{gleiser:1991} M.~Gleiser, E.W.~Kolb, and R.~Watkins,
\NKB {\bf B364}, 411--450 (1991). 

\bibitem{gleiser:1996} M.~Gleiser, R.~Haas,
\PRD {\bf 54}, 1626--1632 (1996).

\bibitem{gleiser:1999} M.~Gleiser and A.~Sornborger,  
{\tt patt-sol/9909002} (1999).

\bibitem{gomez:1992} R.~Gomez, J.~Winicour, and R.~Isaacson,
\JCP {\bf 98}, 11--25 (1992).

\bibitem{gomez:1994} R.~Gomez, P.~Papdopoulos, and J.~Winicour,
\JMP {\bf 35}, 4184--4204 (1994).

\bibitem{griffiths:1987} D.J.~Griffiths,
{\it Introduction to elementary particles}
(Harper \& Row, New York, 1987).

\bibitem{hawking:1973} S.W.~Hawking and G.~Ellis, {\em
The Large Scale Structure of Space-Time}, Cambridge University
Press, Cambridge, 1973.

\bibitem{hubner:1996} P.~H\"ubner,
\PRD {\bf 53}, 701--721 (1996).

\bibitem{kirzhnits:1972} D.A.~Kirzhnits and A.D.~Linde,
\PRL {\bf 42B}, 471 (1972).

\bibitem{krumhansl:1975} J.A.~Krumhansl and J.R.~Schrieffer, 
{\it Phys.\ Rev. B} {\bf 11}, 3535 (1975).

\bibitem{kolb:1990} E.W.~Kolb and M.S.~Turner,
{\it The early universe}
(Addison-Wesley, Reading, MA, 1990).

\bibitem{kreiss:1973} H.O.~Kreiss and J.~Oliger, {\it Methods for the
        approximate solution of time-dependent problems,} GARP
        Publication Series No.~10 (World Meteorological Organization,
        Geneva, 1973).

\bibitem{lee:1981} T.D.~Lee and G.C.~Wick,
\PRD {\bf 9}, 2291--2316 (1974).

\bibitem{lee2:1981} T.D.~Lee,
{\it Particle Physics and introduction to field theory}
(Harwood Academic Publishers, New York, 1981).

\bibitem{lehner:2000} L.~Lehner,
personal communication (2000).

\bibitem{linde:1974} A.D.~Linde,
{\it JETP Lett.} {\bf 19}, 320 (1974).

\bibitem{marsa:1996} R.L.~Marsa and M.W.~Choptuik, 
\PRD {\bf 54}, 4929--4943 (1996).

\bibitem{matzner:2000} R.A.~Matzner and E.~Bonning, 
personal communication (2000).

\bibitem{misner:1973} C.W.~Misner, K.S.~Thorne, and J.A.~Wheeler,
         {\it Gravitation} (W.H.~Freeman, San Francisco, 1973).

\bibitem{morrison:1999} P.J.~Morrison,
personal communication (2000).

\bibitem{penrose:1963} R.~Penrose,
\PRL {\bf 10}, 66 (1963).

\bibitem{rajaraman:1982} R.~Rajaraman, {\it Solitons and instantons}
(North-Holland Pub. Co., Amsterdam, 1982).

\bibitem{richardson:1910} L.F.~Richardson,
{\it Phil. Trans. Roy. Soc.} {\bf 210}, 307--357 (1910).

\bibitem{riotto:1995} A.~Riotto,
\PL {\bf B365}, 64 (1996).

\bibitem{ryder:1996} L.H.~Ryder,
{\it Quantum Field Theory}
(Cambridge University Press, Cambridge, 1996). 

\bibitem{seidel:1986} E.~Seidel and W-M.~Suen, 
\PRL {\bf 57}, 2485 (1986).

\bibitem{seidel:1991} E.~Seidel and W-M.~Suen, 
\PRL {\bf 66}, 1659 (1991).

\bibitem{thomas:1995} J.W.~Thomas, {\it Numerical Partial Differential Equations:
Finite Difference Methods} (Spinger--Verlag, New York, 1995).

\bibitem{umurhan:1998} O.M.~Umurhan, L.~Tao, and E.A.~Spiegel,
``Stellar Oscillons'', {\tt astro-ph/9806209} (1998).

\bibitem{wald:1984} R.~Wald, {\it General Relativity} (The 
University of Chicago Press, Chicago, 1984).

\bibitem{weinberg:1967} S.~Weinberg, \PRL {\bf 19}, 1264 (1967).

\bibitem{weinberg:1972} S.~Weinberg, {\it Gravitation and cosmology: 
      Principles and applications of the general theory of relativity}
      (John Wiley \& Sons, New York, 1972).


\bibitem{winicour:1988} J.~Winicour, 
\JMP {\bf 29}, 2117--2121 (1988).

\bibitem{winicour:1999} J.~Winicour, 
personal communication (1999).

\end{thebibliography}
\end{document}